\documentclass[usenatbib]{mn2e}
\usepackage{amsmath}
\usepackage{natbib}
\usepackage{epsfig}
\usepackage{txfonts}
\usepackage{pict2e}
\usepackage{color}

\newcommand{\UnitInu}{{\rm W\,m^{-2}\,Hz^{-1}\,sr^{-1}}}
\newcommand{\Mpc}{{\rm Mpc}}
\newcommand{\GHz}{{\rm GHz}}

\newcommand{\expf}[1]{{{\rm e}^{#1}}}

\newcommand{\zmu}{{z_{\mu}}}

\newcommand{\nS}{n_{\rm S}}

\newcommand{\nrun}{n_{\rm run}}

\newcommand{\kD}{k_{\rm D}}

\newcommand{\id}{{\,\rm d}}

\newcommand{\beq}{\begin{equation}}   %

\newcommand{\eeq}{\end{equation}}   %

\newcommand{\beqa}{\begin{eqnarray}}   %

\newcommand{\eeqa}{\end{eqnarray}}   %

\newcommand{\beal}{\begin{align}}

\newcommand{\enal}{\end{align}}

\newcommand{\bspl}{\begin{split}}

\newcommand{\espl}{\end{split}}

\newcommand{\bsub}{\begin{subequations}}

\newcommand{\esub}{\end{subequations}}

\newcommand{\bmulti}{\begin{multline}}   %

\newcommand{\beqm}{\begin{mathletters}}   %

\newcommand{\eeqm}{\end{mathletters}}   %

\newcommand{\vek} [1]{\mbox{\boldmath${#1}$\unboldmath}}

\newcommand{\pot}[2]{#1 \times 10^{#2}}
\newcommand{\poterr}[3]{(#1\pm#2)\times10^{#3}}

\definecolor{redwine}{rgb}{.72,0,0}

\usepackage{hyperref}
\usepackage{grffile}
\usepackage{graphics}
\usepackage{booktabs}

\title[Distortion PCA]
{Teasing bits of information out of the CMB energy spectrum}

\author[Chluba and Jeong]{Jens~Chluba$^{1}$\thanks{E-mail:jchluba@pha.jhu.edu} and Donghui Jeong$^{1}$\thanks{E-mail:djeong@pha.jhu.edu}
\\
$^{1}$ Department of Physics and Astronomy, Johns Hopkins University, Bloomberg Center 435, 
3400 N. Charles St., Baltimore, MD 21218
}

\begin{document}

\date{{Accepted 2013 November 30. Received 2013 June 24}}

\maketitle

\begin{abstract}
Departures of the Cosmic Microwave Background (CMB) frequency spectrum from a blackbody -- commonly referred to as spectral distortions -- encode information about the thermal history of the early Universe (redshift $z\lesssim \pot{\rm few}{6}$). While the signal is usually characterized as $\mu$- and $y$-type distortion, a smaller residual (non-$y$/non-$\mu$) distortion can also be created at intermediate redshifts $10^4\lesssim z\lesssim \pot{3}{5}$. Here, we construct a new set of observables, $\mu_k$, that describes the principal components of this residual distortion. The principal components are orthogonal to temperature shift, $y$- and $\mu$-type distortion, and ranked by their detectability, thereby delivering a compression of all valuable information offered by the CMB spectrum. This method provides an efficient way of analyzing the spectral distortion for given experimental settings, and can be applied to a wide range of energy-release scenarios. As an illustration, we discuss the analysis of the spectral distortion signatures caused by dissipation of small-scale acoustic waves and decaying/annihilating particles for a {\it PIXIE}-type experiments. 
We provide forecasts for the expected measurement uncertainties of model parameters and detections limits in each case.
We furthermore show that a {\it PIXIE}-type experiments can in principle distinguish dissipative energy release from particle decays for a nearly scale-invariant primordial power spectrum with small running.
Future CMB spectroscopy thus offers a unique probe of physical processes in the primordial Universe.
\end{abstract}


\section{Introduction}
\label{sec:Intro}
Energy release in the early Universe causes deviations of the cosmic microwave background (CMB) frequency spectrum from a pure blackbody \citep{Zeldovich1969, Sunyaev1970mu, Illarionov1975, Illarionov1975b, Danese1977, Burigana1991, Hu1993}, which we henceforth refer to as spectral distortion (SD). Thus, far no primordial SD was found \citep{Mather1994, Fixsen1996,Fixsen2002, Kogut2006ARCADE, tris1, arcade2}, but technological advances over the past quarter-century since {\it COBE}/{FIRAS} may soon allow much more precise (at least 3 orders of magnitudes improvement in sensitivity) characterization of the CMB spectrum \citep[e.g.,][]{Fixsen2002, Kogut2011PIXIE}. 
This is especially interesting because even for the standard cosmological model, several processes exist that imprint distortion signals at a level within reach of present-day technology \citep[see][for broader overview]{Chluba2011therm, Sunyaev2013, Chluba2013fore}.
{\it PIXIE} \citep{Kogut2011PIXIE} provides one very promising experimental concept for measuring these distortion signals, and more recently {\it PRISM}, an L-class satellite mission with about 10 times the spectral sensitivity of {\it PIXIE}, was put forward \citep{PRISM2013WP}.
These prospects motivated us to further elaborate on what could be learned from measurements of the CMB spectrum, taking another step forward towards the analysis of future distortion data.

Previous works primarily used distortions to rule out various energy-release scenarios (ERSs) on a model-by-model basis. These studies include discussion of {\it decaying} or {\it annihilating particles} \citep{Hu1993b, McDonald2001}, the {\it dissipation of primordial density fluctuations on small scales} \citep{Daly1991, Barrow1991, Hu1994, Hu1994isocurv, Chluba2012, Pajer2012, Dent2012, Ganc2012, Chluba2012inflaton, Powell2012, Khatri2013forecast, Chluba2013iso, Biagetti2013}, {\it cosmic strings} \citep{Ostriker1987, Tashiro2012, Tashiro2012b}, {\it primordial black holes} \citep{Carr2010}, {\it small-scale magnetic fields} \citep{Jedamzik2000} and some new physics examples \citep{Lochan2012, Bull2013, Brax2013}.

Until recently, all constraints were based on simple estimates for the chemical potential, $\mu$, and Compton $y$-parameter \citep{Zeldovich1969, Sunyaev1970mu}. It was, however, shown that the distortion signature from different ERSs generally is {\it not just} given by a superposition of pure $\mu$- and $y$-distortion \citep{Chluba2011therm, Khatri2012mix, Chluba2013Green}. The small residual beyond $\mu$- and $y$-distortion contains information about the time dependence of the energy-release history, which in principle can be used to directly constrain, for instance, the shape of the small-scale power spectrum, measure the lifetime of decaying relic particles, or simply to discern between different energy-release mechanisms \citep{Chluba2013fore}.
In particular, \citet{Chluba2013fore} demonstrated that CMB spectrum measurement with a {\it PIXIE}-type experiment provide a sensitive probe for long-lived particles with lifetimes $t_{\rm X}\simeq 10^9\,{\rm sec}-10^{10}\,{\rm sec}$. Similarly, the shape of the small-scale power spectrum can be directly probed with {\it PIXIE}'s sensitivity if the amplitude of primordial curvature perturbations exceeds $A_\zeta\simeq \pot{\rm few}{-8}$ at wavenumber $k\simeq 45 \, \Mpc^{-1}$ \citep{Chluba2013fore}. 
Future CMB distortion measurements thus provide a unique avenue for studying early-universe models and particle physics.

In \citet{Chluba2013fore}, model parameters (e.g., abundance and lifetime of a decaying particle) were directly translated into the SD signal (the photon intensity in different frequency channels) using a Green's function method \citep{Chluba2013Green}, which was recently added to the cosmological thermalization code {\sc CosmoTherm}\footnote{Available at \url{www.Chluba.de/CosmoTherm}} \citep{Chluba2011therm}. 
Even when explicitly knowing the relation between ERS and SDs, model comparison and forecasts of uncertainties (or detection limits) are still rather involved. This is because (i) different energy-release mechanisms can cause very similar SDs, (ii) the parameters in different models are often unrelated and (iii) in general, the parameter space is non-linear especially close to the detection limit. One natural question therefore is whether the information contained by the CMB spectrum (the intensity in each frequency channel) could be further compressed and described in a model-independent way ($\mu$, $y$, plus additional distortion parameters).

The precise shape of the resulting SD directly depends on the underlying energy-release history. Model dependence is only introduced when asking which physical process caused a specific energy-release history, but this step can be separated from measuring the energy-release history itself. We thus ask, how well future CMB SDs can constrain different energy-release histories, independent of the responsible physical mechanism. For this we perform a principal component analysis \citep[see][for other cosmology-related applications of this method]{Mortonson2008, Finkbeiner2012, Farhang2011, Shaw2013} of the residual (non-$\mu$/non-$y$) distortion signal, in order to identify spectral shapes and their associated energy-release histories that can be best-constrained by future distortion data.
The amplitudes, $\mu_k$, of the signal eigenmodes then define a set of parameters that describes all information encoded by the residual distortion signal. These observables can be measured in a model-independent manner with predictable uncertainties.
The mode amplitudes, by construction, are uncorrelated and the parameter dependence is linear, which greatly simplifies further analysis in this new parameter space.

The principal components depend on experimental setting (number of channels, distribution over frequency, noise in each channel and its correlations; see Sect.~\ref{sec:instrument}) as well as foregrounds and systematic effects. 
Here, we do not consider the effect of foreground contamination, and therefore only focus on what the minimal instrumental sensitivity should be in order to constrain or detect the signatures of different energy-injection scenarios. 
Generalization is straightforward, but we leave a more detailed investigation of foreground issues to future work.
Along similar lines we plan on investigating the optimization of experimental settings for various ERSs using the principal component analysis.

This paper is organized as follows: we start by decomposing the SD signal into temperature shift, $\mu$, $y$ and residual distortion (Sect.~\ref{sec:ortho_Greens}). This decomposition already depends on the experimental settings (we envision a {\it PIXIE}-like experiment), which determines the level to which different spectral shapes are distinguishable. This allows us to obtain {\it visibility functions} in redshift for the different distortion components (Fig.~\ref{fig:R_z}), providing a generalization of the {\it spectral distortion visibility function}\footnote{This name was coined by \citet{Chluba2011therm}, but the original derivation (accounting for the effect of Bremsstrahlung) was given in \citet{Sunyaev1970mu}. \citet{Danese1982} also included the effect of double Compton emission \citep{Lightman1981, Thorne1981, Chluba2007a}, and recent improvements to the shape of $\mathcal{J}_{\rm bb}(z)$ were given by \citet{Khatri2012b}, using semi-analytic approximations.}, $\mathcal{J}_{\rm bb}(z)$ [see Sect.~\ref{sec:Dr_branch} for more details], used in earlier works to account for the suppression of distortions by the efficient thermalization process at redshift $z\gtrsim \pot{\rm few}{6}$ \citep[e.g.,][]{Burigana1991, Hu1993}.
We then construct the energy-release and signal eigenmodes (Sect.~\ref{sec:R_eigenmodes}), and illustrate how they can be used for simple parameter estimation (Sect.~\ref{sec:estimation}).
In Sect.~\ref{sec:mod}, we demonstrate how constraints on different energy-release scenarios can be derived, with particular attention to detectability, errors, and model comparison. 
%

\section{Quasi-orthogonal decomposition of the thermalization Green's function}
\label{sec:ortho_Greens}
The average CMB frequency spectrum, $I_{\nu}^{\rm CMB}$ ($\equiv$ spectral intensity in units $\UnitInu$ as a function of frequency $\nu$), can be broken down as follows:
\beal
\label{eq:I_CMB}
I_{\nu}^{\rm CMB}=B_\nu(T_0) + \Delta I_\nu^{T} + \Delta I^{y}_{\nu}+\Delta I^{\rm prim}_\nu.
\end{align}
The main theme of this paper is to develop an analysis tool for the primordial, pre-recombination distortion signal, $\Delta I^{\rm prim}_\nu$, introduced by different energy-release mechanisms at early times, $z\gtrsim 10^3$ (see Sect.~\ref{sec:primordial_signal}). Because this term is usually small compared to the other contributions to $I_{\nu}^{\rm CMB}$, we seek a scheme to optimize the search for this signal.
The first term in Eq.~\eqref{eq:I_CMB} describes the CMB blackbody part, $B_\nu(T_0)=\frac{2h\nu^3}{c^2}/(\expf{x}-1)$, where $T_0$ is the CMB monopole temperature $T_0=(2.726\pm0.001)\,{\rm K}$ \citep{Fixsen1996, Fixsen2009} and $x\equiv h\nu/k T_0$. The {\it exact} value of the CMB monopole temperature, $T$, is not known down to the accuracy that can be reached by future experiments ($\Delta T\simeq {\rm few}\times{\rm nK}$). It thus has to be determined in the analysis. This  is captured by the second term in Eq.~\eqref{eq:I_CMB}, which is obtained by shifting a blackbody from one temperature $T_0$ to $T$, causing a signal
\beal
\label{eq:DI_T}
&\Delta I_\nu^{T}= G_{T}(\nu)\,\Delta_{T} [1+\Delta_{T}]+Y_{\rm SZ}(\nu)\,\Delta_{T}^2/2 +\mathcal{O}(\Delta^3_T),
\end{align}
where $\Delta_T=(T-T_0)/T_0\ll 1$. Here, we defined the spectrum of a temperature shift $G_{T}(\nu)=[T\,\partial_T B_\nu(T)]|_{T=T_0}\equiv \frac{2h\nu^3}{c^2} \frac{x\expf{x}}{(\expf{x}-1)^2}$ at lowest order in $\Delta_T$. At second order in $\Delta_T$, a correction related to the superposition of blackbodies \citep{Zeldovich1972, Chluba2004} appears, having a spectrum that is similar to a Compton $y$-distortion, $Y_{\rm SZ}(\nu)\equiv G_{\rm T}\left[x\coth(x/2)-4\right]$, also known in connection with the thermal Sunyaev-Zeldovich effect caused by galaxy clusters \citep{Zeldovich1969}.

Finally, in Eq.~\eqref{eq:I_CMB} we also added a $y$-distortion, $\Delta I^{y}_{\nu}=y\,Y_{\rm SZ}(\nu)$, that is created at low redshifts ($z\lesssim10^3$) but is not directly accounted for by the primordial distortion, $\Delta I^{\rm prim}_\nu$.
One strong source of late-time $y$-distortions stems from reionization and structure formation, giving rise to an effective $y$-parameter $y_{\rm re}\simeq 10^{-7}-10^{-6}$ \citep{Sunyaev1972b, Hu1994pert, Cen1999, Miniati2000, Refregier2000, Oh2003, Zhang2004}. 
The aim of this section is to find an operational decomposition of the spectral signal caused by early energy release ($z\gtrsim 10^3$) from the non-primordial signatures such as temperature shift and late-time $y$-distortion.

\subsection{Instrumental aspects}
\label{sec:instrument}
For our analysis we envision an experiment similar to {\it PIXIE}, which is based on a Fourier transform spectrometer \citep{Kogut2011PIXIE}. {\it PIXIE} covers the frequency range $\nu=30\,\GHz-6\,{\rm THz}$, with synthesized channels of constant frequency resolution $\Delta \nu_{\rm c}=15\,\GHz$, depending on the mirror stroke\footnote{Excursion of the modulating (dihedral) mirror of the Fourier transform spectrometer around the zero-point.}. The noise in each channel over the mission's duration is $\Delta I_{\rm c}\simeq \pot{5}{-26}\,\UnitInu$. We assume the noise to be constant and uncorrelated between channels (diagonal covariance matrix $C_{ij}=\Delta I_{\rm c}^2\,\delta_{ij}$), with bandpass given by top-hat functions. 
The SD signal we are after is important only at $\nu\simeq 30\,\GHz-1\,{\rm THz}$, which for $\Delta \nu_{\rm c}=15\,\GHz$ means about $65$ channels. The remaining $\simeq 335$ channels at $\nu\gtrsim 1\,{\rm THz}$ are used to construct a detailed model for the dust and cosmic infrared background (CIB) component, which we assume is subtracted down to the noise level for the lower frequency channels.
In the text we refer to these specifications as {\it PIXIE}-settings. We also consider cases with improved channel noise $\Delta I_{\rm c}$, as specified.

Detailed foreground modeling could make use of the high-resolution maps obtained with {\it Planck} \citep{Planck2013components}, allowing to separate bright clusters \citep{Planck2013ymap}, and providing spatial templates for the CO emission \citep{Planck2013CO}, the CIB \citep{Planck2013CIBlens} and Zodiacal light \citep{Planck2013Zodi}, but a more in depth analysis is left to future work.

\subsection{Defining the residual distortion}
\label{sec:primordial_signal}
Information about the thermal history before recombination is encoded by $\Delta I^{\rm prim}_\nu$ in Eq.~\eqref{eq:I_CMB}. The problem is to disentangle all spectral functions in of Eq.~\eqref{eq:I_CMB}, with the aim to isolate the primordial signal.
A small\footnote{At all times, the distortion has to be small compared to the CMB blackbody, since otherwise non-linear effects become important and the Green's function approach is inapplicable.} primordial distortion, $\Delta I^{\rm prim}_\nu$, caused by some energy-release history, $\id (Q/\rho_\gamma)/\id z$, can be computed using a Green's function method\footnote{An alternative method is described in \citet{Khatri2012mix}.} \citep{Chluba2013Green}:
\beq\label{eq:final_dist}
\Delta I^{\rm prim}_\nu(z=0)
\equiv \int G_{\rm th}(\nu, z') \, \frac{\id (Q/\rho_\gamma)}{\id z'} \id z'.
\eeq
Here, $\rho_\gamma\simeq 0.26\,(1+z)^4\,{\rm eV\,cm^{-3}}$ is the CMB blackbody energy density and $Q$ has dimensions of energy density.
The Green's function, $G_{\rm th}(\nu, z)$, contains all the physics of the thermalization problem. The accuracy of the Green's function method simply relies on the condition that the thermalization problem can be linearized, i.e., that the distortion remains small. It describes the observed SD response for single energy injection at $10^3\lesssim z$, and can be tabulated prior to the computation to accelerate the calculation. 

At very early times ($z\gtrsim \pot{2}{6}$), thermalization processes are extremely efficient, and the Green's function has the shape of a simple temperature shift, $G_{\rm th}(\nu, z)\propto G_{\rm T}\equiv \frac{2h\nu^3}{c^2} \frac{x\expf{x}}{(\expf{x}-1)^2}$. 
Later ($\pot{3}{5}\lesssim z\lesssim \pot{2}{6}$), photon production by double Compton and Bremsstrahlung at low frequencies becomes less efficient, while redistribution of photons over frequency by Compton scattering is still very fast. In this regime the distortion assumes the shape of a pure $\mu$-distortion,
$M(\nu)= G_{\rm T}\left[ x/\beta -1\right]/x$, with $\beta=3\zeta(3)/\zeta(2)\approx 2.1923$. 
At late times ($z\lesssim 10^4$), even Compton scattering becomes inefficient and the distortion is very close to a pure $y$-distortion, $Y_{\rm SZ}\equiv G_{\rm T}\left[x\coth(x/2)-4\right]$. 

At all intermediate redshifts, the Green's functions is given by a superposition of these extreme cases with some correction, $R(\nu, z)$, which we call {\it residual distortion} \citep[see][for similar discussion]{Chluba2013Green}:
\beq
\label{eq:G_approx}
G_{\rm th}(\nu, z) = 
\frac{G_{\rm T}(\nu)}{4} \,\mathcal{J}_{T}(z) 
+ \frac{Y_{\rm SZ}(\nu)}{4} \,\mathcal{J}_{y}(z)  
+\alpha\,M(\nu) \,\mathcal{J}_{\mu}(z) + R(\nu, z).
\eeq
Here, we used the identities $\int G_{\rm T}(\nu)\id \nu=\int Y_{\rm SZ}(\nu)\id \nu=4 \rho_\gamma$ and $\int M(\nu)\id \nu=\rho_\gamma/\alpha$ with $\alpha= [4\zeta(2)/[3\zeta(3)]-\zeta(3)/\zeta(4)]^{-1}\approx 1.401$ to re-normalize terms.
The redshift-dependent function, $\mathcal{J}_k(z)$, for $k\in\{T,y,\mu\}$, define the {\it branching ratios} of energy going into different components of the signal (see Sect.~\ref{sec:Dr_branch}). These ratios are {\it not} unique but depend on the experimental settings, which determine the orthogonality between different spectral components.
To obtain these functions, we use {\it PIXIE}-like instrumental specification (Sect.~\ref{sec:instrument}), where 
the CMB spectrum is sampled over some range of frequencies $\nu\in [\nu_{\rm min}, \nu_{\rm max}]$ with constant bandwidth $\Delta \nu_{\rm c}$ and constant sensitivity $\Delta I_{\rm c}$ per channel. 
This turns Eq.~\eqref{eq:G_approx} into $G_{i, \rm th}(z) = G_{i, \rm T} \,\mathcal{J}_{T}(z)/4 + Y_{i, \rm SZ} \,\mathcal{J}_{y}(z)/4 +\alpha\,M_i \,\mathcal{J}_{\mu}(z) + R_i(z)$, where the subscripts indicate the individual signals in the $i^{\rm th}$ channel. 
Then, we can interpret $G_{i,\rm th}(z)$, $G_{i,\rm T}$, $Y_{i, \rm SZ}$, $M_i$ and $R_i(z)$ $(i=1,\cdots,N)$ as $N$-dimensional ($N\equiv$ number of frequency channels) vectors\footnote{Henceforth, we shall denote vectors with bold font.}.

\begin{figure}
\centering
\includegraphics[width=\columnwidth]{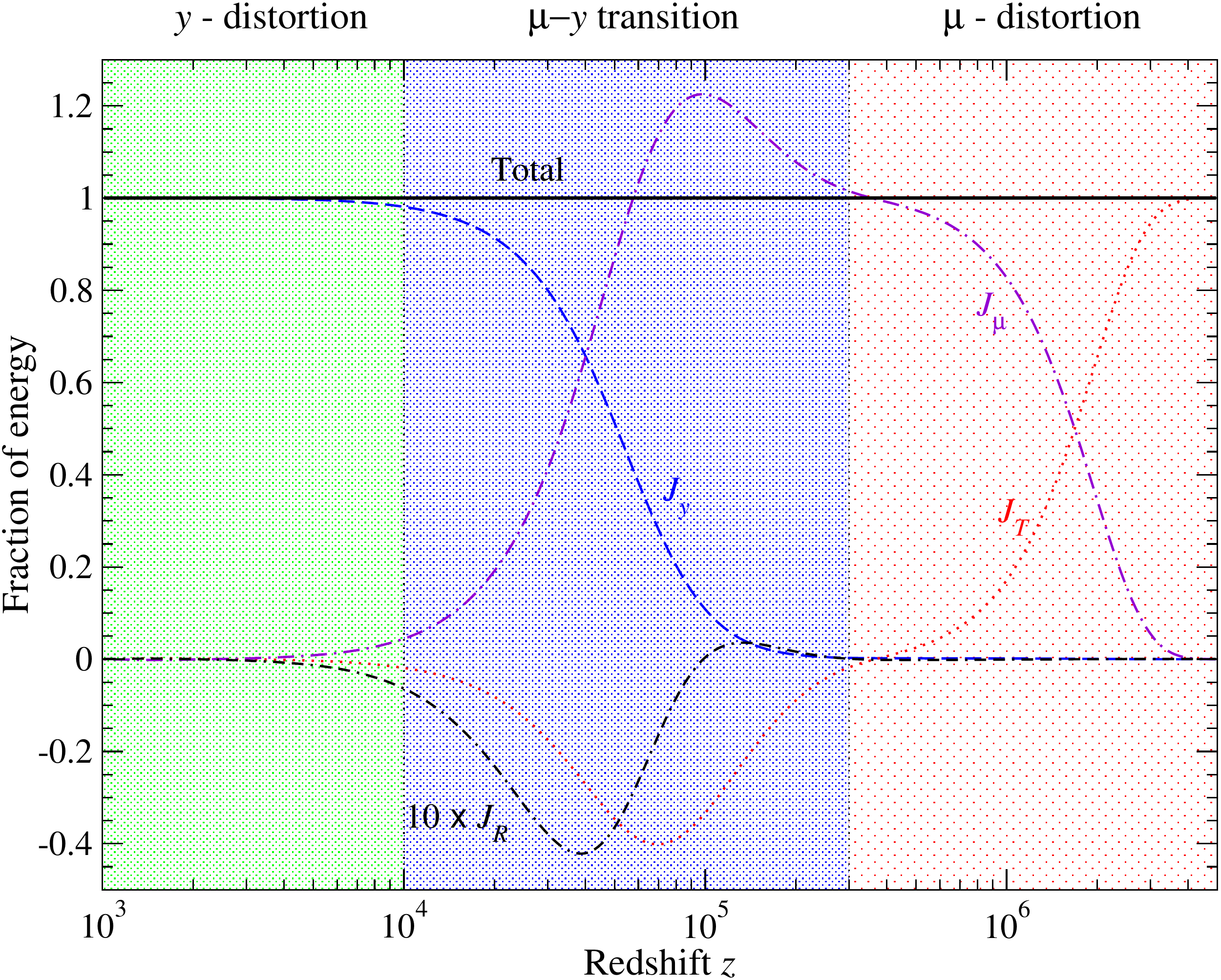}
\caption{Energy branching ratios, $\mathcal{J}_{k}(z)$ according to Eq.~\eqref{eq:R_Jfuncs} [in the figure the symbol $J\equiv \mathcal{J}$]. We multiplied $\mathcal{J}_{R}(z)$ by 10 to make it more visible.  For the construction we assumed $\{\nu_{\rm min}, \nu_{\rm max},\Delta\nu_{\rm s}\}=\{30, 1000, 1\}\,{\rm GHz}$ and diagonal noise covariance.}
\label{fig:J_z}
\end{figure}

In this vector space, the decomposition problem reduces to finding the residual distortion $\vek{R}(z)$ such that it is perpendicular to the space spanned by $\vek{G}_{\rm T}$, $\vek{Y}_{\rm SZ}$, and $\vek{M}$ (see Appendix~\ref{app:basis} for details). Once the residual distortion is identified, we obtain all energy branching ratios, $\mathcal{J}_{k}(z)$, of Eq.~\eqref{eq:G_approx} by projecting the rest of the Green's function on to 
$\vek{G}_{\rm T}$, $\vek{Y}_{\rm SZ}$ and $\vek{M}$, respectively.
The results are shown in Fig.~\ref{fig:J_z}. We also defined $\mathcal{J}_{R}(z)=1-\mathcal{J}_{T}(z)-\mathcal{J}_{y}(z)-\mathcal{J}_{\mu}(z)$, which determines the amount of energy found in the residual distortion only.
At redshift $z\lesssim \pot{4}{4}$, most of the energy release produces a $y$-distortion, while at $\pot{4}{4}\lesssim z \lesssim \pot{1.7}{6}$ most of the energy goes into a $\mu$-distortion. At $\pot{1.7}{6}\lesssim z$ the thermalization process, mediated by Compton scattering, double Compton emission and Bremsstrahlung, is so efficient that practically all energy just increases the average CMB temperature.

\begin{figure}
\centering
\includegraphics[width=1.02\columnwidth]{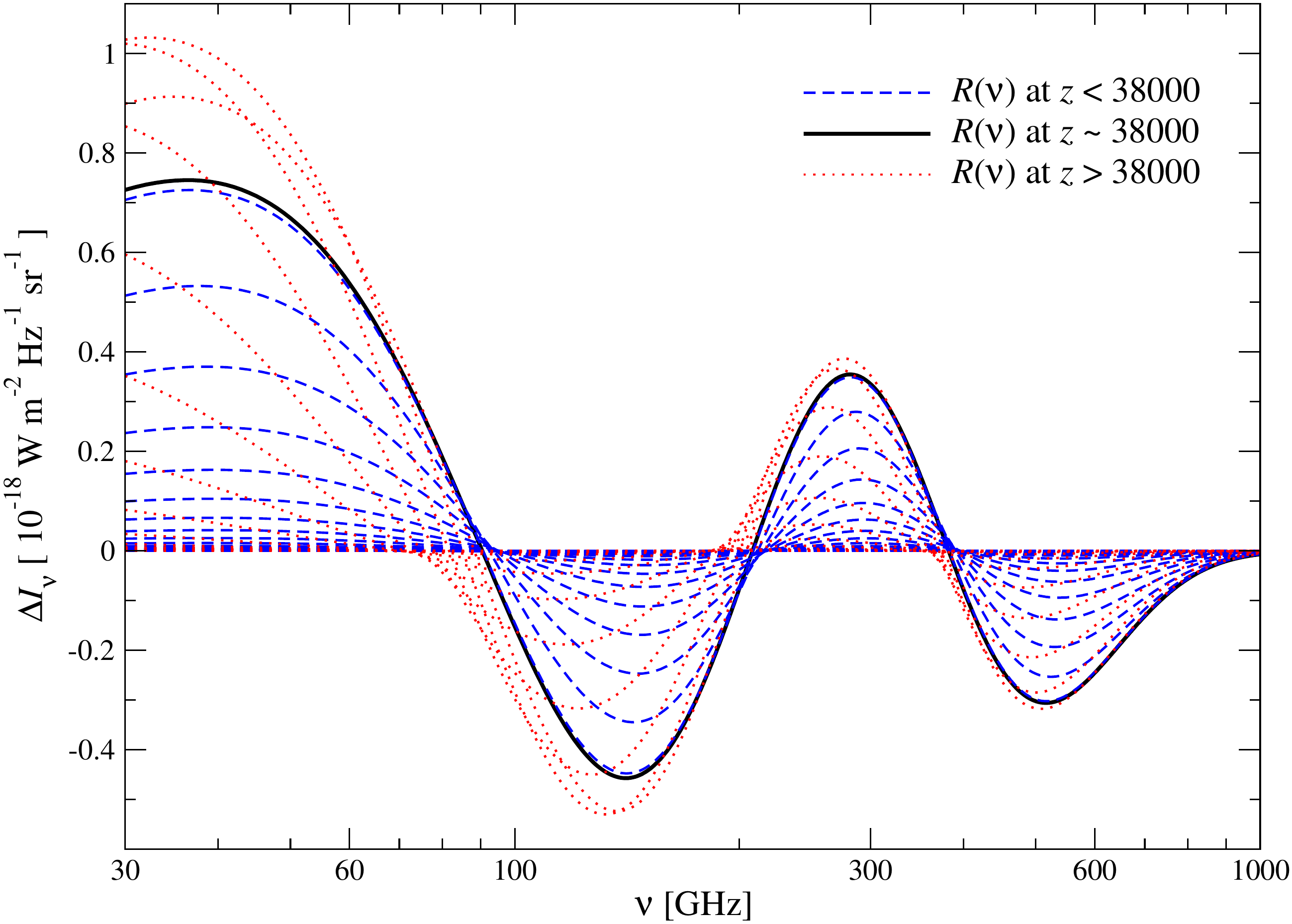}
\caption{Residual SD at different redshifts. For the construction we assumed $\{\nu_{\rm min}, \nu_{\rm max},\Delta\nu_{\rm s}\}=\{30, 1000, 1\}\,{\rm GHz}$ and diagonal noise covariance.}
\label{fig:R_z}
\end{figure}

Around $z\simeq \pot{4}{4}$, a few percent of the energy is stored by the residual distortion, and the amplitude of this signal depends strongly on redshift (see Fig.~\ref{fig:R_z}). Although small in terms of energy density, the residual distortion reaches $\simeq 10\%-20\%$ of $M(\nu)$ and $Y_{\rm SZ}(\nu)$ at high frequencies, 
and can even be comparable to $M(\nu)$ at $\nu\lesssim100\,{\rm GHz}$.
The fraction of energy release to the residual distortion is extremal around $z\simeq \pot{3.8}{4}$ (see Fig.~\ref{fig:J_z}), while the low-frequency amplitude of the residual distortion is largest at $z\simeq \pot{6.2}{4}$ (see Fig.~\ref{fig:R_z}). 
In Figure~\ref{fig:R_z}, we can also observe a small dependence of the phase of the residual distortion on the redshift of energy release. The redshift-dependent phase shift of the residual distortion provides 
model-independent information about the time dependence of the energy-release process,
while analysis of the superposition between $\mu$- and $y$-distortion can only be interpreted in a model-dependent way.

Figure~\ref{fig:J_z} also shows that $\mu$-distortion and temperature shift have a significant overlap around $z\simeq 10^5$. There $\mathcal{J}_{\mu}(z)$ exceeds unity, while $\mathcal{J}_{T}(z)$ is negative. Similarly, for the chosen experimental setting $\mathcal{J}_{R}(z)$ is negative, ensuring energy conservation.
Although below $z\simeq 10^5$ photon production becomes very weak and the thermalization of distortions to a temperature shift ceases, the shape of the distortion still projects on to $\vek{G}_{\rm T}$, leading to $\mathcal{J}_{T}(z)\neq 0$. When thinking about the different contributions to the total distortion signal these points should be kept in mind.

Another way to define the temperature shift is to integrate the distortion over all frequencies. Scattering terms, to which the $\mu$- and $y$-distortion are related, conserve photon number density, so that any deviation from zero should be caused by contributions from a temperature shift, related to $\vek{G}_{\rm T}(\nu)$. 
This approach was used by \citet{Chluba2013Green}, where by construction $0<\mathcal{J}_{k}(z)<1$ for $k\in\{T, y, \mu, R\}$.
In practice, i.e., with contaminations from foregrounds, this procedure may not be applicable, and simultaneous fitting of different spectral components is expected to work better. We therefore did not further follow this path.

\subsubsection{Dependence on experimental settings}
\label{sec:PCA_settings}
\begin{figure}
\centering
\includegraphics[width=\columnwidth]{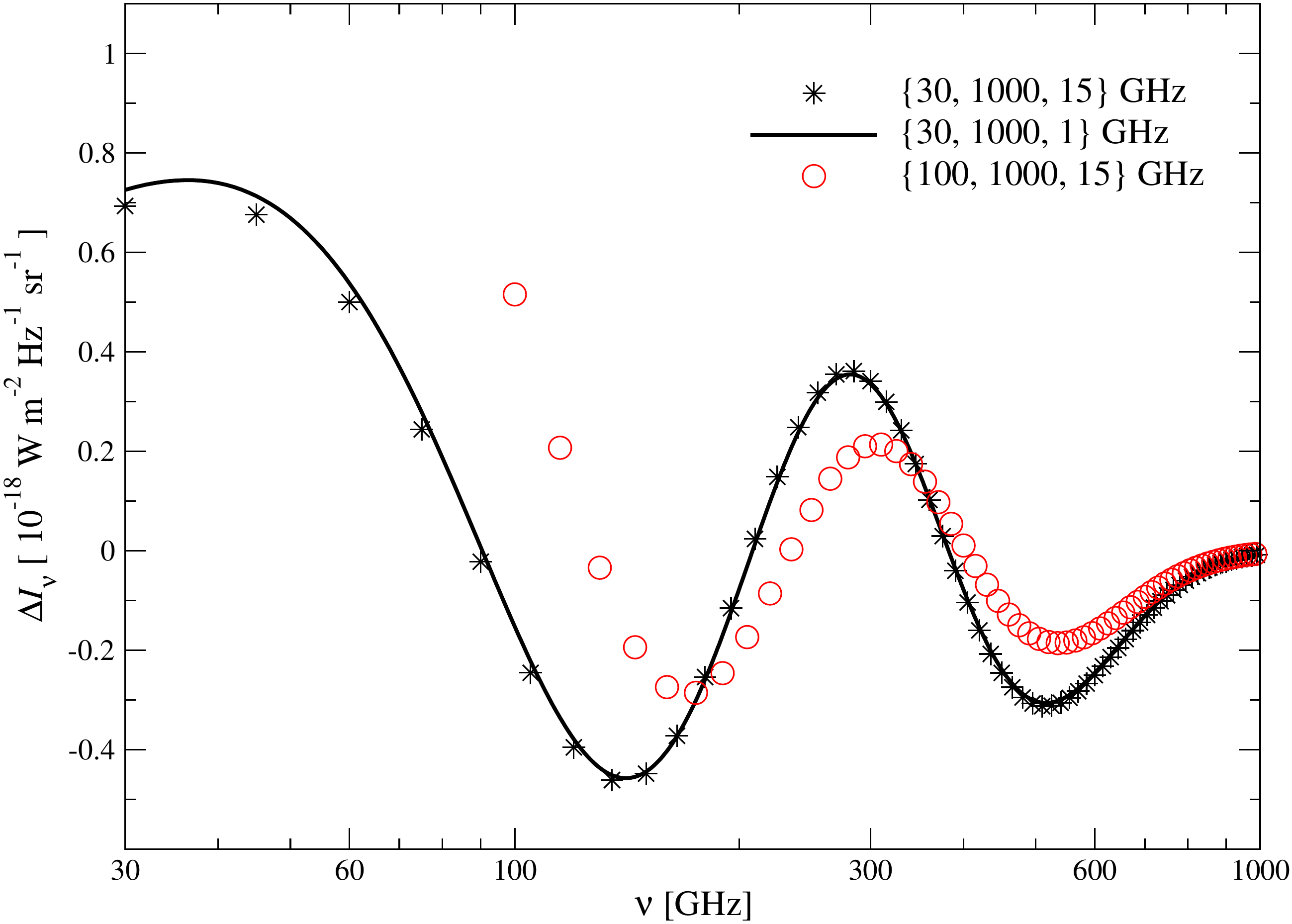}
\caption{Residual function at redshift $z\simeq 38000$ but for different instrumental settings. The annotated values are $\{\nu_{\rm min}, \nu_{\rm max},\Delta\nu_{\rm s}\}$ and we assumed diagonal noise covariance.}
\label{fig:R_z_change}
\end{figure}

It is clear that the decomposition [$R(\nu, z)$ and $\mathcal{J}_k(z)$] presented above depends on the chosen values for $\{\nu_{\rm min}, \nu_{\rm max},\Delta\nu_{\rm s}\}$. Changing the frequency resolution has a rather small effect, while changing $\nu_{\rm min}$ is more important (see Fig.~\ref{fig:R_z_change}). 
The differences are therefore mainly driven by the way the distortion projects on to $G_{\rm T}, M$ and $Y_{\rm SZ}$ between $\nu_{\rm min}$ and $\nu_{\rm max}$ rather than how precisely the channels are distributed over this interval.
%

Also, so far we assumed uniform and uncorrelated noise in the different channels. In this case, the construction of the modes becomes independent of the value of $\Delta I_{\rm c}$, but more generally one has to include this into the eigenmode analysis. This can be achieved by redefining the scalar product of two frequency vectors, e.g., $\vek{a}\cdot\vek{b}\equiv \sum_{ij} a_i \,C^{-1}_{ij}\,b_j $, where $C_{ij}$ is the full noise covariance matrix. 
Similarly, signals related to foregrounds can be included when performing the decomposition of the Green's function.
These are expected to lead to a degradation of the signal towards both lower and higher frequencies, 
however, these aspects are beyond the scope of this paper and will be explored in another work.

\subsection{Energy release and branching ratios}
\label{sec:Dr_branch}
The amplitude of the SD is directly linked to the total energy that was released over the cosmic history. 
One way, which has been widely applied in the cosmology community, to make this connection is to use the \textit{effective} $\mu$ and $y$-parameter to characterize the associated distortion, $\mu\simeq 1.4\,\Delta\rho_\gamma/\rho_\gamma|_\mu$ and $y\simeq (1/4)\,\Delta\rho_\gamma/\rho_\gamma|_y$ \citep{Zeldovich1969, Sunyaev1970mu}.  The total energy release causing distortions is $\Delta \rho_\gamma/\rho_\gamma|_{\rm dist}=\Delta\rho_\gamma/\rho_\gamma|_y+\Delta\rho_\gamma/\rho_\gamma|_\mu$, with the partial contributions, $\Delta\rho_\gamma/\rho_\gamma|_y$ and $\Delta\rho_\gamma/\rho_\gamma|_\mu$, from the $y$- and $\mu$-era, respectively. 
In terms of the energy-release history, $\mathcal{Q}(z')=\id (Q/\rho_\gamma)/\id \ln z'\approx (1+z')\,\id (Q/\rho_\gamma)/\id z'$, the {\it effective} $y$- and $\mu$-parameters can be written as
\beal
\label{eq:Drho_rho_old}
y&\approx \frac{1}{4}\,\int^{z_{\mu, y}}_0 \mathcal{Q}(z') \id \ln z' 
\nonumber \\
\mu&\approx 1.4\,\int_{z_{\mu, y}}^\infty \mathcal{J}_{\rm bb}(z') \, \mathcal{Q}(z') \id \ln z',
\end{align}
where we introduced the {\it spectral distortion visibility function}, $\mathcal{J}_{\rm bb}(z)\approx \expf{-(z/z_\mu)^{5/2}}$, with thermalization redshift $z_\mu\simeq \pot{2}{6}$ \citep[e.g., see][]{Hu1993}. The visibility function accounts for efficient thermalization process for redshifts $z\gtrsim \zmu$, at which only the average temperature of the CMB is increased and no distortion is created.
In Eq.~\eqref{eq:Drho_rho_old}, the transition between the $\mu$- and $y$-era is modeled as step-function at $z_{\mu, y}\simeq \pot{5}{4}$.

The decomposition, Eq.~\eqref{eq:Drho_rho_old}, into $\mu$- and $y$-distortion is only rough and has to be refined for the future generation of CMB experiments. Our approach described in this section provides a natural extension.
By inserting Eq.~\eqref{eq:G_approx} into Eq.~\eqref{eq:final_dist} and integrating over all $\nu$ we find that the total change of the CMB photon energy density, $\rho_\gamma$, caused by energy release is given by
\beal
\label{eq:Drho_rho}
\left.\frac{\Delta \rho_\gamma}{\rho_\gamma}\right|_{\rm tot} &= 
\left.\frac{\Delta \rho_\gamma}{\rho_\gamma}\right|_{T}
+\left.\frac{\Delta \rho_\gamma}{\rho_\gamma}\right|_{y}
+\left.\frac{\Delta \rho_\gamma}{\rho_\gamma}\right|_{\mu}
+\left.\frac{\Delta \rho_\gamma}{\rho_\gamma}\right|_{R}
\nonumber\\
&\equiv 4\Delta_T + 4y + \mu/\alpha+ \varepsilon
\nonumber\\
\Delta_T&=
\frac{1}{4}\left.\frac{\Delta \rho_\gamma}{\rho_\gamma}\right|_{T}
=\frac{1}{4}\int \mathcal{J}_{T}(z')\,\mathcal{Q}(z') \id \ln z'
\nonumber\\
y&=\frac{1}{4}\left.\frac{\Delta \rho_\gamma}{\rho_\gamma}\right|_{y}
=\frac{1}{4}\int \mathcal{J}_{y}(z')\,\mathcal{Q}(z') \id \ln z'
\nonumber\\
\mu&=\alpha\left.\frac{\Delta \rho_\gamma}{\rho_\gamma}\right|_{\mu}
=\alpha\int \mathcal{J}_{\mu}(z')\,\mathcal{Q}(z') \id \ln z'
\nonumber\\
\varepsilon&=\left.\frac{\Delta \rho_\gamma}{\rho_\gamma}\right|_{R}
=\int\,\mathcal{J}_{R}(z') \,\mathcal{Q}(z') \id \ln z',
\end{align}
with $\alpha\simeq 1.401$. In addition to $\Delta_T=\Delta T /T_0$ (defining a relative temperature shift), $y$- and $\mu$-parameter, we defined $\varepsilon$ to characterize the energy stored in the residual distortion. For a given energy-release history or mechanism these numbers can be directly computed, but only $y$, $\mu$, and $\varepsilon$ can be used to study the energy-release mechanism. The integrals can be carried out as a simple inner product in the discretized {\it redshift} vector space, making parameter estimation very efficient.

The expressions, Eq.~\eqref{eq:Drho_rho}, for $\mu$ and $y$ are very similar to the usual formulae, Eq.~\eqref{eq:Drho_rho_old}. 
The main difference is that here the origin of the redshift-dependent window functions, $\mathcal{J}_k(z)$, becomes apparent, being related to the representation of the different quasi-orthogonal components to the SD. Equations~\eqref{eq:Drho_rho} are thus a generalization, introducing visibility functions, or branching ratios $\mathcal{J}_k$, for $k=\mu$, $y$, $T$ and residual distortion, $R(\nu)$, respectively.
They are, however, dependent on the experimental settings (Sect.~\ref{sec:PCA_settings}).

\section{Principal component decomposition for the residual distortion}
\label{sec:R_eigenmodes}
In the previous section we showed that for a given experimental setting the Green's function can be decomposed into quasi-orthogonal parts. The primordial distortion is then fully described by the parameters $p=\{\Delta_T, y, \mu\}$ and a residual distortion
\beq
\label{eq:final_dist_R}
\Delta I^R_{i}
\equiv \int R_i(z') \, \mathcal{Q}(z') \id \ln z',
\eeq
which can be computed knowing the function $R_i(z)$. To constrain the energy-release history, $\Delta_T$, can be omitted, while interpretation of $y$ and $\mu$ only give model-dependent constraints on $\mathcal{Q}(z')$ [we discuss this point below]. 
We now ask how much can be learned about the redshift dependence of $\mathcal{Q}(z')$ by analyzing $\Delta I^R_{i}$. Since the overall signal is only a correction to the main superposition of $\mu$ and $y$-distortion signals, the experimental sensitivity has to be high or the overall energy release ought to be large.
By construction, $\Delta I^R_{i}$ is orthogonal to the space spanned by $y$ and $\mu$-distortion. We can thus perform a simple principal component decomposition for $\Delta I^R_{i}$ to get a handle on $\mathcal{Q}(z')$.  
For this we discretize the energy-release integral, Eq.~\eqref{eq:final_dist_R}, as a sum: 
\beal
\label{eq:final_dist_binned}
\Delta I^R_{i} &\approx \sum_a \hat{R}_{i}(z_a) \, \mathcal{Q}_a.
\end{align}
where $\hat{R}_{i}(z_a)=R_{i}(z_a)\,\Delta\ln z$ and $\mathcal{Q}_a=\mathcal{Q}(z_a)$. 
For our computations we distributed the bins logarithmically between $z_{\rm min}=10^3$ and $z_{\rm max}=\pot{5}{6}$ with log-spacing $\Delta\ln z=\pot{2.135}{-2}$, i.e., 400 grid points, using the mid-point integration rule. While only accurate at the level of $\simeq 0.1\%$, this approximation is sufficient for deriving the basis functions. When computing the SD from a given energy-release history we still explicitly carry out the full integral, Eq.~\eqref{eq:final_dist}, using Patterson quadrature rules \citep{Patterson1968}.
The Fisher-information matrix for measuring energy-release history, $\mathcal{Q}_a$, from the observed residual intensities $\Delta I_i^R$ is
\beal
\label{eq:Fisher}
\mathcal{F}_{ab}=
\frac{1}{\Delta I_{\rm c}^2}\sum_i\,
\frac{\partial \Delta I^R_{i}}{\partial \mathcal{Q}_a}\,
\frac{\partial \Delta I^R_{i}}{\partial \mathcal{Q}_b}
=\frac{1}{\Delta I_{\rm c}^2}\sum_i\,\hat{R}_{i}(z_a)\,\hat{R}_{i}(z_b),
\end{align}
where we assumed that the frequency channels, represented by index $i$, are independent. The eigenvectors of $\mathcal{F}_{ab}$ determine the principal components, $\vek{E}^{(k)}$, of the problem. The eigenvalues, $\lambda_k$, furthermore determine how well one might be able to recover $\mathcal{Q}(z)$ for a given sensitivity $\Delta I_{\rm c}$. 

The eigenmodes are vectors in discretized-redshift space, which we normalize as $\vek{E}^{(k)}\cdot \vek{E}^{(l)} =\delta_{kl}$. The energy-release history, $\mathcal{Q}(z)$, and the residual distortion, $\Delta I^R_{i}$, can then be written as
\beal
\label{eq:eigenmodes}
\mathcal{\vek{Q}}&\approx \sum_k \vek{E}^{(k)} \,\mu_k,
&\Delta I^R_{i}&\approx \sum_k S^{(k)}_{i} \,\mu_k,
&S^{(k)}_{i}&=\sum_a \hat{R}_{i}(z_a) \, E^{(k)}_a,
\end{align}
where $\mu_k$ and $S^{(k)}_{i}$ are the amplitude and distortion signal of the $k^{\rm th}$ eigenmode, respectively. 
By construction, the eigenvectors, $\vek{E}^{(k)}$, span an ortho-normal basis, while all $\vek{S}^{(k)}$ only define an orthogonal basis (generally $\vek{S}^{(k)}\cdot \vek{S}^{(l)} \geq \delta_{kl}$). 
We furthermore defined the energy-release vector $\mathcal{\vek{Q}}=(\mathcal{Q}(z_0), \mathcal{Q}_R(z_1), ..., \mathcal{Q}(z_{n}))^{T}$ of $\mathcal{Q}(z)$ in different redshift bins and the mode amplitudes $\mu_k=\vek{E}^{(k)}\cdot \mathcal{\vek{Q}}$. 
%
The expected absolute error in the recovered mode amplitudes $\mu_k$ is determined by $\Delta \mu_k=1/\sqrt{\lambda_k}\propto \Delta I_{\rm c}$. This scaling implies that for a given frequency range and resolution the eigenvalue problem only has to be solved once.
This is possible because we assume the same sensitivity in each channel, but generalization is straightforward.

\begin{figure}
\centering
\includegraphics[width=1.0\columnwidth]{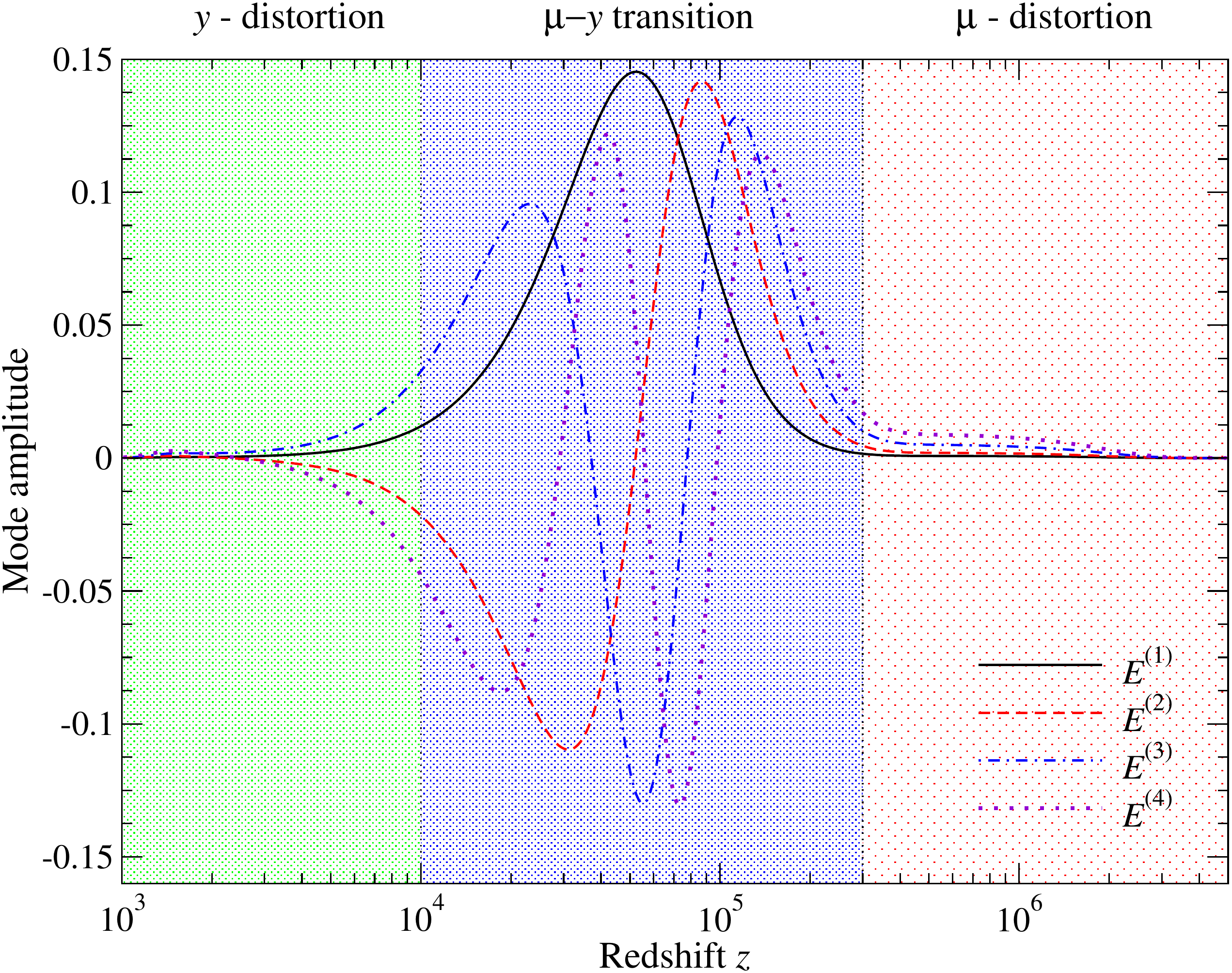}
\\[3mm]
\includegraphics[width=1.02\columnwidth]{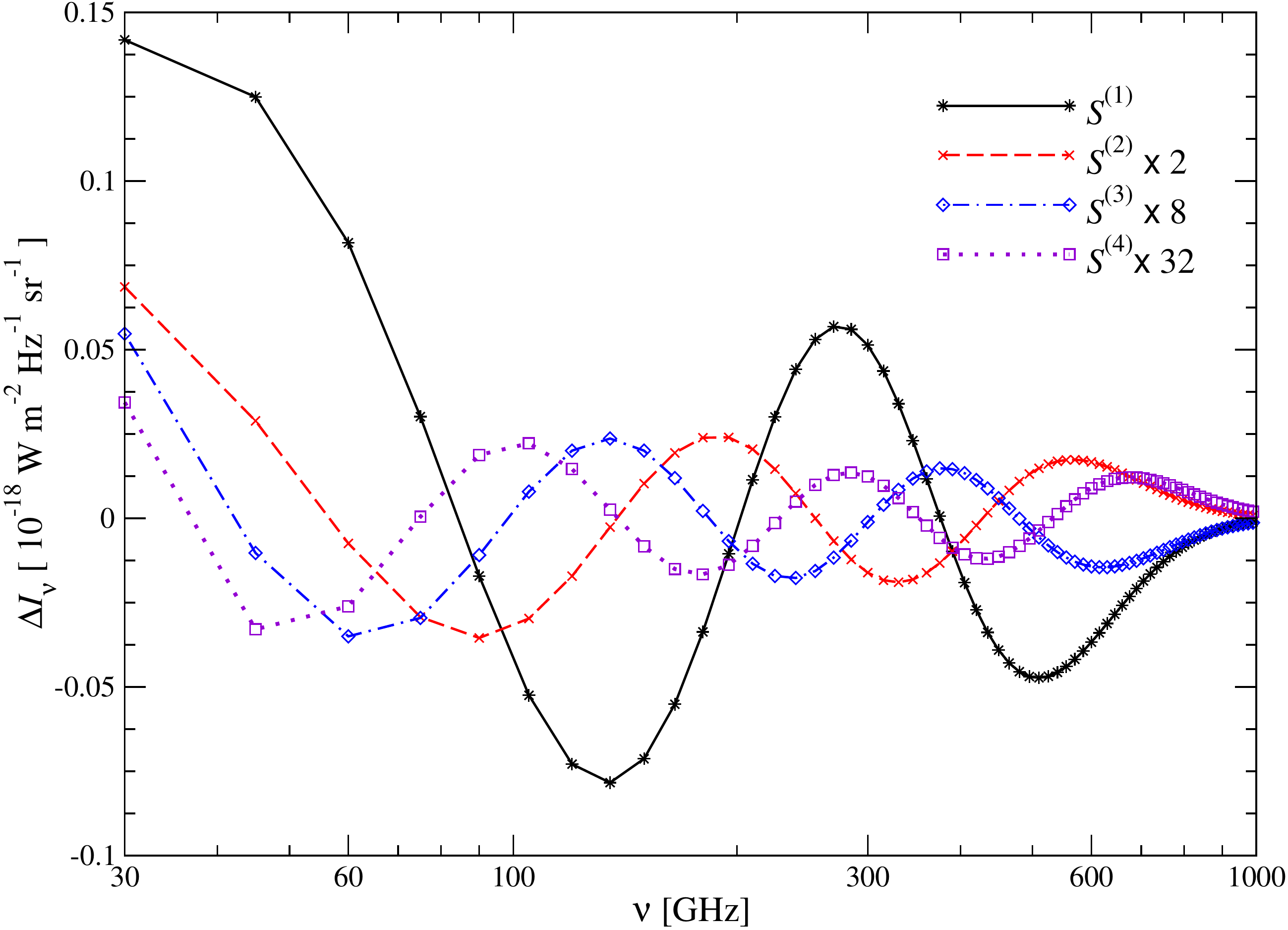}
\caption{First few eigenmodes $\vek{E}^{(k)}$ and $\vek{S}^{(k)}$ for {\it PIXIE}-type settings ($\nu_{\rm min}=30\,{\rm GHz}$, $\nu_{\rm max}=1000\,{\rm GHz}$ and $\Delta\nu_{\rm s}=15\,{\rm GHz}$). In the mode construction we assumed that energy release only occurred at $10^3 \leq z \leq \pot{5}{6}$.}
\label{fig:PCA}
\end{figure}

\subsection{Results for the eigenvectors and eigenvalues}
\label{sec:modes_0}
In Fig.~\ref{fig:PCA}, we show the first few $\vek{E}^{(k)}$ and $\vek{S}^{(k)}$ for a {\it PIXIE}-like experiment. We defined the signs of the modes such that the mode amplitudes are positive for $\mathcal{Q}={\rm const}>0$.
The first energy-release mode, $\vek{E}^{(1)}$, has a maximum at $z\simeq \pot{5.3}{4}$, while higher modes show more variability, extending both towards lower and higher redshift.
The corresponding distortion modes, $\vek{S}^{(k)}$, show increasing variability and decreasing overall amplitude with growing $k$.
They capture all corrections to the simple superposition of pure $\mu$- and $y$-distortion, needed to  morph between these two extreme cases.

In Table~\ref{tab:one}, we summarize the projected errors for the first six mode amplitudes. The errors, $\Delta\mu_k$, increase rapidly with mode number (this is how we order the eigenmodes), meaning that for a fixed amplitude of the distortion signal the information in the higher modes can only be accessed at higher spectral sensitivity.

Knowing the signal eigenvectors, $\vek{S}^{(k)}$, we can directly relate the mode amplitudes, $\mu_k$, to the fractional energy, $\varepsilon$, stored by the residual distortion. 
It thus allows us to estimate how much information is contained by the residual distortion.
Since integration over frequency can be written as a sum over all frequency bins, with $\varepsilon_k=4\sum_i S^{(k)}_i/\sum_i G_{i, T}$ we have 
$\varepsilon\approx \sum_k \varepsilon_k\,\mu_k$.
The first six $\varepsilon_k$ are given in Table~\ref{tab:one}. The signal modes, $\vek{S}^{(1)}$ and $\vek{S}^{(2)}$, contribute most to the energy, while energy release into the higher modes is suppressed by an order of magnitude or more.

Even if individual mode amplitudes cannot be separated, the total energy density contained in the residual distortion might still be detectable. The error of $\varepsilon$ can be found using Gaussian error propagation, $\Delta \varepsilon\approx (\sum_k \varepsilon_k^2 \Delta \mu_k^2)^{1/2}\simeq \{\pot{3.68}{-9}, \pot{3.53}{-9}, \pot{3.14}{-9},\pot{2.84}{-9}\}$, where the numbers show, respectively, uncertainties when all modes, all but $\mu_1$, all but $\mu_k$ with $k\leq2$ and all but $\mu_k$ with $k\leq3$ are included. 
Another estimator for the residual distortion is the modulus of the residual distortion vector $|R|^2\approx \sum_k \vek{S}^{(k)}\cdot \vek{S}^{(k)} \,\mu^2_k$. The required scalar product amplitudes are also given in Table~\ref{tab:one}. Similar to $\varepsilon$, the error of $|R|^2$ scales like $\Delta |R|^2\approx 2 (\sum_k [\vek{S}^{(k)}\cdot \vek{S}^{(k)} \mu_k]^2 \Delta \mu_k^2)^{1/2}$.
Both $\varepsilon$ and $|R|^2$ can be used to estimate how much information is left in the residual when the mode hierarchy is truncated at some fixed value $k$. If the signal-to-noise ratio is larger than unity, more modes should be added.

\begin{table}
\centering
\caption{Forecasted $1\sigma$ errors of the first six eigenmode amplitudes, $\vek{E}^{(k)}$. We also give $\varepsilon_k=4\sum_i S^{(k)}_i/\sum_i G_{i, T}$, and the scalar products $\vek{S}^{(k)}\cdot \vek{S}^{(k)}$ (in units of $[10^{-18}\,\UnitInu]^2$). 
The fraction of energy release to the residual distortion and its uncertainty are given by $\varepsilon\approx\sum_k \varepsilon_k\, \mu_k$ and $\Delta\varepsilon\approx(\sum_{k}\varepsilon_k^2\Delta\mu_k^2)^{1/2}$, respectively.
For the mode construction we used {\it PIXIE}-settings ($\{\nu_{\rm min},\nu_{\rm max},\Delta\nu_{\rm s}\} =\{30, 1000, 15\}\,{\rm GHz}$ and channel sensitivity $\Delta I_{\rm c}=\pot{5}{-26}\,{\rm W\,m^{-2}\,Hz^{-1}\,sr^{-1}}$). The errors roughly scale as $\Delta \mu_k\propto \Delta I_{\rm c}/\sqrt{\Delta\nu_{\rm s}}$.}
\begin{tabular}{c cccc}
\hline
$k$ & $\Delta \mu_k$ & $\Delta\mu_k/\Delta \mu_1$ & $\varepsilon_k$ & $\vek{S}^{(k)}\cdot \vek{S}^{(k)}$
\\
\hline
$1$ & $\pot{1.48}{-7}$ & $1$ & $\pot{-6.98}{-3}$ & $\pot{1.15}{-1}$
\\[1pt]
$2$ & $\pot{7.61}{-7}$ & $5.14$ & $\pot{2.12}{-3}$ & $\pot{4.32}{-3}$
\\
$3$ & $\pot{3.61}{-6}$ & $24.4$ & $\pot{-3.71}{-4}$ & $\pot{1.92}{-4}$
\\[1pt]
$4$ & $\pot{1.74}{-5}$ & $\pot{1.18}{2}$ & $\pot{8.29}{-5}$ & $\pot{8.29}{-6}$
\\[1pt]
$5$ & $\pot{8.52}{-5}$ & $\pot{5.76}{2}$ & $\pot{-1.55}{-5}$ & $\pot{3.45}{-7}$
\\[1pt]
$6$ & $\pot{4.24}{-4}$ & $\pot{2.86}{3}$ & $\pot{2.75}{-6}$ & $\pot{1.39}{-8}$
\\
\hline
\label{tab:one}
\end{tabular}
\end{table}

\section{Parameter estimation using energy-release eigenmodes}
\label{sec:estimation}
In the previous sections, we created a set of orthogonal signal modes that can be constrained by future SD experiments and used to recover part of the energy-release history in a model-independent way. We derived a set of energy-release eigenmodes that describes the residual distortion signal that cannot be expressed as simple superposition of temperature shift, $\mu$- and $y$-distortion. 

As explained above, nothing can be learned from the change in the value of the CMB temperature caused by energy release. Thus, the useful part of the primordial signal is determined by the parameters $p_{\rm prim}=\{y, \mu, \mu_k\}$. The number of residual modes, $\mu_k$, that can be constrained depends on the typical amplitude of the distortion and instrumental aspects.
To the primordial signal, we need to add $y_{\rm re}$ to describe the late-time $y$-distortion, and $\Delta_T$ to parametrize the uncertainty in the exact value of the CMB monopole. The total distortion signal therefore takes the form
\beal
\label{eq:DI_tot}
&\Delta I_i=  \Delta I_i^{T} + \Delta I_i^{y} +  \Delta I_i^{\mu} + \Delta I_i^{R}
\nonumber\\
&\Delta I_i^{T}= G_{i, \rm T}\Delta_{T} [1+\Delta_{T}]+Y_{i, \rm SZ}\,\Delta_{T}^2/2 
\nonumber\\
&\Delta I_i^{y}=Y_{i, \rm SZ} \, (y_{\rm re}+y)
\nonumber\\
&\Delta I_i^{\mu}=M_{i} \, \mu
\end{align}
where $G_{i, \rm T}$, $Y_{i, \rm SZ}$ and $M_i$ are the average signals of $G_{\rm T}$, $Y_{\rm SZ}$ and $M$ over the $i^{\rm th}$ channel. 
The dependence of $\Delta I_i^{T}$ on $\Delta_{T}$ is quadratic, but since $\Delta_{T}\ll 1$, the problem remains quasi-linear, with the second-order term leading to a negligible correction to the covariance matrix, once expanded around the best-fitting value for $\Delta_{T}$. For estimates one can thus set $\Delta I_i^{T}\approx G_{i,\rm T}\,\Delta_{T}$ without loss of generality. 
This defines the parameter set $p=\{\Delta_{T}, y^\ast, \mu, \mu_k\}$, where $y^\ast=y_{\rm re}+y$. 
Note that because of the low-$z$ contribution, it is hard to disentangle the primordial components of $\Delta_T$ and $y^\ast$. The primordial energy release, therefore, is best constrained with $\mu$ and the $\mu_k$s.

\subsection{Errors of $\Delta_{T}$, $y^\ast$ and $\mu$}
\label{sec:TYM_errors}
As a first step, we estimate the errors on the values of $\Delta_{T}$, $y^\ast$ and $\mu$ assuming {\it PIXIE}-like settings. The relevant projections to construct the Fisher matrix, analogous to Eq.~\eqref{eq:Fisher}, are
\beal
\label{eq:dot_prods}
&\vek{G}_{\rm T}\cdot (\vek{G}_{\rm T},\vek{Y}_{\rm SZ}, \vek{M})
  = (\pot{2.46}{3},\pot{1.23}{3}, \pot{4.60}{2})
\nonumber\\
 &\vek{Y}_{\rm SZ} \cdot (\vek{Y}_{\rm SZ}, \vek{M}) = (\pot{5.37}{3}, \pot{5.62}{2})
\nonumber\\
&\vek{M}\cdot \vek{M} = \pot{1.23}{2} 
\end{align}
all in units of $[10^{-18}\,{\rm W\,m^{-2}\,Hz^{-1}\,sr^{-1}}]^2$. 
Defining $\alpha=\Delta I_{\rm c}/[\pot{5}{-26}\,{\rm W\,m^{-2}\,Hz^{-1}\,sr^{-1}}]$ we expect errors $\Delta \Delta_{T} \approx \pot{2.34}{-9}\,\alpha$ (or $\Delta T \simeq 6.4\, \alpha\,{\rm nK}$), $\Delta y^\ast\approx \pot{1.20}{-9}\,\alpha$ and $\Delta \mu\approx \pot{1.37}{-8}\,\alpha$ at $1\sigma$ level. These numbers are close to the estimates given by \citet{Kogut2011PIXIE} for the expected $1\sigma$ errors on $y$- and $\mu$-parameter, and show that a huge improvement over {\it COBE}/{FIRAS} ($\Delta y^\ast\approx \pot{7.5}{-6}$ and $\Delta \mu\approx \pot{4.5}{-5}$ at $1\sigma$ level) can be expected. 
Adding the residual distortion eigenmodes to the parameter estimation should not affect these estimates as they are constructed to be orthogonal to the signals from $\Delta_T$, $y$ and $\mu$.

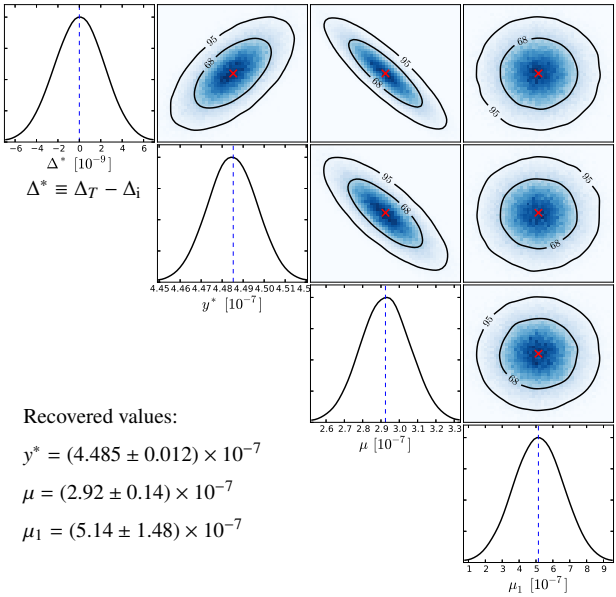
\begin{figure}
\centering
\setlength{\unitlength}{0.05\columnwidth}
  \begin{picture}(20,20)
    \put(0,0){\includegraphics[width=\columnwidth]{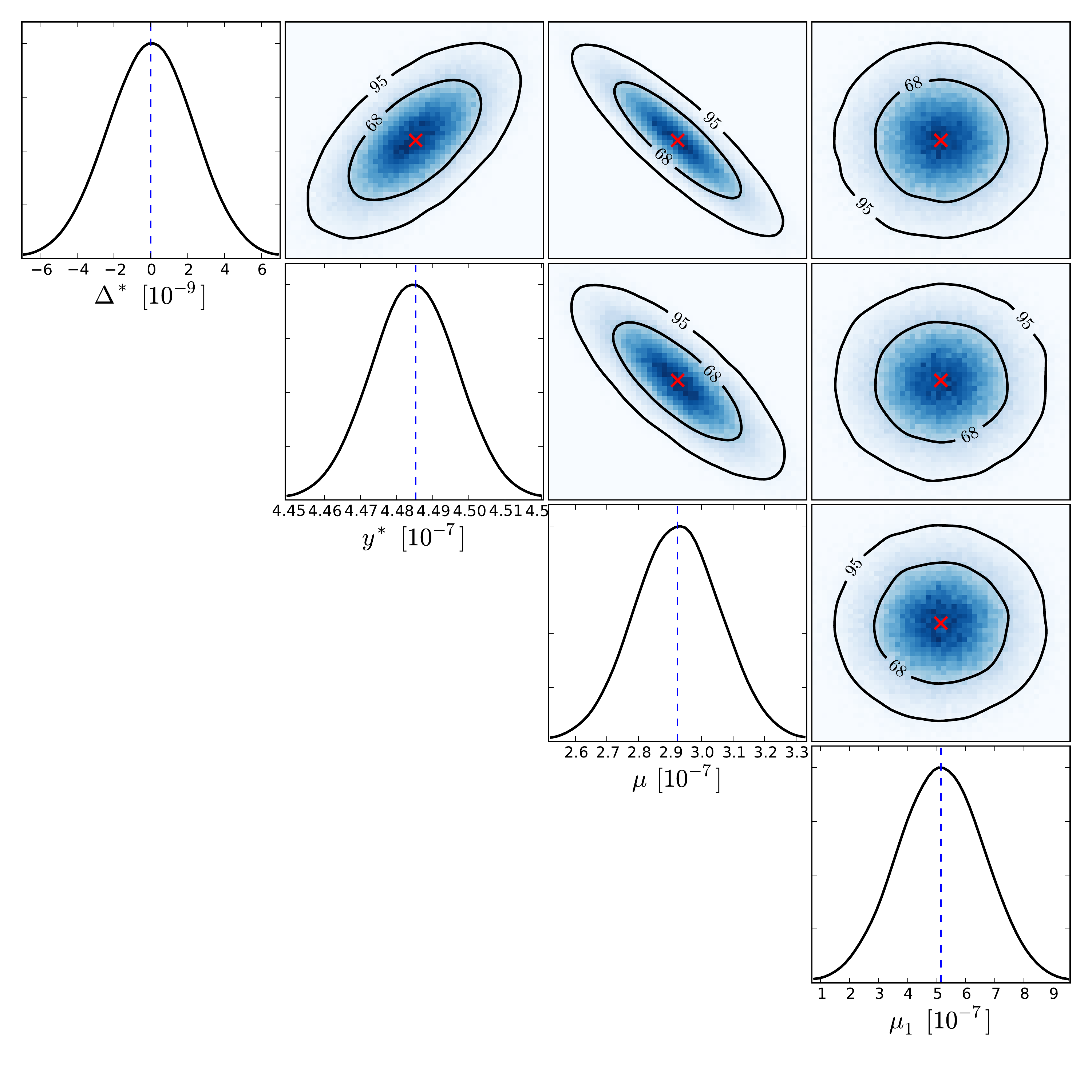}}
    \put(1.1,13.6){$\Delta^\ast\equiv\Delta_{T}-\Delta_{\rm i}$}
    \put(1.0,6.4){Recovered values:}
    \put(1.0,5.2){$y^\ast=\poterr{4.485}{0.012}{-7}$}
    \put(1.0,4.0){$\mu=\poterr{2.92}{0.14}{-7}$}
    \put(1.0,2.8){$\mu_1=\poterr{5.14}{1.48}{-7}$}
  \end{picture}
\caption{Analysis of energy-release history with $\mathcal{Q}(z)=\pot{5}{-8}$ in the redshift interval $10^3 < z < \pot{5}{6}$ using signal eigenmode, $\vek{S}^{(1)}$ (Fig.~\ref{fig:PCA}). We assumed $\{\nu_{\rm min},\nu_{\rm max},\Delta\nu_{\rm s}\} =\{30, 1000, 15\}\,{\rm GHz}$ and channel sensitivity $\Delta I_{\rm c}=\pot{5}{-26}\,{\rm W\,m^{-2}\,Hz^{-1}\,sr^{-1}}$. The dashed blue lines and red crosses indicate the expected recovered values. Contours are for $68\%$ and $95\%$ confidence levels. All errors and recovered values agree with the Fisher estimates. We shifted $\Delta_{T}$ by $\Delta_{\rm i}=\Delta_{\rm f}+\Delta_{\rm prim}$ with $\Delta_{\rm f} = \pot{1.2}{-4}$ and $\Delta_{\rm prim}\simeq\pot{-8.46}{-9}$, where $\Delta_{\rm prim}$ is the primordial contribution.
}
\label{fig:example_I}
\end{figure}

\subsection{Simple parameter estimation example: proof of concept}
\label{sec:simple_example}
To illustrate how the modes can be used to constrain the energy-release history, let us consider $\mathcal{Q}(z)\equiv \pot{5}{-8}$ in the redshift interval $10^3 < z < \pot{5}{6}$. Using Eq.~\eqref{eq:Drho_rho}, this implies a total energy release of $\Delta \rho_\gamma/\rho_\gamma=\pot{4.26}{-7}$, with $\Delta \rho_\gamma/\rho_\gamma|_{\rm dist}=4y+\mu/\alpha+\varepsilon\approx\pot{4.00}{-7}$ going into distortions. 
We also expect $y\simeq \pot{4.85}{-8}$, $\mu\simeq \pot{2.93}{-7}$, and $\Delta_{\rm prim}\simeq\pot{-8.46}{-9}$ for the primordial distortion. 
The first three mode amplitudes are $\mu_1=\pot{5.14}{-7}$, $\mu_2=\pot{4.34}{-9}$, and $\mu_3=\pot{3.38}{-7}$, and thus $\mu_1$ should be detectable with a {\it PIXIE}-like experiment (see the $\Delta\mu_k$ in Table \ref{tab:one}). 
For illustration, we furthermore assume that the value of the monopole temperature is $T_0=2.726\,{\rm K} (1+\Delta_{\rm f})$ with $\Delta_{\rm f} = \pot{1.2}{-4}$, and that a low redshift $y$-distortion with $y_{\rm re}=\pot{4}{-7}$ is introduced.

We implemented a simple Markov Chain Monte Carlo (MCMC) simulation of this problem using {\sc CosmoTherm}. To compute the primordial distortion signal we used Eq.~\eqref{eq:final_dist}, i.e., we did not decompose the signal explicitly, but included all contributions to the distortion. We then added a temperature shift with $\Delta_{\rm f} = \pot{1.2}{-4}$ and a $y$-distortion with $y_{\rm re}=\pot{4}{-7}$ to the input signal, and analyzed it using the model, Eq.~\eqref{eq:DI_tot}, with only $\mu_1$ included.
Figure~\ref{fig:example_I} shows the results of this analysis. All the recovered values and errors agree with the predictions. We can furthermore see that $\mu_1$ does not correlate to any of the standard parameters $p_{\rm s}=\{\Delta_T, y^\ast, \mu\}$, as ensured by construction. The standard parameters are slightly correlated with each other, since in the analysis we used $G_{i, \rm T}$, $Y_{i, \rm SZ}$ and $M_i$ which themselves are not orthogonal. Alternatively, one could use the orthogonal basis $G_{i, \rm T, \perp}$, $Y_{i, \rm SZ}$ and $M_{i, \perp}$ (see Appendix~\ref{app:basis}), but since the interpretation of the results is fairly simple we preferred to keep the well-known parametrization. 
We confirmed that adding more distortion eigenmodes to the estimation problem does not alter any of the constraints on the other parameters. 
This demonstrates that the eigenmodes constructed above can be directly used for model-independent estimations and compression of the useful information provided by the CMB spectrum.

\begin{figure}
\centering
\includegraphics[width=\columnwidth]{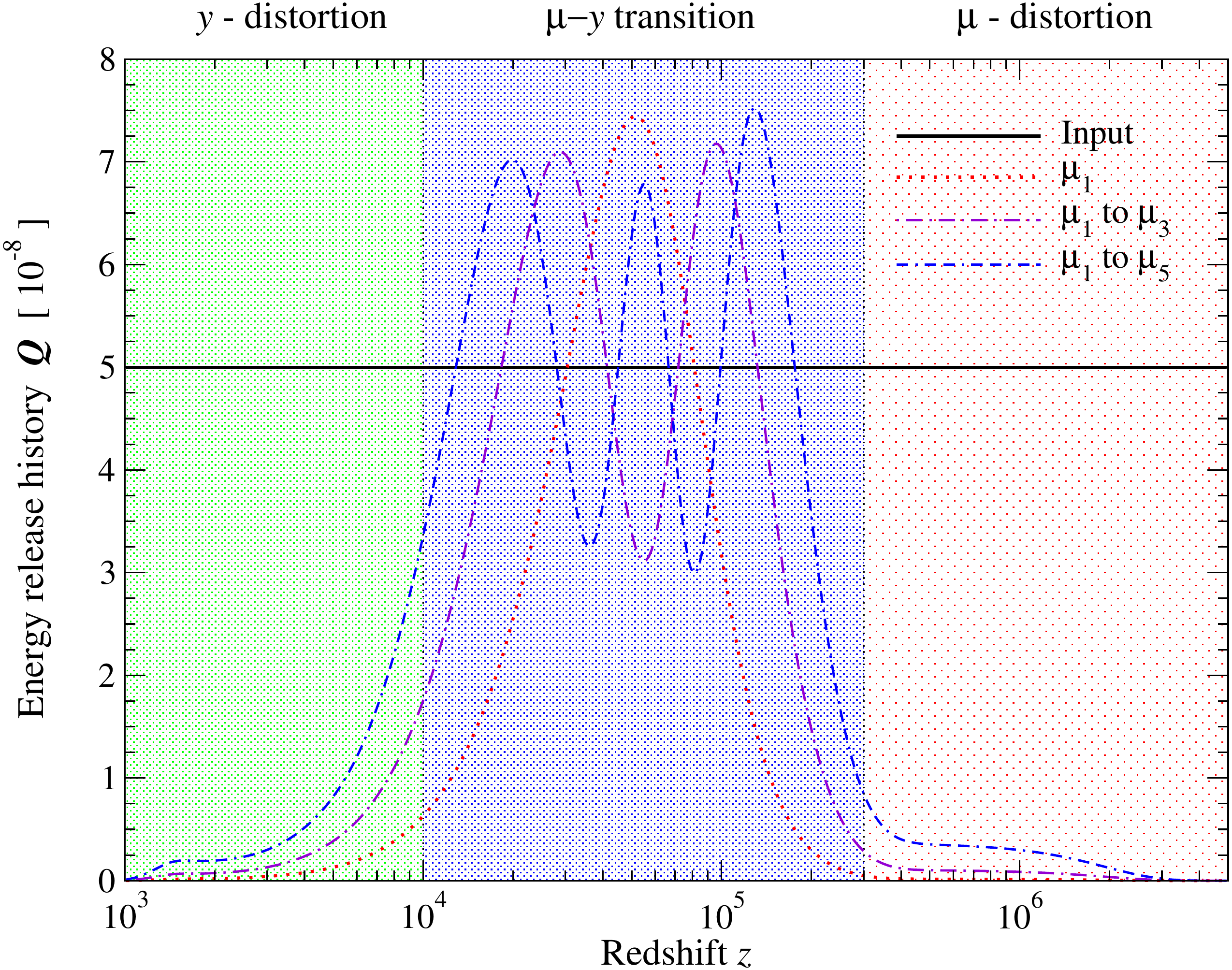}
\caption{Partial recovery of the input energy-release history, $\mathcal{Q}=\pot{5}{-8}$.}
\label{fig:Q_const}
\end{figure}
\subsection{Partial recovery of the energy-release history}
\label{sec:modes_recovery}
The energy-release eigenmodes define an ortho-normal basis to describe the energy-release history over the considered redshift range. In the limit of extremely high sensitivity and very fine spectral coverage ($\equiv$ all modes can be measured) a complete reconstruction of the input history would be possible. 
Since realistically only a finite number of energy-release eigenmodes (2 or 3 really) might be measured, this means that a partial but model-independent reconstruction of the input energy-release history can be derived. 

Considering the simple example, $\mathcal{Q}=\pot{5}{-8}$, in Fig.~\ref{fig:Q_const} we show the comparison of input history and the corresponding reconstruction if one, three or five modes can be measured. Clearly, the SD signal can only probe energy release around $z\simeq \pot{5}{4}$, providing the means to obtain a wiggly recovery of the input history. The SD signal created by an energy-release history that is constant, or has the other shapes is virtually indistinguishable from the observational point of view, because the energy release from the oscillatory parts does not leave any significant traces. Still, the trajectories of energy-release histories from different scenarios are directly constrained once the set of $\mu_k$ is known. This is one of the interesting model-independent ways of interpreting CMB SD results.

\subsection{Overall picture and how to apply the eigenmodes}
\label{sec:picture}
We now have all the pieces together to explain how to interpret and use the eigenmode decomposition presented above. Given the distortion data, $\Delta I^{\rm d}_{i}$ (we assume that foregrounds have been removed perfectly), in different frequency channels we can estimate the spectral model parameters $p_{\rm m}=\{\Delta_T, y^\ast, \mu, \mu_k\}$. Using the signal eigenvectors, $\vek{S}^{(k)}$, we can directly obtain the mode amplitudes by $\mu_k\approx \sum_i \Delta I^{\rm d}_{i}\,\vek{S}^{(k)}_i/|\vek{S}^{(k)}|^2$. Similarly, we can compute $y^\ast$ and $\mu$ as simple scalar products of the data vector with $\vek{Y}_{\rm SZ}$ and $\vek{M}_{\perp}$. The errors can be deduced using Table~\ref{tab:one} and Sect.~\ref{sec:TYM_errors}. At this point, we have compressed all the useful information contained by the CMB spectrum into a few numbers, $p_{\rm m}$. 
The number of operations needed to compute the SD from a given ERS also roughly reduces by a factor of $\eta\simeq (m+2)/N_{\rm freq}$, where $m$ is the included number of eigenmodes and $N_{\rm freq}$ the number of channels. For {\it PIXIE}, this means $\eta^{-1}\simeq 15-20$ times improvement of the performance, when using the signal eigenmodes for parameter estimation.

The $\mu$-parameter provides an integral constraint on the energy release, with redshift-dependent weighting function, $\mathcal{J}_\mu(z)$ (see Fig.~\ref{fig:J_z}). Many energy-release histories can give rise to exactly the same value of $\mu$. Still any specific scenario has to reproduce this number, although an interpretation becomes model dependent at this point.
Similarly, the recovered $y$-parameter can only be interpreted in a model-dependent way. Since only the combination $y^\ast=y_{\rm re}+y$ can be constrained, the model-dependent step allows us to deduce an estimate for $y_{\rm re}$, but otherwise does not help constraining the energy-release history unless $y_{\rm re}$ is known (precisely) by another method. Conversely, $y_{\rm re}$ remains uncertain, since a large contribution to $y^\ast$ could be caused by pre-recombination energy release.
This compromises our ability to learn about reionization and structure formation by studying the average CMB spectrum.

On the other hand, the recovered eigenmode amplitudes $\mu_k$ allow us to constrain the energy-release history, $\mathcal{Q}(z)$, in a model-independent way (Sect.~\ref{sec:modes_recovery} and Fig.~\ref{fig:Q_const}). Since we can only expect the first few modes to be measured, from Fig.~\ref{fig:PCA} it is clear that one is most sensitive to energy release around $z\simeq \pot{5}{4}$. ERSs with little activity during that epoch will project weakly on to $\mu_k$.
Different ERSs are furthermore expected to have specific eigenspectra, $\mu_k$, which in principle allows distinguishing them and constraining their specific model parameters. Computing the eigenspectra as a function of parameters can thus be used to quickly explore degeneracies between models. It is also clear that for ERSs with $m$ parameters, at least $m$ distortion parameters (excluding $y$) have to be observable. To distinguish between different types of models generally one additional parameter has to be measured and the eigenspectra of the scenarios have to be sufficiently orthogonal with respect to the experimental sensitivity.
We find that even for optimistic setting typically no more than the first three eigenmodes plus $\mu$ can be measured, so that in the foreseeable future energy-release models with more than 4 parameters cannot be constrained without providing additional information.

\begin{table*}
\centering
\caption{Eigenspectra for different energy-release scenarios. The mode amplitudes were scaled by the variable $A$, as indicated. An asterisk $(\ast)$ indicates that the parameter can be detected at more than $1\sigma$ with {\it PIXIE}-like sensitivity, while a dagger ($\dagger$) shows that 5 times the sensitivity is required for a $1\sigma$ detection. The last few rows are $\rho_k=[\mu_k/\Delta\mu_k]/[\mu/\Delta \mu]$, which give a representation that shows how the difficulty of a measurement relative to $\mu$ increases. Also, by comparing the numbers between models one can directly estimate how hard it is to distinguish them experimentally.}
\begin{tabular}{c cccccccc}
\hline
 & Dissipation & Dissipation 
 & Annihilation 
 & Annihilation 
 & Decay & Decay & Decay
\\
\hline
\hline
Shape  & $\nS=1$ & $\nS=0.96$ 
& $\left<\sigma{\rm v}\right>={\rm const}$ 
&  $\left<\sigma{\rm v}\right>\propto (1+z)$
& $z_{\rm X}=\pot{2}{4}$
& $z_{\rm X}=\pot{5}{4}$
& $z_{\rm X}=10^{5}$
\\
parameters & $\nrun=0$ & $\nrun=-0.02$ 
& (s-wave) 
& (p-wave, rel.) 
& $(t_{\rm X}=\pot{5.8}{10}\,{\rm sec})$
& $(t_{\rm X}=\pot{9.2}{9}\,{\rm sec})$
& $(t_{\rm X}=\pot{2.3}{9}\,{\rm sec})$
\\
\hline
\hline
$A$ 
& $\frac{A_\zeta}{\pot{2.2}{-9}}$ 
& $\frac{A_\zeta}{\pot{2.2}{-9}}$
& $\frac{f_{\rm ann, s}}{\pot{2}{-23}\,{\rm eV \, sec^{-1}}}$
& $\frac{f_{\rm ann, p}}{10^{-27}\,{\rm eV \, sec^{-1}}}$
%
& $\frac{f_{\rm X}/z_{\rm X}}{1\,{\rm eV}}$
& $\frac{f_{\rm X}/z_{\rm X}}{1\,{\rm eV}}$
& $\frac{f_{\rm X}/z_{\rm X}}{1\,{\rm eV}}$
\\
\hline
\hline
$y/A$ 
& $\pot{4.70}{-9}\,\ast$ 
& $\pot{3.52}{-9}\,\ast$
& $\pot{5.18}{-10}\,\dagger$
& $\pot{4.84}{-10}\,\dagger$
& $\pot{1.41}{-7}\,\ast$
& $\pot{8.47}{-8}\,\ast$
& $\pot{2.96}{-8}\,\ast$
\\[1pt]
$\mu/A$ 
& $\pot{3.11}{-8}\,\ast$ 
& $\pot{1.16}{-8}\,\dagger$
& $\pot{3.99}{-9}\,\dagger$
& $\pot{8.35}{-8}\,\ast$
& $\pot{2.27}{-7}\,\ast$
& $\pot{7.07}{-7}\,\ast$
& $\pot{1.01}{-6}\,\ast$
\\[1pt]
\hline
$\mu_1/A$ 
& $\pot{5.42}{-8}\,\dagger$ 
& $\pot{2.95}{-8}$
& $\pot{6.84}{-9}$
& $\pot{2.10}{-8}$
& $\pot{1.59}{-6}\,\ast$
& $\pot{3.34}{-6}\,\ast$
& $\pot{2.36}{-6}\,\ast$
\\[1pt]
$\mu_2/A$ 
& $\pot{1.01}{-9}$ 
& $\pot{-5.19}{-9}$
& $\pot{2.61}{-10}$
& $\pot{2.03}{-8}$
& $\pot{-1.74}{-6}\,\ast$
& $\pot{-4.95}{-7}\,\dagger$
& $\pot{2.47}{-6}\,\ast$
\\
$\mu_3/A$ 
& $\pot{3.53}{-8}$ 
& $\pot{1.91}{-8}$
& $\pot{4.39}{-9}$
& $\pot{3.04}{-8}$
& $\pot{1.66}{-6}\,\dagger$
& $\pot{-3.83}{-7}$
& $\pot{7.1}{-7}\,\dagger$
\\[1pt]
$\mu_4/A$ 
& $\pot{2.26}{-9}$ 
& $\pot{-5.13}{-9}$
& $\pot{4.39}{-10}$
& $\pot{3.71}{-8}$
& $\pot{-1.38}{-6}$
& $\pot{6.85}{-7}$
& $\pot{-1.23}{-8}$
\\[1pt]
\hline
\hline
%
$y/A$ 
& $3.9\,\sigma\,\ast$ 
& $2.9\,\sigma\,\ast$
& $0.43\,\sigma\,\dagger$
& $0.40\,\sigma\,\dagger$
& $117\,\sigma\,\ast$
& $70.6\,\sigma\,\ast$
& $24.7\,\sigma\,\ast$
\\[1pt]
$\mu/A$ 
& $2.3\,\sigma\,\ast$ 
& $0.85\,\sigma\,\dagger$
& $0.29\,\sigma\,\dagger$
& $6.1\,\sigma\,\ast$
& $16.6\,\sigma\,\ast$
& $51.6\,\sigma\,\ast$
& $73.8\,\sigma\,\ast$
\\[1pt]
\hline
$\rho_1$ & $0.161\,\dagger$ & $0.235$ & $0.159$ & $\pot{2.33}{-2}$
& $0.648\,\ast$ & $0.437\,\ast$ & $0.216\,\ast$
\\[1pt]
$\rho_2$ & $\pot{5.86}{-4}$ & $\pot{-8.07}{-3}$ & $\pot{1.18}{-3}$ & $\pot{4.37}{-3}$
& $-0.138\,\ast$ & $\pot{-1.26}{-2}\,\dagger$ & $\pot{4.41}{-2}\,\ast$
\\[1pt]
$\rho_3$ & $\pot{4.31}{-3}$ & $\pot{6.25}{-3}$ & $\pot{4.17}{-3}$ & $\pot{1.38}{-3}$
& $\pot{2.78}{-2}\,\dagger$ & $\pot{-2.06}{-3}$ & $\pot{2.66}{-3}\,\dagger$
\\[1pt]
$\rho_4$ & $\pot{5.72}{-5}$ & $\pot{-3.49}{-4}$ & $\pot{8.66}{-5}$ & $\pot{3.49}{-4}$
& $\pot{-4.79}{-3}$ & $\pot{7.63}{-4}$ & $\pot{-9.55}{-6}$
\\
\hline
\label{tab:two}
\end{tabular}
\end{table*}

\section{Constraints on different scenarios and model comparison}
\label{sec:mod}
The signal decomposition and residual eigenmodes developed in the previous sections provide new insight into the primordial energy-release analysis. This is because we collapse the multi-frequency data (order $\simeq 100$ numbers) to lower dimensions, with only a handful number of parameters required to describe the distortion signal. 
In this section, we shall present a few illustrative examples to illustrate how to use the signal eigenmodes in the analysis. In particular, we consider three different classes of early ERSs: dissipation of acoustic modes, particle annihilation and decaying particles.We summarize the parameters and eigenspectra for some examples in Table~\ref{tab:two}. 

In the following we precede in a step-by-step manner: we first give details about the parametrizations of the different ERSs (Sect.~\ref{sec:parametrization}). In Sect.~\ref{sec:distortion_signal}, we illustrate the general dependence of the distortion signals on the model parameters, while in Sect.~\ref{sec:detection} we discuss future detection limits for $\mu$ and $\mu_k$. We close our analysis in Sect.~\ref{sec:model_comparison} by providing details about direct model comparisons.

\subsection{Parametrization of the energy-release scenarios}
\label{sec:parametrization}
The two cases for the dissipation of small-scale acoustic modes presented in Table~\ref{tab:two} are computed according to \citet{Chluba2012}, using the standard parametrization of the primordial curvature power spectrum, $\mathcal{P}_\zeta(k)\equiv A_\zeta \,(k/k_0)^{\nS-1+\frac{1}{2} n_{\rm run} \ln(k/k_0)}$, with the pivot-scale $k_0=0.05\,\Mpc^{-1}$. 
The associated SD is thus a family of three parameters ($A_\zeta$, $n_S$, $n_{\rm run}$), with heating rate defined by \citep[cf.][]{Chluba2012, Chluba2013iso}
\beal
\label{eq:Q_ac_eff}
\left.\frac{\id (Q/\rho_\gamma)}{\id z}\right|_{\rm ac}
& \approx 
2 D^2  \int^\infty_{k_{\rm cut}} \mathcal{P}_\zeta(k) \, \partial_z e^{-2k^2/\kD^2} \id\ln k,
\end{align}
where $\kD(z)$ is the dissipation scales, $k_{\rm cut}\simeq 1\,\Mpc^{-1}$ denotes the $k$-space cut-off scale\footnote{The exact value of $k_{\rm cut}$ does not matter much, since it only affects the primordial $y$-distortion contribution, which we do explicitly not use constrain the underlying ERS.}, and $D^2\simeq 0.81$ is the heating efficiency for adiabatic modes (assuming the standard value for the effective number of relativistic neutrino species $N_{\rm eff}=3.046$). 
The distortion depends on the type of initial conditions (adiabatic versus isocurvature); however, as shown by \citet{Chluba2013iso}, the differences can be captured by redefining the heating efficiency, the spectral index and its running. Thus, a discussion of the SD caused by adiabatic modes sweeps the whole parameter space.
For a scale-invariant power spectrum $\id (Q/\rho_\gamma)/\id z|_{\rm ac} \propto z^{-1}$ so that $\mathcal{Q}_{\rm ac}\simeq {\rm const}$.

The two annihilation scenarios given in Table~\ref{tab:two} are for s-wave and p-wave annihilation cross-section with redshift dependence $\left<\sigma{\rm v}\right>={\rm const}$ and $\left<\sigma{\rm v}\right>\propto (1+z)$, respectively. The heating rate can be parametrized as \citep[see also][]{Chluba2011therm}
\beal
\label{eq:Q_ann}
\left.\frac{\id (Q/\rho_\gamma)}{\id z}\right|_{\rm ann}
& \approx f_{\rm ann}\,\frac{N_{\rm H}(z) (1+z)^{2+\lambda}}{H(z)\,\rho_\gamma(z)},
\end{align}
where $\lambda=0$ for s-wave and $\lambda=1$ for p-wave annihilation. Furthermore, $N_{\rm H}(z)\simeq \pot{1.9}{-7} (1+z)^3 \, {\rm cm^{-3}}$ denotes the number density of hydrogen nuclei, and $H(z)\simeq \pot{2.1}{-20} (1+z)^2\,{\rm sec^{-1}}$ is the Hubble rate, assuming radiation domination.
Thermally produced dark matter particles are expected to have s-wave annihilation cross-section with possible amplification due to Sommerfeld-enhancement \citep[e.g., see][]{Hannestad2011}. 
The p-wave scenario corresponds to a Majorana particle which either is still relativistic after freeze out [e.g., a sterile neutrino with low abundance \citep{Ho2013}], or shows ${\rm v}^{-1}\propto(1+z)^{-1}$ Sommerfeld-enhanced annihilation cross-section \citep[e.g., see][]{Chen2013}. For the non-relativistic case the cross-section drops even faster towards lower redshifts, $\left<\sigma \rm v\right>\simeq (1+z)^2$, causing practically no energy release at late times.
The annihilation efficiency, $f_{\rm ann}$, parametrizes all the dependences of the energy-release rate on the mass of the particle, its abundance, and overall annihilation cross-section.
We have $\mathcal{Q}_{\rm ann, s}\simeq {\rm const}$ and $\mathcal{Q}_{\rm ann, p}\propto (1+z)$. Fixing the redshift dependence of the annihilation cross-section (more elaborate scenarios are possible but beyond the scope of this work), the distortion is a one parameter family that only depends on $f_{\rm ann}$.

Finally, in Table~\ref{tab:two} we consider three decaying particle scenarios. The total energy release in all these cases is $\Delta \rho_\gamma/\rho_\gamma\simeq \pot{6.4}{-7}$ and the energy-release rate is parametrized as \citep[cf.][]{Chluba2011therm}
\beal
\label{eq:Q_decay}
\left.\frac{\id (Q/\rho_\gamma)}{\id z}\right|_{\rm dec}&\approx 
\epsilon_{\rm X}\,\frac{N_{\rm H}(z) (1+z_{\rm X}) \Gamma_{\rm X}}{H(z)\rho_\gamma(z)\,(1+z)}
\,\exp\left(- \Gamma_{\rm X}\, t\right)
\end{align}
with $\epsilon_{\rm X}=f_{\rm X}/z_{\rm X}$ parametrizing the energy-release efficiency, and $\Gamma_{\rm X}\simeq 2 H(z_{\rm X})$ denoting the particle decay rate. 
The efficiency factor $f_{\rm X}$ depends on the mass and abundance of the decaying particle and the efficiency of energy transfer to the baryons. In the radiation dominated era one has $\mathcal{Q}_{\rm dec}\propto \epsilon_{\rm X} \, z^{-3} \, \exp(-[z_{\rm X}/z]^2)$. The distortion is thus a two parameter family. 
Well-motivated candidates comprise excited states of dark matter \citep[e.g.,][]{Finkbeiner2007, Pospelov2007}, or other, dynamically unimportant relic particles \citep[see][for more references]{Kawasaki2005, Kohri2010, Feng2010, Pospelov2010}.

\begin{figure}
\centering
\includegraphics[width=0.99\columnwidth]{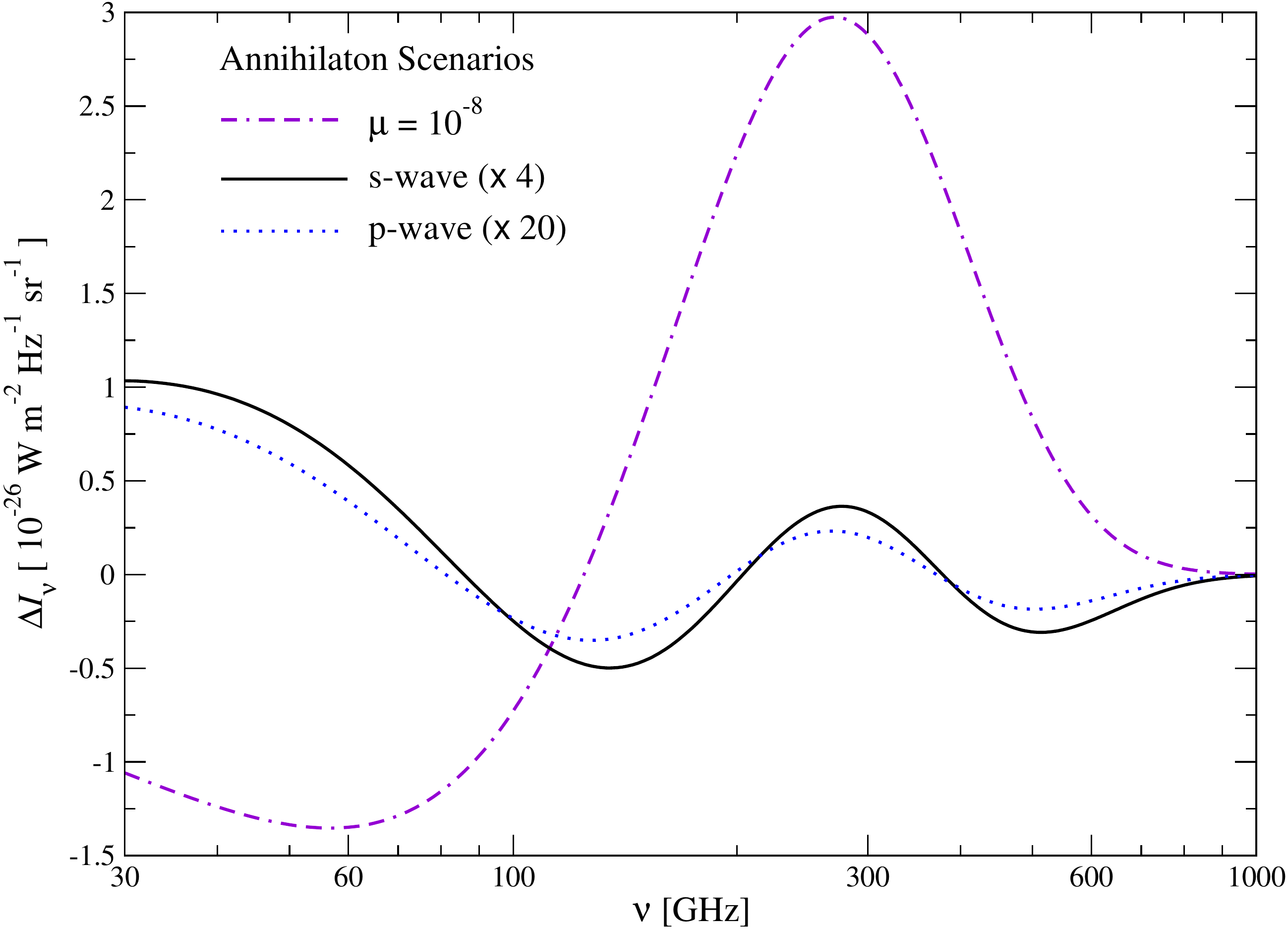}
\\[2mm]
\includegraphics[width=0.99\columnwidth]{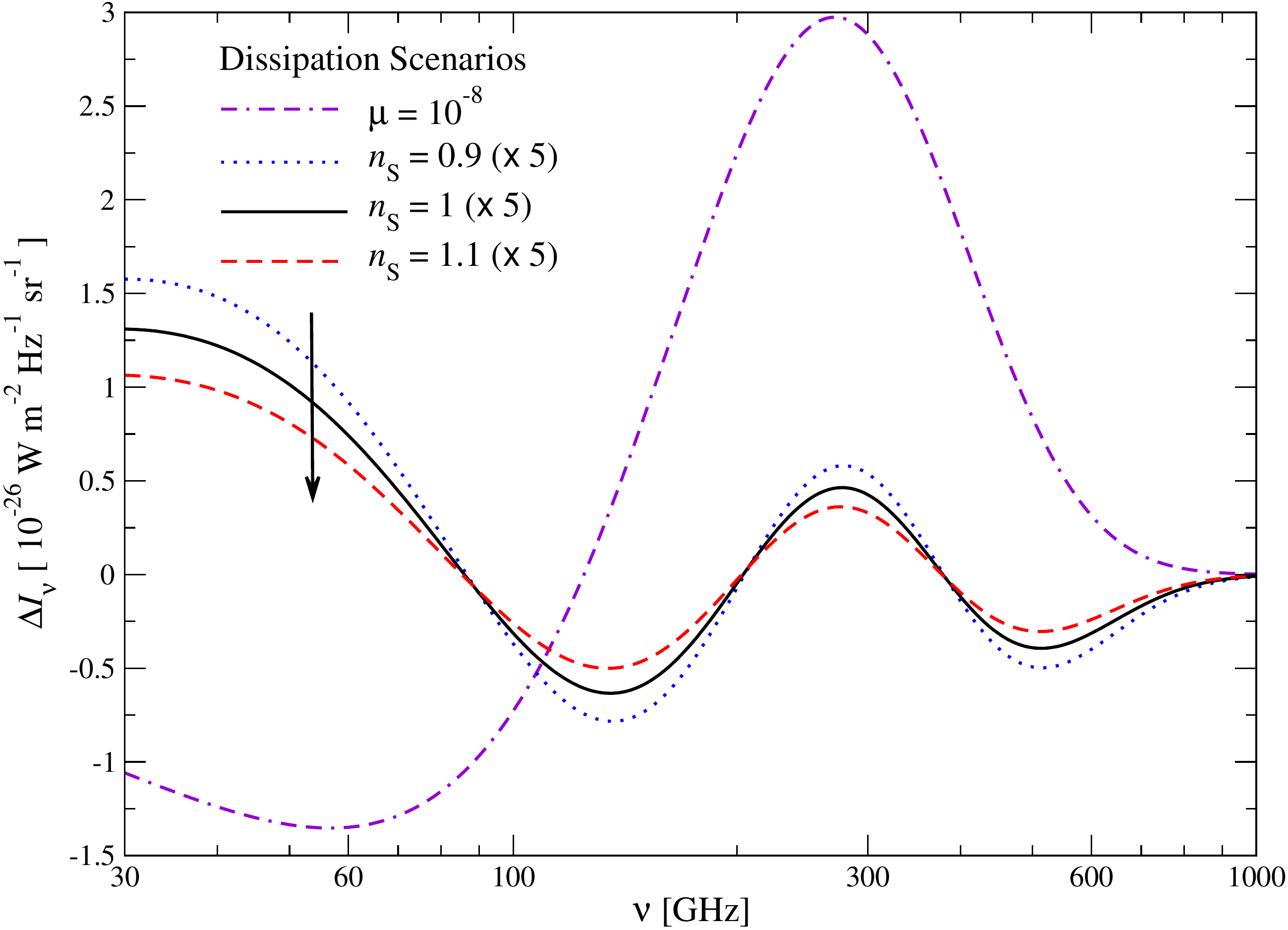}
\\[2mm]
\includegraphics[width=0.99\columnwidth]{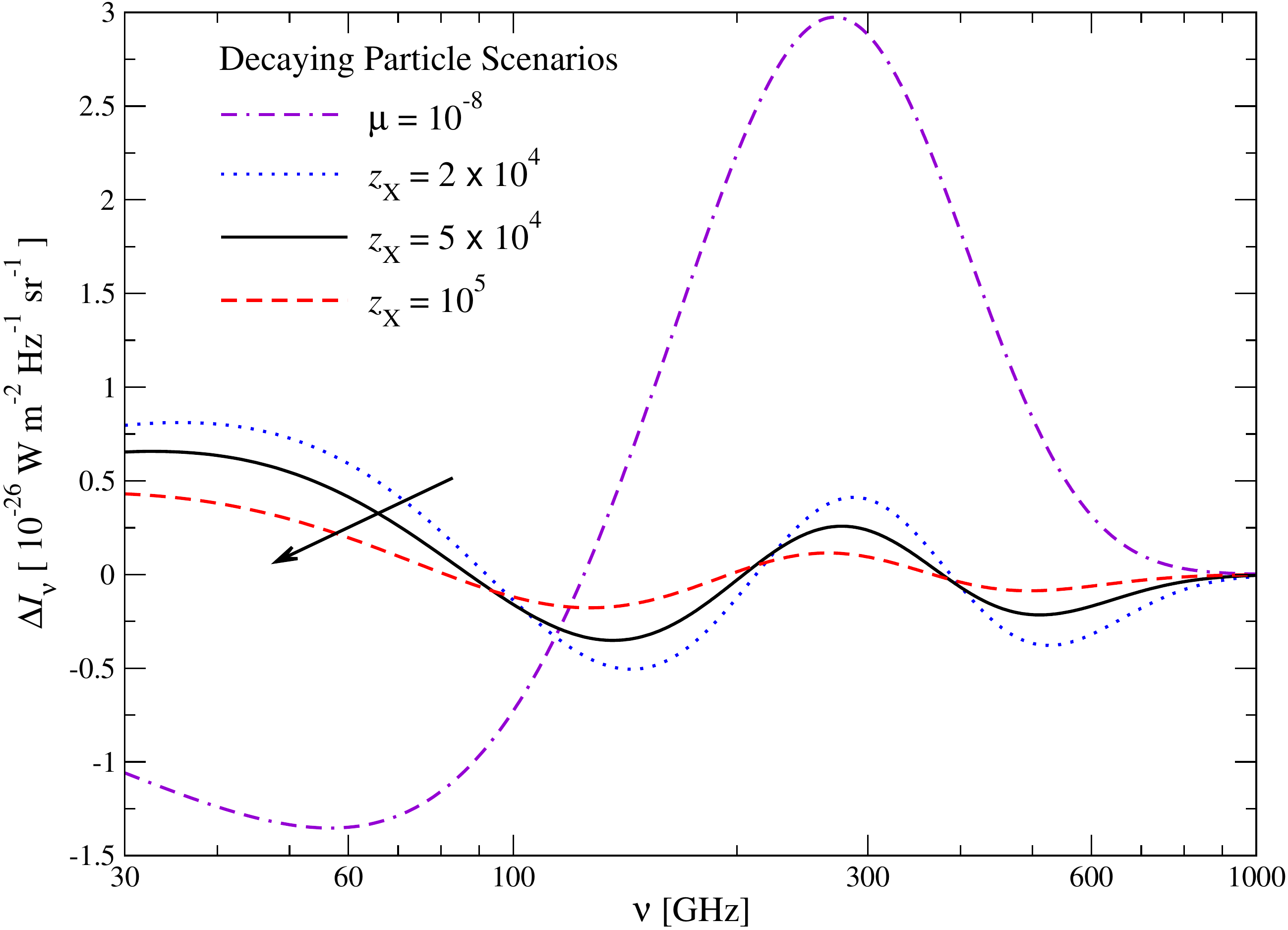}
\caption{Comparison of $\mu$-distortion with the residual distortion for different ERSs. We rescaled all cases to have $\mu=10^{-8}$. For the decomposition we used $\{\nu_{\rm min}, \nu_{\rm max},\Delta\nu_{\rm s}\}=\{30, 1000, 1\}\,\GHz$. The arrows indicate the direction of increasing parameter. For the dissipation scenarios we set $\nrun=0$. We furthermore rescaled the residual signal by the annotated values to make the distortion more visible.}
\label{fig:Signals_annihilation}
\end{figure}

\subsection{Shape of the distortion signal}
\label{sec:distortion_signal}

\subsubsection{Annihilating particles}
We start with the annihilation scenarios, for which the distortion has a fixed shape and only the overall amplitude changes, depending on the annihilation efficiency, $f_{\rm ann}$. The residual distortion signals are illustrated in Fig.~\ref{fig:Signals_annihilation} (upper panel). We scaled the {\it total} distortion such that in both cases $\mu=10^{-8}$. This emphasizes the differences in the shape of the distortion rather than its overall amplitude.
The residual distortion is significantly smaller for the p-wave scenario, showing that most of the energy is released during the $\mu$-era ($y, \mu_1 < \mu$, see Table~\ref{tab:two}). The small difference in the phase and amplitude of the residual distortion relative to $\mu$ in principle allows discerning the s- and p-wave cases, however, a detection of $\mu_1$ is required to break the degeneracy.
The values of $y$, $\mu$ and $\mu_k$ given in Table~\ref{tab:two} fully specify the shape of the distortion for s- and p-wave annihilation scenarios and all other cases can be obtained by rescaling the overall amplitude appropriately.

\begin{figure}
\includegraphics[width=\columnwidth]{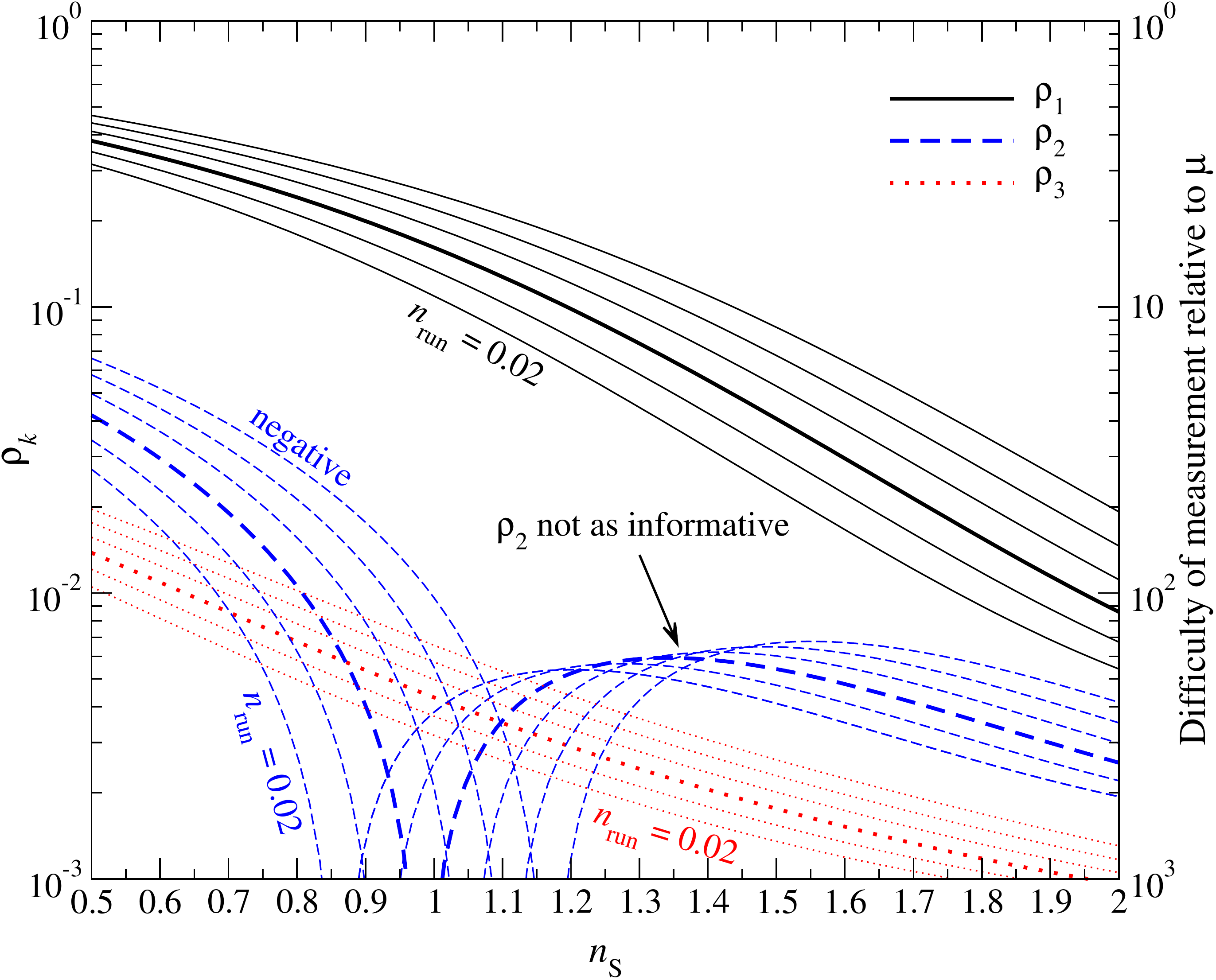}
\caption{Dependence of $\rho_k$ on $\nS$ and $\nrun$. The heavy lines are for $\nrun=0$, while all other lines are for $\nrun=\{-0.03, -0.02, -0.01, 0.01, 0.02\}$  in each group. For reference, we marked the case $\nrun=0.02$. We also indicated parts of the curves that are negative.}
\label{fig:Comp_nS_nrun}
\end{figure}

\subsubsection{Dissipation of small-scale acoustic modes}
The central panel of Fig.~\ref{fig:Signals_annihilation} illustrates the $\nS$-dependence of the residual distortion for the dissipation scenario. For different values of $\nS$, mainly the amplitude of the distortion changes, while the shape and phase of the residual distortion is only mildly affected. For $\nS>1$, relative to the scale-invariant case more energy is released at earlier times. This increase the value of $\mu$ relative to the $\mu_k$s, implying that the amplitude of the residual distortion decreases.
By measuring $\mu$ and $\mu_1$ one can thus constrain $A_\zeta$ and $\nS$ independently. However, when allowing $\nrun$ to vary, also $\mu_2$ (which is significantly harder to access) is required to distinguish these cases. Running again dominantly affects the amplitude of the residual distortion, while changes in the phase of the signal are weaker.

We can represent the dependence of the distortion on the parameters by specifying the amplitude of $\mu(A_\zeta, \nS, \nrun)$ and the ratios $\mu_k/\mu$, which are only function of $p=\{\nS, \nrun\}$. To also rank the variables in terms of the level of difficulty that is met to measure them, we furthermore weight them by their $1\sigma$-errors. This defines the new variable
\beal
\label{eq:rho_k}
\rho_k=\frac{\mu_k/\Delta \mu_k}{\mu/\Delta \mu},
\end{align}
and the distortion parameter set $p_{\rm d}=\{y, \mu, \rho_k\}$. To give an example, having $\rho_k\equiv 1$ means the value of $\mu_1$ is as hard to measure as $\mu$, while $\rho_k<1$ means it is $\rho_k^{-1}$ times harder. If $\mu$ is observable with significance, $s_\mu>1$, then $\rho_k<1$ only implies non-detections of $\mu_k$ if also $\rho_k s_k < 1$. The real advantage of this variable is that it parametrizes the {\it shape} of the energy-release history without depending on the overall amplitude. Its error is simply $\Delta \rho_k\approx (1+\rho_k^2)^{1/2}\,\Delta \mu/\mu\approx \Delta \mu/\mu$, where in the last step we assumed $|\rho_k|\ll 1$. Especially for model comparisons, this parametrization is very useful (see Sect.~\ref{sec:model_comparison}).

In Figure~\ref{fig:Comp_nS_nrun}, we show the dependence of the first three $\rho_k$ on $\nS$ and $\nrun$. 
We emphasize that in the considered range of parameters for fixed $\nrun$ the vector $\vek{\rho}=(\rho_1, \rho_2\, \rho_3)$ is uniquely linked to $\nS$. The different curves have, however, very similar shapes when varying $\nrun$. The equivalent shift in $\nS$ for each $\rho_k$ differs slightly and also depends on $\nS$, so that sensitivity to $p=\{A_\zeta, \nS, \nrun\}$ can be expected. 
Both $\rho_1$ and $\rho_3$ vary rather slowly, while $\rho_2$ changes sign around $\nS\simeq 1$. This indicates that if $\mu$, $\mu_1$ and $\mu_2$ are measurable, most sensitive constraints on $p=\{A_\zeta, \nS, \nrun\}$ are expected around $\nS\simeq 1$. However, since in the considered range $\rho_2\simeq 10^{-3}-10^{-2}$, it is already clear that pretty high precision for the measurement of $\mu$ is needed (see Fig.~\ref{fig:Limit_nS_nrun}). 
We can furthermore see that around $\nS\simeq 1.2-1.4$, the dependence of $\rho_2$ on $\nrun$ is rather weak, and degenerate with $\nS$. This indicates that high sensitivity is required to discern different cases in this regime.

\subsubsection{Decaying relic particles}
The lower panel of Fig.~\ref{fig:Signals_annihilation} illustrates the dependence of the distortion signal caused by a decaying particle on its lifetime. Shorter lifetime means most energy is released at earlier times so that the distortion is closer to a pure $\mu$-distortion with a smaller residual distortion.
Increasing the lifetime (lowering $z_{\rm X}$), the overall amplitude of the residual distortion increases and shows a small phase shift towards higher frequencies. These are the main signatures that allow measuring the particle lifetime, and in principle only $\mu$ and $\mu_1$ are needed to achieve this goal. 

\begin{figure}
\centering
\includegraphics[width=\columnwidth]{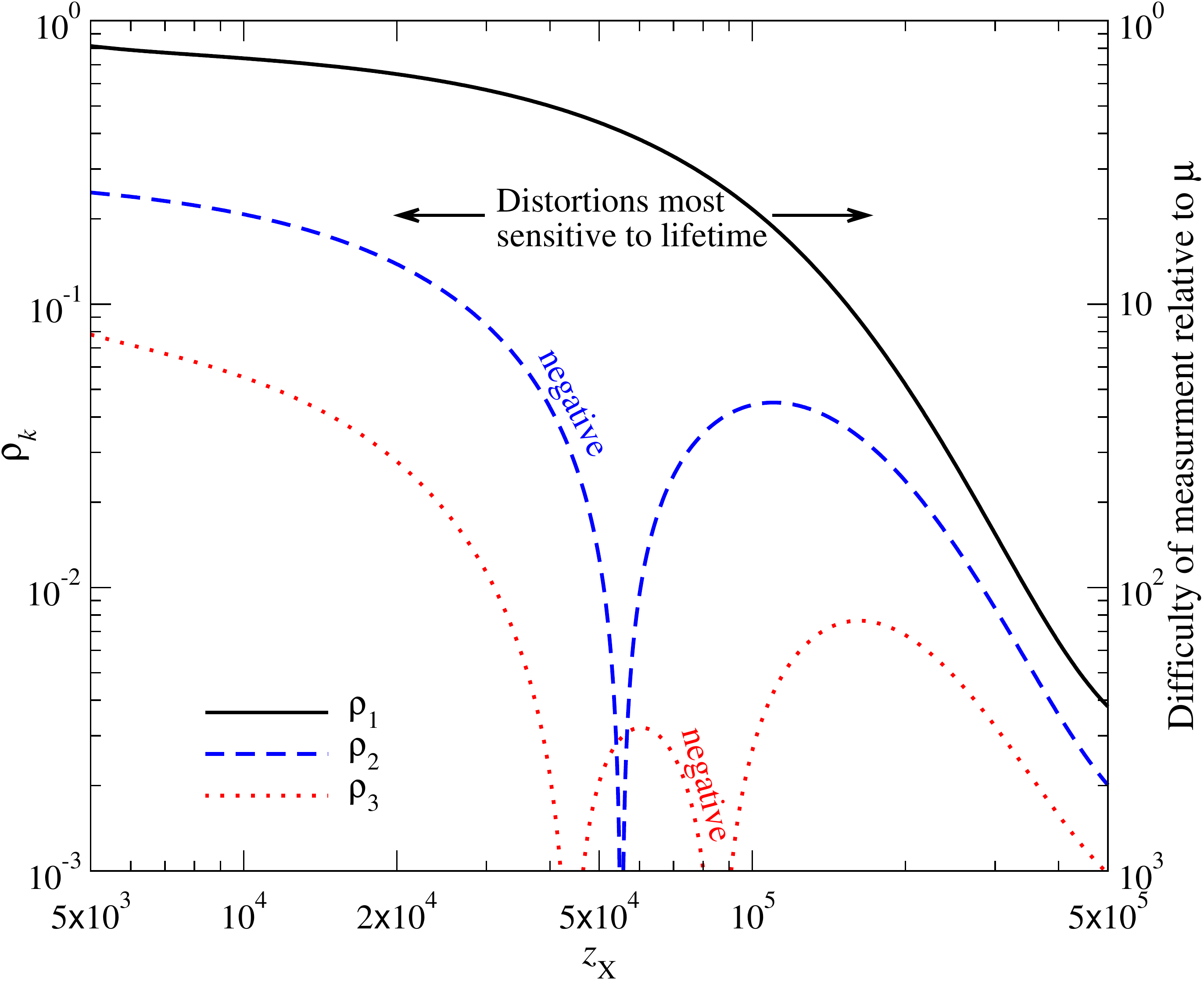}
\caption{Dependence of $\rho_k$ on the lifetime of the decaying particle. We indicated parts of the curves that are negative.}
\label{fig:Comp_dec}
\end{figure}

In Fig.~\ref{fig:Comp_dec}, we show the eigenspectra, $\rho_k$, for decaying particle scenarios as a function of $z_{\rm X}$. In the considered range, the vector $\vek{\rho}$ is uniquely linked to $z_{\rm X}$, e.g., since $\rho_1$ never shows any degeneracy. One can thus hope to be able to constrain $p=\{\epsilon_{\rm X}, z_{\rm X}\}$ using SD measurements. For $z_{\rm X}\lesssim 10^4$, all curves become rather flat, so that CMB distortions are less sensitive to the precise lifetime of the particle and a large uncertainty in $z_{\rm X}$ is expected. Similarly, at high redshift ($z\gtrsim \pot{2}{5}$) the amplitude $\rho_1$ decreases so that sensitivity to the particle lifetime is diminished \citep[see also][]{Chluba2013fore}.
Around $z\simeq \pot{5}{4}$, $\rho_1$ and $\rho_2$ show the largest variation with $z_{\rm X}$ and thus the highest sensitivity to the particle lifetime. In particular, $\rho_1$ and $\rho_2$ both change sign in this range. This also suggests that finding $\rho_2<0$ provides indication for a decaying particle over a dissipation scenario, giving one possible criterion for model selection (see Sect.~\ref{sec:model_comparison}).
 
\subsection{Detectability of the distortion signal}
\label{sec:detection}

\subsubsection{Annihilating particles}
The signal caused by the s-wave annihilation scenario given in Table~\ref{tab:two} is undetectable with a {\it PIXIE}-like experiment, but could be detected at $\simeq 3\sigma$ with {\it PRISM}. The distortion signal depends on only one free parameter, $f_{\rm ann, s}$, for which we chose a value that is close to the upper $1\sigma$ bound derived from current CMB anisotropy measurements \citep{Galli2009, Huetsi2009, Slatyer2009, Huetsi2011, Planck2013params}. The spectral sensitivity needs to be increased $\simeq 4$ times over {\it PIXIE} to detect the s-wave $\mu$-distortion signature, while a factor of $\simeq 22$ improvement is needed to recover the first distortion eigenmode, $\mu_1$.  

The considered p-wave scenario illustrates how the eigenspectra change when the redshift scaling of the energy-release rate is modified. Since most of the energy is liberated at early times, the distortion signal is dominated by $\mu$ (see Sect.~\ref{sec:distortion_signal}). With a {\it PIXIE}-like experiment the first distortion eigenmode remains undetectable and even {\it PRISM} will not suffice to measure this number values. Again, the distortion is just determined by $f_{\rm ann, p}$, but since the eigenspectrum differs from the one of the s-wave scenario, by measuring the first eigenmode these are distinguishable. For $f_{\rm ann, s}\simeq \pot{2}{-23}\,{\rm eV\,sec^{-1}}$ and $f_{\rm ann, p}\simeq \pot{4.8}{-29}\,{\rm eV\,sec^{-1}}$, s- and p-wave scenarios both give rise to $\mu\simeq \pot{4}{-9}$. In this case, $\mu_{1, \rm p}\simeq \pot{9.7}{-10}$ for the p-wave, and $\mu_{1, \rm s}\simeq \pot{6.4}{-9}$ for the s-wave case. Thus, by increasing the sensitivity $\simeq 22$ times over {\it PIXIE}, the s- and p-wave scenarios in principle become distinguishable ($\mu_1$ from the s-wave scenario would be detected at $1\sigma$, while for a p-wave case $\mu_1$ should be consistent with zero).
These findings are in good agreement with those of \citet{Chluba2013fore}, where an MCMC analysis was used.

A {\it PIXIE}-type experiment could place independent $1\sigma$-limits of $f_{\rm ann, s}\lesssim \pot{6.9}{-23}\,{\rm eV \, sec^{-1}}$ and $f_{\rm ann, p}\lesssim \pot{1.6}{-28}\,{\rm eV \, sec^{-1}}$ on the annihilation efficiency, with practically all  the information coming from $\mu$ itself \citep[see also][]{Chluba2013fore}.
Using the parametrization according to the recent {\it Planck} papers \citep{Planck2013params}, this implies\footnote{$f_{\rm ann, s}\equiv \pot{1.5}{-17}\,{\rm eV\,kg \,m^{-3}}\,p_{\rm ann, s}$} $p_{\rm ann, s}<\pot{9.2}{-6} \,{\rm m^3 \, kg^{-1} \,s^{-2}}$ ($95\%$ c.l.), which is several times weaker than the current CMB anisotropy limit obtained with {\it Planck} ($p_{\rm ann}<\pot{3.1}{-6} \,{\rm m^3 \, kg^{-1} \,s^{-2}}$). Uncertainties in the modeling of the energy-deposition rates indicate that this limit is in fact slightly weaker, but still once the full polarization data from {\it Planck} is included, one does expect an improvement of this bound to $p_{\rm ann}<\pot{1.7}{-7} \,{\rm m^3 \, kg^{-1} \,s^{-2}}$ \citep{Galli2013}.
Thus, only an increase of the spectral sensitivity by a factor of $\simeq 50$ over {\it PIXIE} could make future CMB distortion measurements a competitive probe for annihilating dark matter particles, although one should emphasize that SD would still give an independent measurement, suffering from very different systematics.
In the future, {\it PRISM} might allow direct detection of a dark matter annihilation signature, if $p_{\rm ann, s}\gtrsim \pot{4.6}{-7} \,{\rm m^3 \, kg^{-1} \,s^{-2}}$.

\subsubsection{Dissipation of small-scale acoustic modes}
From measurements of the CMB anisotropies at large scales we have $A_\zeta\simeq \pot{2.2}{-9}$, $\nS\simeq 0.96$ and $\nrun\simeq -0.02$ \citep{Planck2013params}. Using these values and extrapolating all the way to wavenumber $k\simeq \pot{\rm few}{4}\,\Mpc^{-1}$, we obtain the distortion parameters given in Table~\ref{tab:two}. For comparison, we also show the case with no running and scale-invariant power spectrum.
For these two dissipation scenarios the $y$-parameter will contribute at a few $\sigma$-level to $y^\ast=y_{\rm re}+y$ for a {\it PIXIE}-like experiment, while no information can be extracted from the residual distortion (none of the $\mu_k$ can be detected). For a scale-invariant power spectrum also a non-vanishing $\mu$-parameter could be found ($\simeq 2.3\sigma$) with a {\it PIXIE}-like experiment \citep[see also][]{Chluba2012}. 
For {\it PRISM}, a more than $20\,\sigma$ detection of $\mu$ for a scale-invariant power spectrum should be feasible, while for $A_\zeta\simeq \pot{2.2}{-9}$, $\nS\simeq 0.96$ and $\nrun\simeq 0$ we expect a $\simeq 17\,\sigma$ detection of $\mu$.

Since the $y$-parameter is degenerate with $y_{\rm re}$, only $\mu$ can be used to place constraints in these cases, however, the degeneracy among model parameters is very large. For example, the small difference in the value of $\mu$ for the two considered cases can be compensated by adjusting $A_\zeta$ at small scales.
Increasing the sensitivity 10 times over {\it PIXIE} will allow an additional detection of the first eigenmode ($\simeq 3.7\sigma$ and $\simeq 2.0\sigma$ for the two dissipation scenarios given Table~\ref{tab:two}, respectively). In this case, the parameter degeneracies ($A_\zeta$, $\nS$, and $\nrun$) can be partially broken (two numbers, $\mu$ and $\mu_1$, are used to limit three variables). Improvement by another factor of $10$ allows marginal detections of the second mode amplitude, but to truly constrain the shape of the small-scale power spectrum (assuming the standard parametrization) using SD data alone an overall factor $\gtrsim 200$ over {\it PIXIE} will be necessary, making this application of SDs rather futuristic \citep[see also][]{Chluba2013fore}.

\begin{figure}
\raggedright
\includegraphics[width=1.25\columnwidth, ]{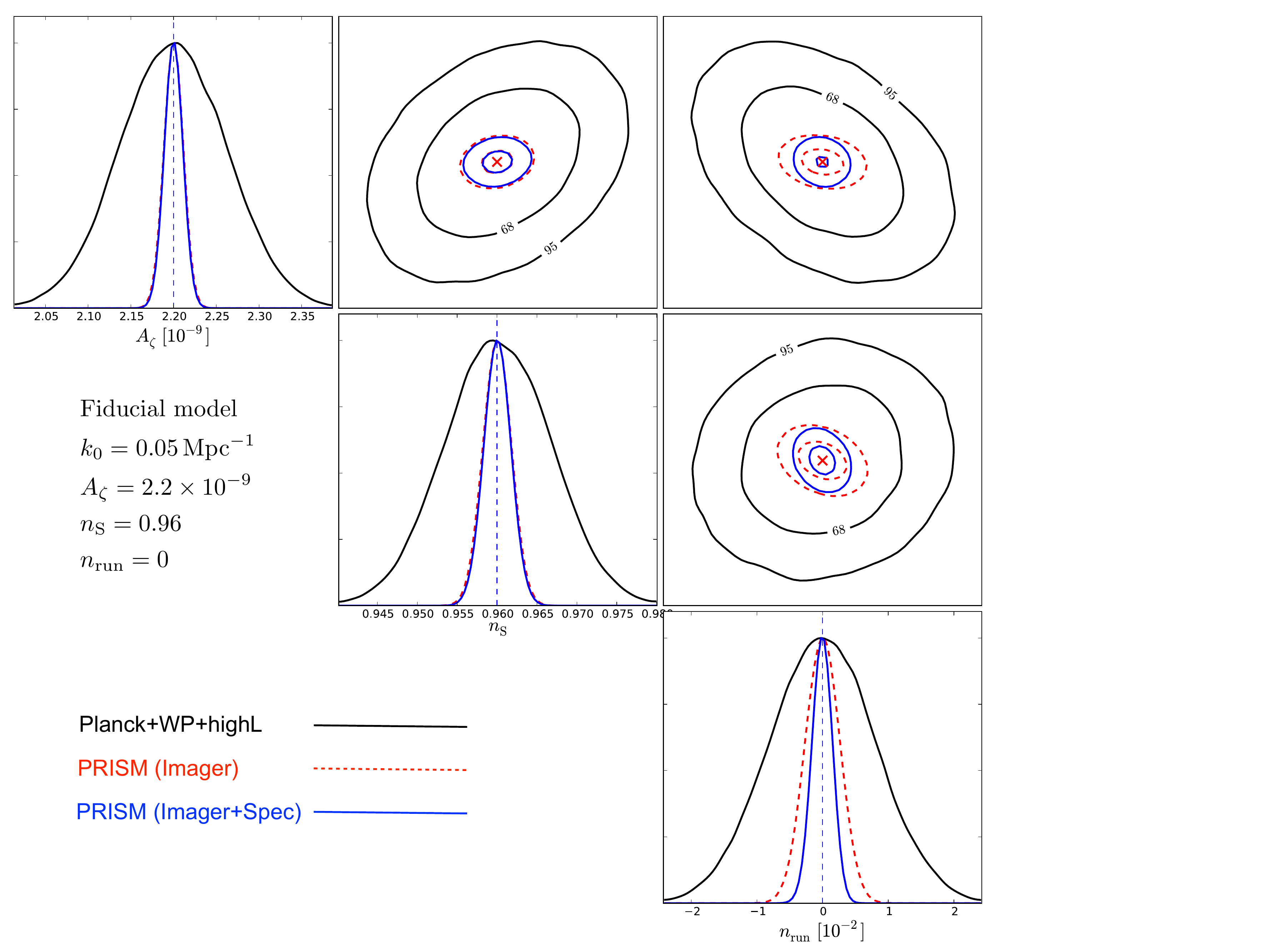}
\\[2mm]
\includegraphics[width=1.13\columnwidth]{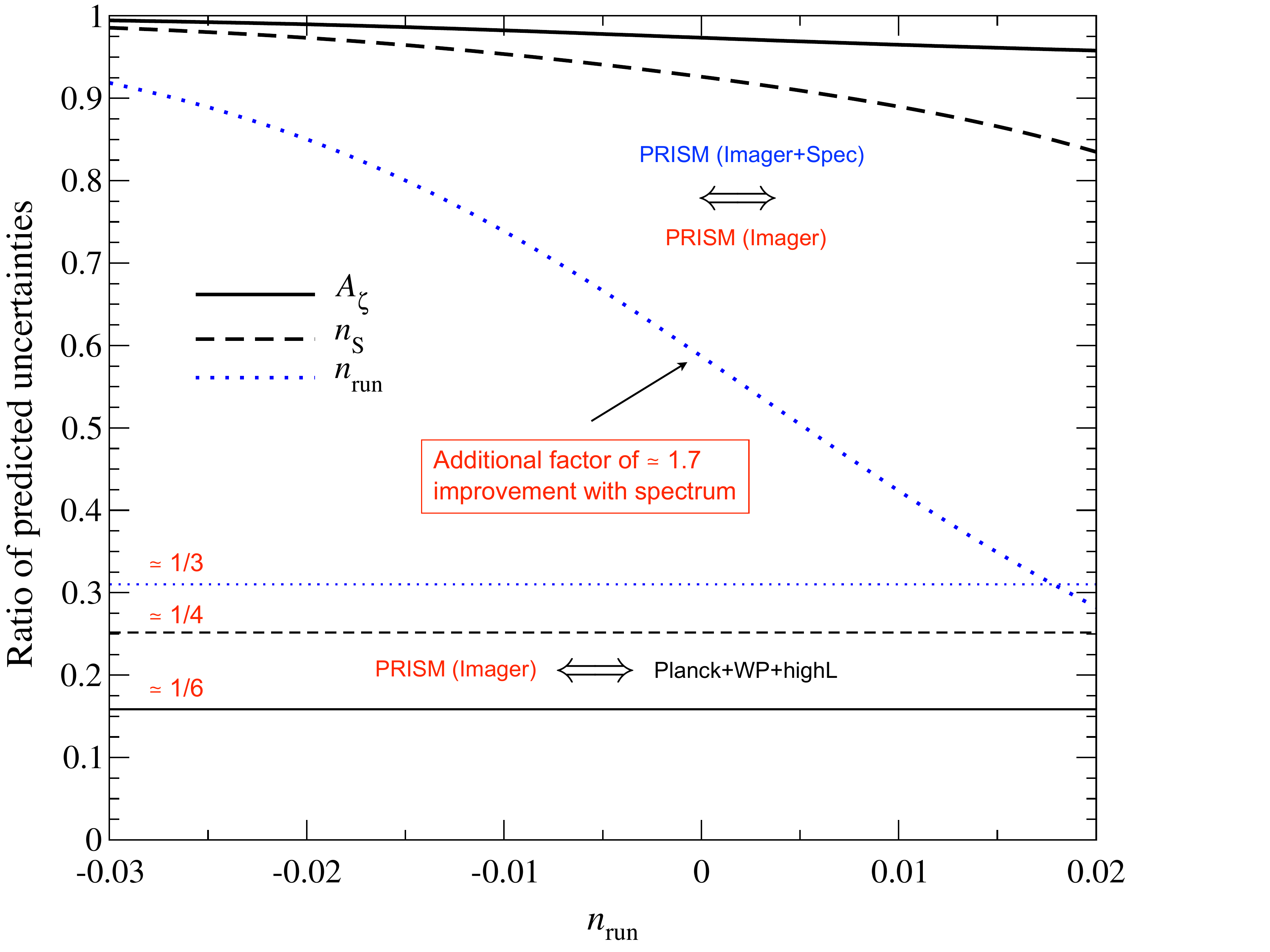}
\caption{Forecasted constraints on $A_\zeta$, $\nS$ and $\nrun$. The case labeled Planck+WP+highL uses the published  covariance matrix of Planck with inclusion of WMAP polarization data and the high $\ell$ data from ACT and SPT. The case labeled {\it PRISM} is based on estimates given in \citet{PRISM2013WPII} for the {\it PRISM} imager and spectrometer part. The upper panel shows the 2D contours and marginalized distributions for $A_\zeta$, $\nS$ and $\nrun$, while the lower panel illustrates the expected improvement (decrease) in the measurement uncertainty of the {\it PRISM} imager over Planck (horizontal lines) and the additional gain when adding the {\it PRISM} SD data. Note that the {\it PRISM} spectrometer is about one order of magnitude more sensitive that {\it PIXIE}.}
\label{fig:combination_constraints}
\end{figure}

These simple estimates indicate that SD alone only provide competitive constraints on $A_\zeta$, $\nS$ and $\nrun$ for much higher spectral sensitivity; however, SD data can help to slightly improve the constraint on $\nrun$ when combined with future CMB anisotropy measurements \citep[see][for similar discussion]{Powell2012, Khatri2013forecast}. This is simply because both $A_\zeta$ and $\nS$ can be tightly constrained with the CMB anisotropy measurement, while the long lever arm added with SD measurements improves the sensitivity to running of the power spectrum. We illustrate this in Fig.~\ref{fig:combination_constraints} for {\it PRISM} and current constraints from Planck, WMAP \citep[e.g.,][]{Komatsu2010} and high $\ell$ data from ACT \citep[e.g.,][]{Dunkley2010} and SPT \citep[e.g.,][]{spt}. For the standard power spectrum, SD data add little with respect to $A_\zeta$ and $\nS$, but do improve the constraint on $\nrun$ for $\nrun>-0.02$. However, similar improvements can also be expected from future small-scale (Stage IV) CMB measurements \citep{Abazajian2013}. At {\it PIXIE}'s sensitivity, we do not find any significant improvement of power spectrum constraints derived from CMB anisotropy measurements when adding the SD data.

\subsubsection{Dissipation of small-scale acoustic modes: generalization}
The above statements assume that the three-parameter Ansatz for the primordial curvature power spectrum holds for more than six to seven decades in scales. Strictly speaking, the exact shape and amplitude of the small-scale power spectrum are unknown and a large range of viable early-universe models \citep[e.g.,][]{1989PhRvD..40.1753S, 1992JETPL..55..489S, 1994PhRvD..50.7173I, 1996NuPhB.472..377R, Stewart1997b, 1998PhRvD..58f3508C,1998GrCo....4S..88S, 2000PhRvD..62d3508C, Hunt:2007dn, 2008PhRvD..77b3514J, Neil2009I, Barnaby2010, Ido2010, 2011JCAP...01..030A,2012arXiv1201.4848C} producing enhanced small-scale power exist \citep[see,][for more examples and simple SD constraints]{Chluba2012inflaton}. 
Observationally, the amplitude of the primordial small-scale power spectrum is limited to $A_\zeta\lesssim 10^{-7}-10^{-6}$ at wavenumber $3\,\Mpc^{-1}\lesssim k \lesssim \pot{\rm few}{4}\,\Mpc^{-1}$ (the range that is of most interest for CMB distortions) using ultra compact mini haloes \citep{BSA11, Scott2012}. Although slightly model-independent, this still leaves a lot of room for non-standard dissipation scenarios, with enhanced small-scale power.

\begin{figure}
\centering
\includegraphics[width=1.02\columnwidth]{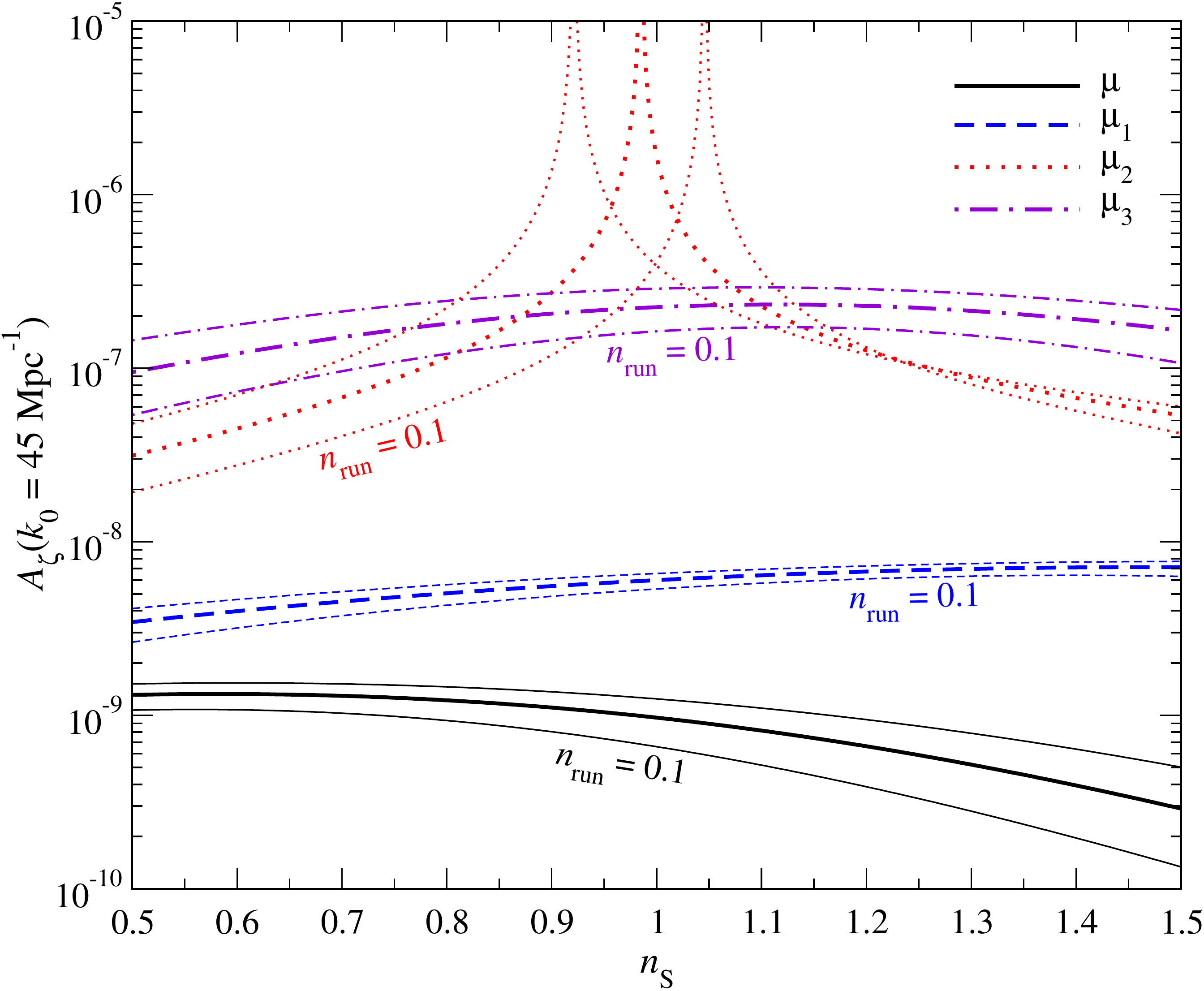}
\caption{$1\sigma$-detection limits for $\mu$, $\mu_1$, $\mu_2$ and $\mu_3$ caused by dissipation of small-scale acoustic modes for {\it PIXIE}-like settings. We used the standard parametrization for the power spectrum with amplitude, $A_\zeta$, spectral index, $\nS$, and running $\nrun$ around pivot scale $k_0=45\,\Mpc^{-1}$. 
The heavy lines are for $\nrun=0$, while all other lines are for $\nrun=\{-0.1, 0.1\}$  in each group. For reference we marked the case $\nrun=0.1$.}
\label{fig:Limit_nS_nrun}
\end{figure}

To study how well the small-scale power spectrum might be constrained by future SD measurements, it is convenient to consider the shape and amplitude of the curvature power spectrum at $3\,\Mpc^{-1}\lesssim k \lesssim \pot{\rm few}{4}\,\Mpc^{-1}$ independent of the large-scale power spectrum. 
We therefore change the question as follows: by shifting the pivot scale to $k_0=45\,\Mpc^{-1}$ (corresponding to heating around $z_{\rm diss}\simeq \pot{4.5}{5}[k/10^3\,\Mpc^{-1}]^{2/3}\simeq \pot{5.7}{4}$) and using the standard parametrization for the power spectrum, how large does the power spectrum amplitude, $A_\zeta(k_0=45\,\Mpc^{-1})$, have to be to obtain a $1\sigma$-detection of $\mu$, $\mu_1$, $\mu_2$ and $\mu_3$, respectively? The results of this exercise are shown in Fig.~\ref{fig:Limit_nS_nrun} for {\rm PIXIE} settings.
Around $\nS\simeq 1$, a detection of $\mu$ is possible for $A_\zeta\gtrsim 10^{-9}$, while $A_\zeta\gtrsim \pot{6}{-9}$ is necessary to also have a detection of $\mu_1$. In this case, two of the three model parameters can in principle be constrained independently. To also access information from $\mu_2$ and $\mu_3$ one furthermore needs $A_\zeta\gtrsim 10^{-7}$. In this case, we could expect to break the degeneracy between all three parameters with a {\it PIXIE}-type experiment. 

The detection limits depend both on the value of $\nS$ and $\nrun$. For $\nrun<0$, in total less energy is released so that  larger $A_\zeta$ is required for a detection. For $\nS>1$, more power is found at $k>45\,\Mpc^{-1}$, so that more energy is released in the $\mu$-era. Consequently, the $\mu$-distortion can be detected for lower $A_\zeta$. Similarly, when increasing $\nS$, less energy is released around $z\simeq \pot{5}{4}$, so that the value of $\mu_1$ decreases. Thus, larger $A_\zeta$ is required to warrant a detection of $\mu_1$.

The above statements can be phrased in another way. Assuming $A_\zeta\simeq 10^{-9}$ and $\nS\simeq 1$, at least a factor of $5$ improvement over {\it PIXIE} sensitivity is needed to allow constraining combinations of two power spectrum parameters. To determine all $p=\{A_\zeta, \nS, \nrun\}$ independently an overall factor of $\gtrsim 200$ improvement over {\it PIXIE} sensitivity is required, although in this (very conservative) case the corresponding constraints would still not be competitive with those obtained using large-scale CMB anisotropy measurements.

\begin{figure}
\includegraphics[width=\columnwidth]{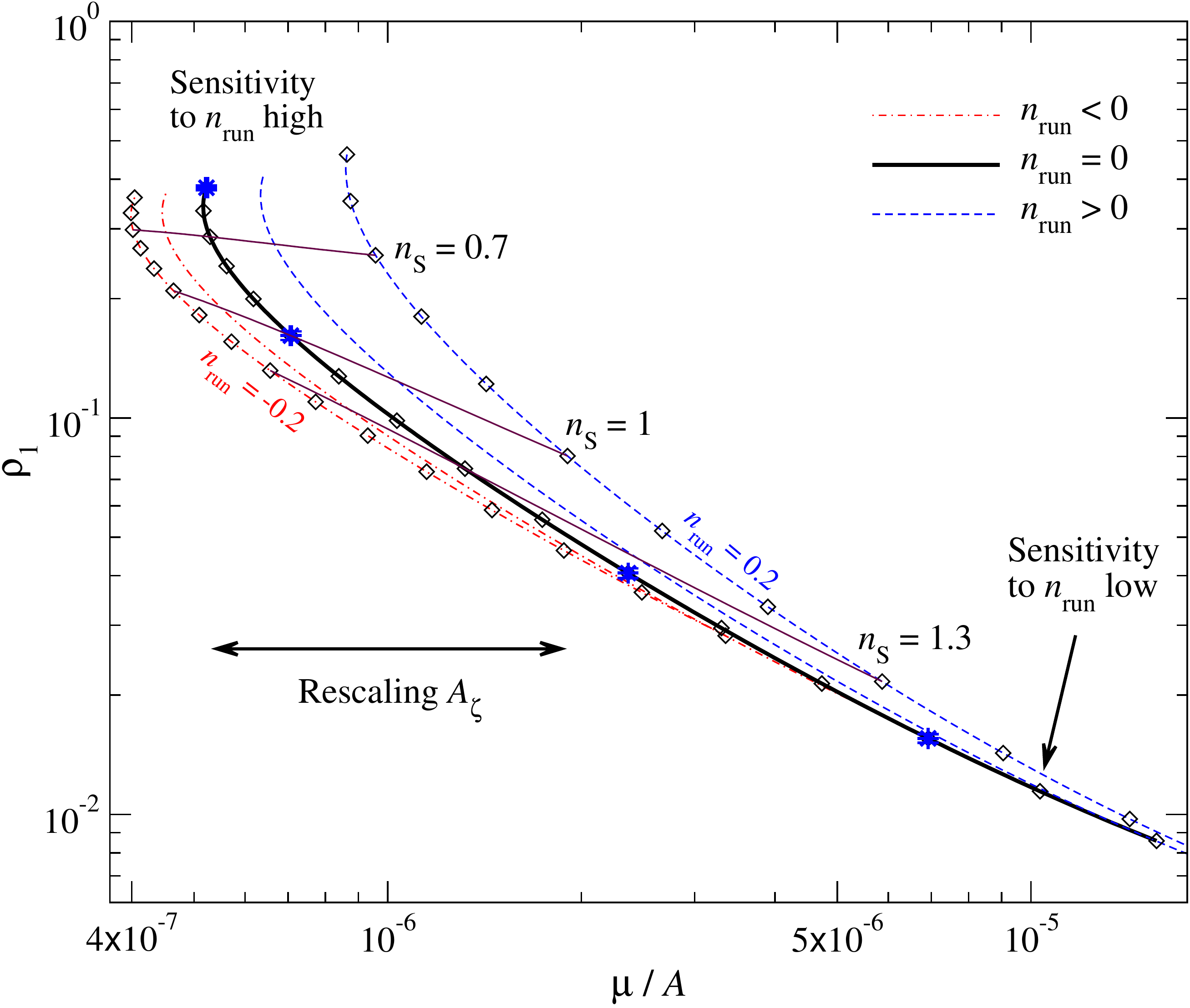}
\\[2mm]
\includegraphics[width=1.02\columnwidth]{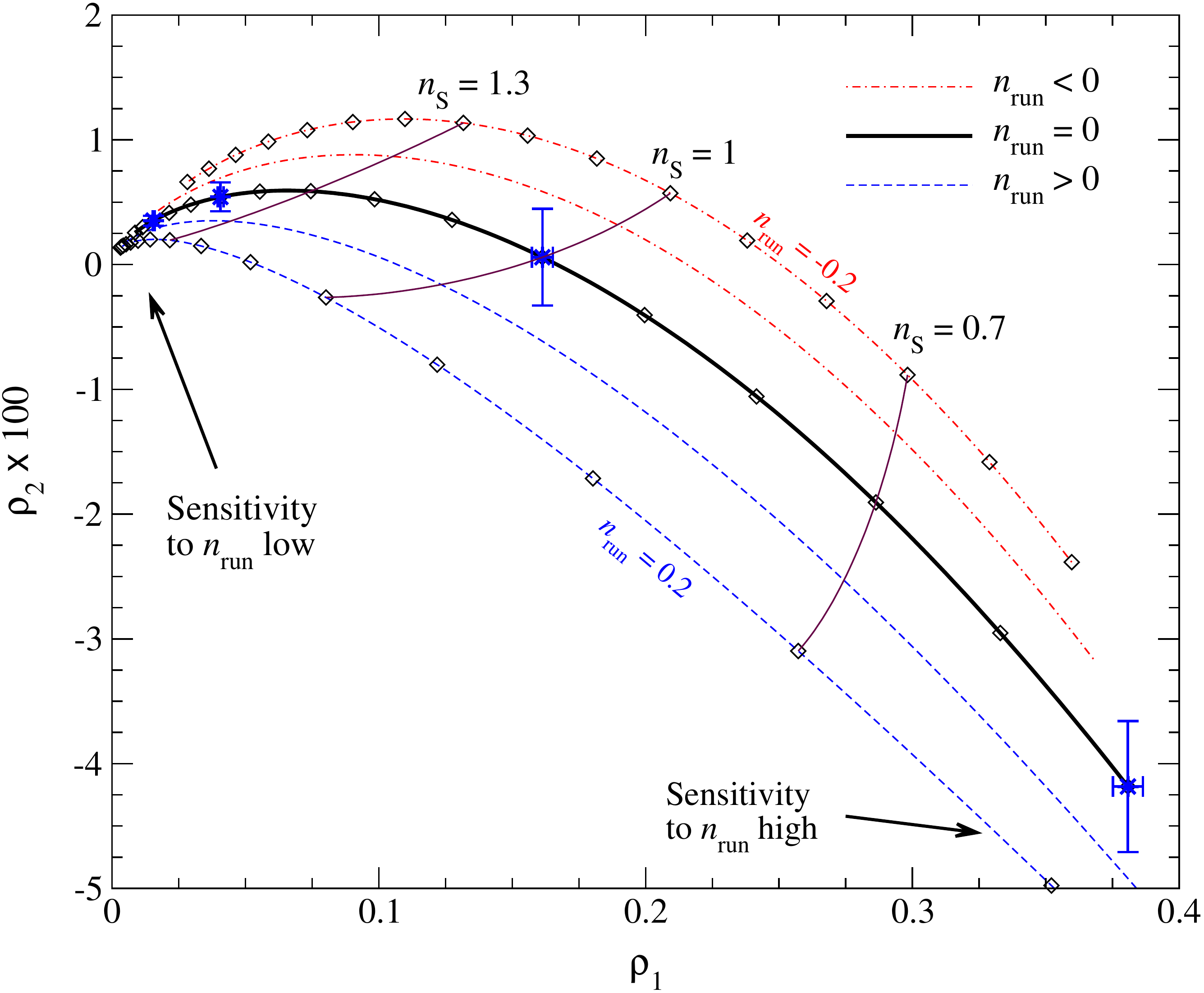}
\caption{Parameter range of $\mu$, $\mu_1$ and $\mu_2$ for dissipation scenarios. We assumed {\it PIXIE} settings with 5 times its sensitivity, and a power spectrum amplitude $A_\zeta(k_0=45\,\Mpc^{-1})=\pot{5}{-8}$ (i.e. $A\equiv A_\zeta/\pot{5}{-8}$). 
The heavy solid black lines are for $\nrun=0$, while the thin solid brown lines indicate $\nS={\rm const}$. The other light lines are for $\nrun=\{-0.2, -0.1, 0.1, 0.2\}$. The open symbols mark $\nS$ in steps $\Delta \nS=0.1$. The blue symbols with error bars (tiny in the upper panel) are for $\nS=\{0.5, 1, 1.5, 1.8\}$ and $\nrun=0$. They illustrate how the error scales in different regions of the parameter space.
Measurements in the $\mu-\rho_1$ plane can be used to fix the overall amplitude of the small-scale power spectrum for a given pair $\nS$ and $\nrun$, but no independent constraint on $\nS$ and $\nrun$ can be deduced. The constraints on $\rho_1$ and $\rho_2$ allow us to partially break the remaining degeneracy.
}
\label{fig:Diss_mu_rho_1_rho_2}
\end{figure}

\begin{figure}
\centering
\includegraphics[width=\columnwidth]{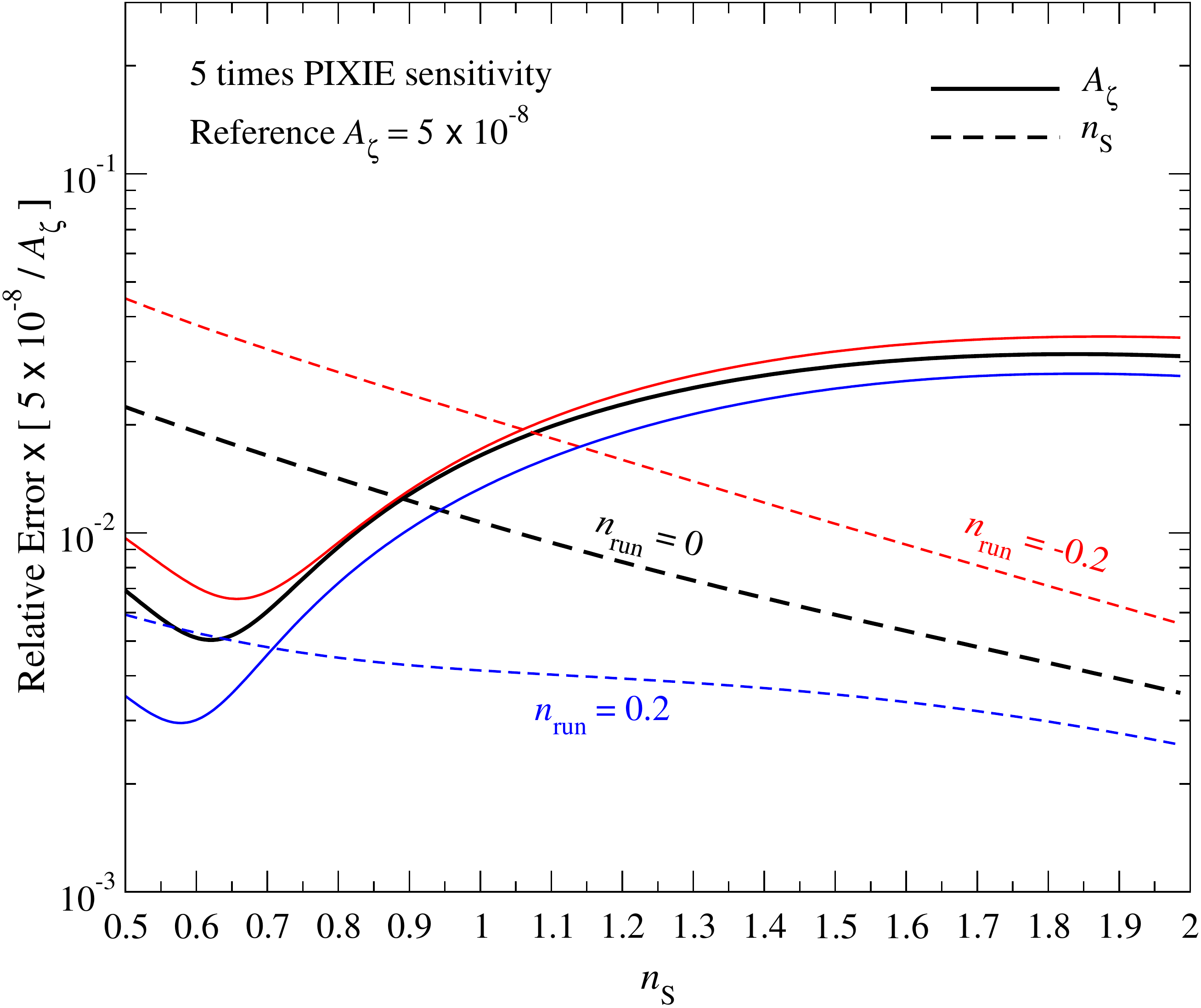}
\\[2mm]
\includegraphics[width=\columnwidth]{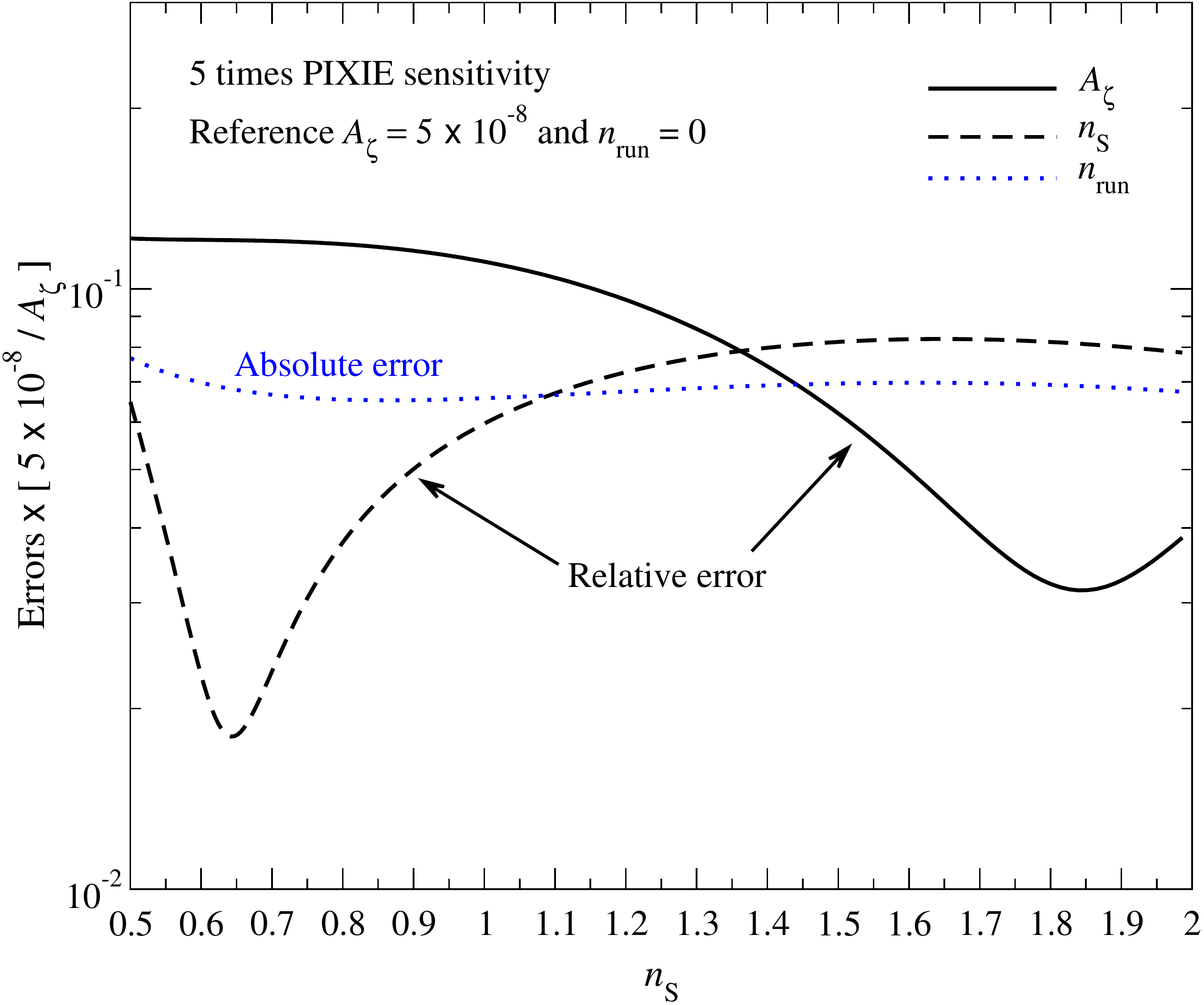}
\caption{Expected uncertainties of $A_{\zeta}(k_0=45\,\Mpc^{-1})$, $\nS$, and $\nrun$ using measurements of $\mu$, $\mu_1$, and $\mu_2$. We assumed 5 times the sensitivity of {\it PIXIE} and $A_{\zeta}=\pot{5}{-8}$ as reference value (other cases can be estimated by simple rescaling). 
For the upper panel we also varied $\nrun$ as indicated, while in the lower panel it was fixed to $\nrun=0$.
}
\label{fig:Diss.error}
\end{figure}

We can also ask the question of how well the power spectrum parameters can be constrained for different cases. If only $\mu$ is available, then the corresponding constraints on small-scale power spectrum parameters remain rather weak, but could still be used to limit the parameters space \citep[e.g.,][]{Chluba2012, Chluba2012inflaton}. 
If $\mu$ and $\mu_1$ can be accessed, we can limit the overall amplitude of the power spectrum for given pairs of $\nS$ and $\nrun$. 
This can be seen from the upper panel of Fig.~\ref{fig:Diss_mu_rho_1_rho_2}, where we illustrate the possible parameter space of $\mu$, $\rho_1\propto \mu_1/\mu$ and $\rho_2\propto \mu_2/\mu$ in some range of $\nS$ and $\nrun$. For the considered sensitivity and fiducial value of $A_\zeta$, the errors on $\mu$ and $\rho_1$ are very small, but since $A_\zeta$ can be adjusted without affecting $\rho_1$, the measurement is not independent of $\nS$ and $\nrun$. 

If in addition $\mu_2$ can be constrained, then the degeneracy can be broken. 
For {\it PIXIE}-settings and $\nS\simeq 0.96$, this is only conceivable if the amplitude of the small-scale power spectrum is $A_\zeta\gtrsim 10^{-7}-10^{-6}$ (see Fig.~\ref{fig:Limit_nS_nrun}).
As the lower panel of Fig.~\ref{fig:Diss_mu_rho_1_rho_2} indicates, the relative dependence on $\nrun$ seems rather similar in all parts of parameter space: although the absolute distance between the lines varies relative to the error bars they seem rather constant. 
To show this more explicitly, from $\mu$, $\mu_1$, and $\mu_2$ we compute the expected $1\sigma$-errors on $A_{\zeta}(k_0=45\,\Mpc^{-1})$, $\nS$, and $\nrun$ around the fiducial value using the Fisher information matrix, $\mathcal{F}_{ij}=\Delta\mu^{-2}\,\partial_{p_i} \mu\,\partial_{p_j} \mu+\sum_k \Delta\mu_k^{-2}\,\partial_{p_i} \mu_k\partial_{p_j} \mu_k$, with $p\equiv \{A_{\zeta}, \nS, \nrun\}$.
Figure~\ref{fig:Diss.error} shows the corresponding forecasts assuming {\it PIXIE}-setting but with 5 times its sensitivity. 
If only $p\equiv \{A_{\zeta}, \nS\}$ are estimated for fixed $\nrun$, the errors of $A_{\zeta}$ and $\nS$ are only a few percent. When also trying to constrain $\nrun$, we see that the uncertainties in the values of $A_{\zeta}$ and $\nS$ increase by about one order of magnitude, with an absolute error $\Delta \nrun\simeq 0.07$ rather independent of $\nS$. 

Little information is added when also $\mu_3$ can be measured (we find small differences in the constraints for small $\nS$ when $\nrun$ is varied), although for model comparison $\mu_3$ could become important.
Also, for power spectra that result in $\mu_2\simeq 0$, the detection limit of $\mu_3$ is much lower (see Fig.~\ref{fig:Limit_nS_nrun}), so that the combination of $\mu_2$ consistent with zero but $\mu_3>0$ provides a useful confirmation of the dissipation scenario.

We can also use the results of Figure~\ref{fig:Diss.error} to estimate the expected uncertainties for other cases.
Adjusting the spectral sensitivity by a factor $f=\Delta I_{\rm c}/[10^{-26}\,\UnitInu]$, all curves can be rescaled by this factor to obtain the new estimates for the errors. Similarly, if $A_\zeta(k_0=45\,\Mpc^{-1})$ differs by $f_\zeta=A_\zeta/\pot{5}{-8}$, we have to rescale the error estimates by $f^{-1}_\zeta$.
We checked the predicted uncertainties for some representative cases using the MCMC method of \citet{Chluba2013fore}, finding excellent agreement.
Overall, our analysis shows that CMB SD measurement provide an unique probe of the small-scale power spectrum, which can be utilized to directly constraint inflationary models. Especially, if the small-scale power spectrum is close to scale-invariant with small running, very robust constraints can be expected from {\it PIXIE} and {\it PRISM}, if $A_\zeta(k_0=45\,\Mpc^{-1})\simeq 10^{-8}-10^{-7}$.

\begin{figure}
\centering
\hspace{-3mm}
\includegraphics[width=1.05\columnwidth]{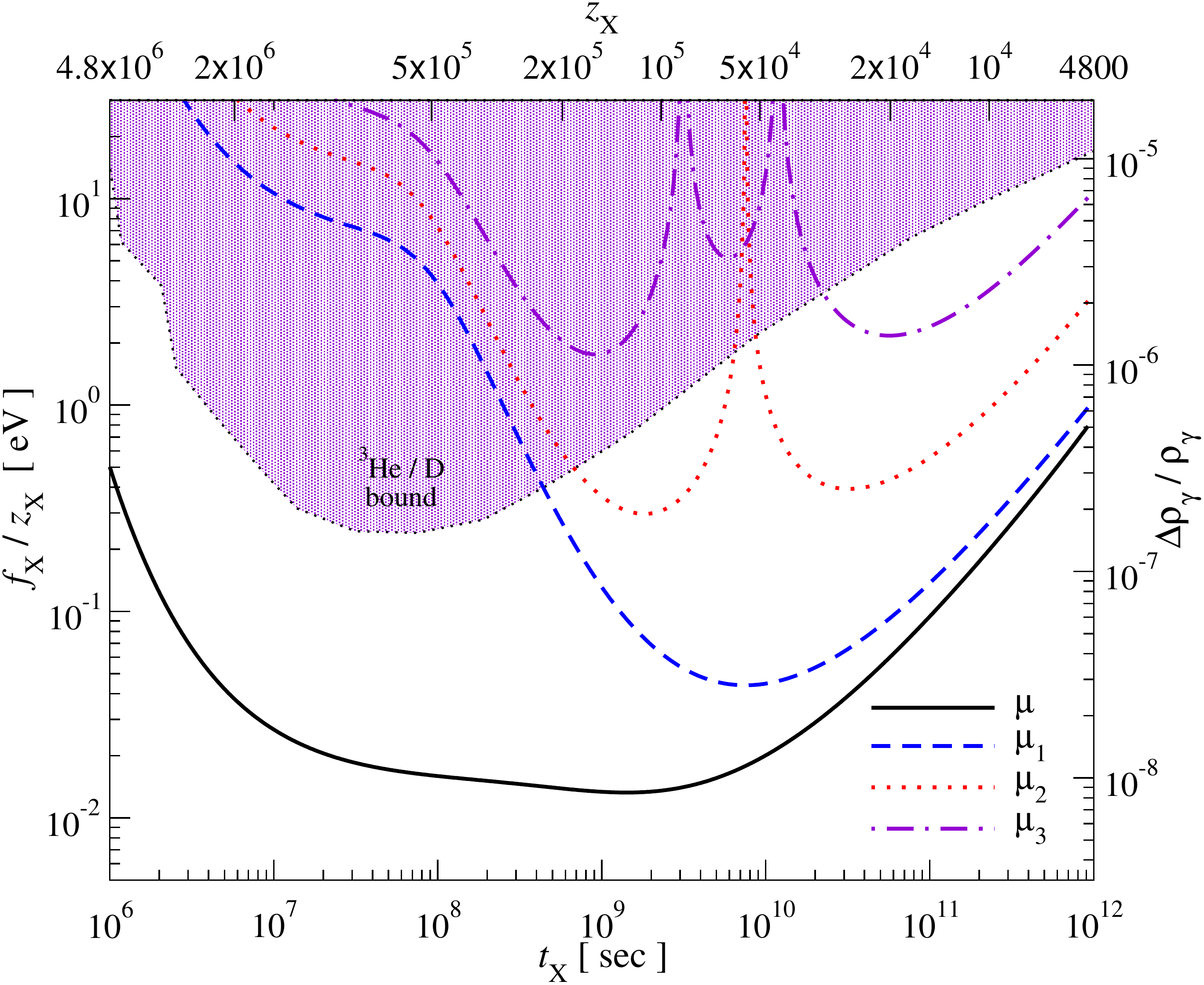}
\\[3mm]
\hspace{-2mm}
\includegraphics[width=0.99\columnwidth]{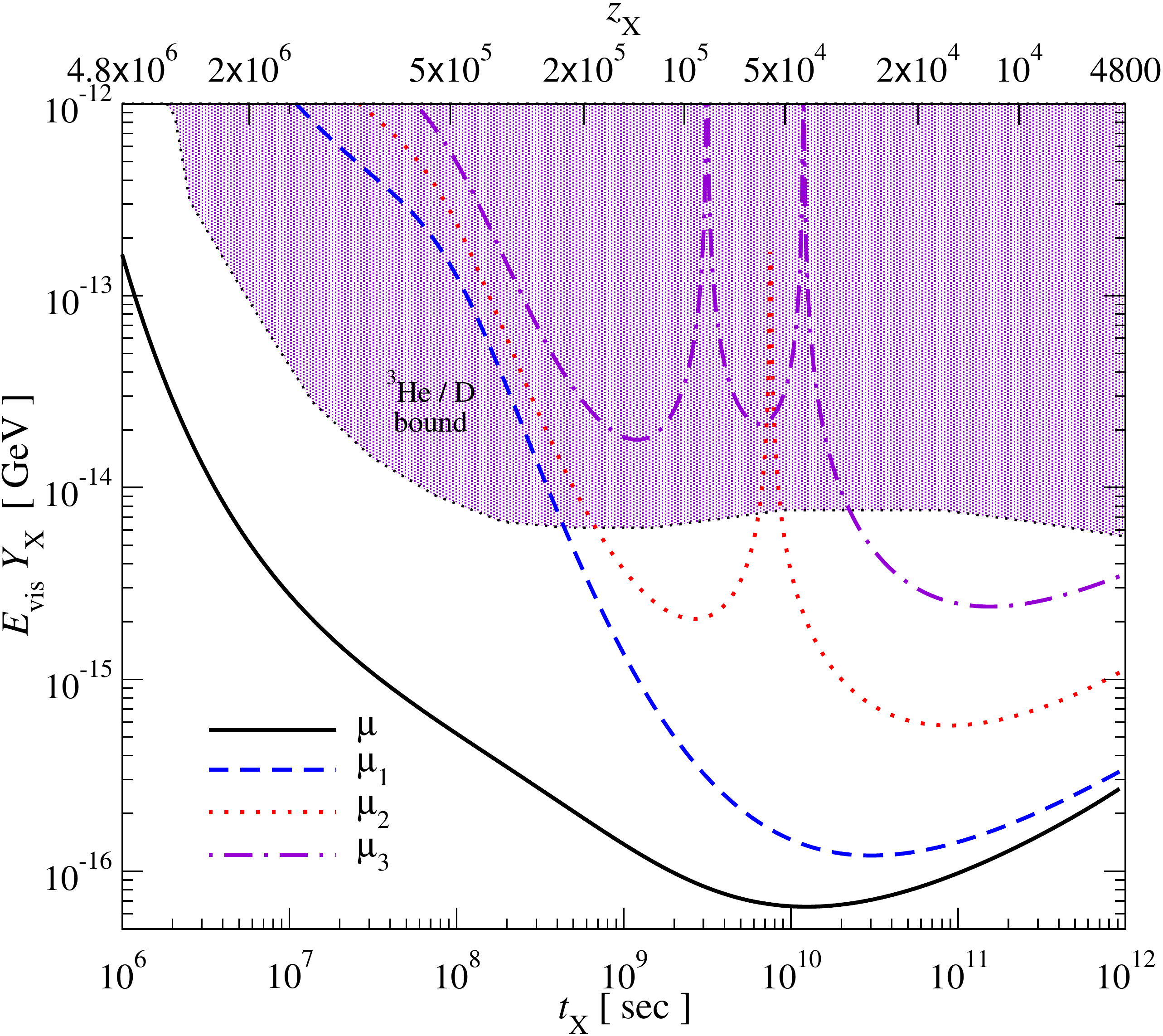}
\caption{Detectability of $\mu$, $\mu_1$, $\mu_2$ and $\mu_3$. The upper panel shows the limits for $\epsilon_{\rm X}=f_{\rm X}/z_{\rm X}$, while the lower panel uses the standard yield variable, $E_{\rm vis} Y_{\rm X}$ \citep[cp.,][]{Kawasaki2005}.
For a given particle lifetime, we compute the required value of $\epsilon_{\rm X}$ for which a $1\sigma$-detection of the corresponding variable is possible with {\it PIXIE}. The violet shaded area is excluded by measurements of the primordial $^3{\rm He}/{\rm D}$ abundance ratio \citep[$1\sigma$-level, adapted from Fig.~42 of][]{Kawasaki2005}.}
\label{fig:fX_limit}
\end{figure}

\subsubsection{Decaying relic particles}
The distortion signals for the three decaying particle scenarios presented in Table~\ref{tab:two} will all be detectable with a {\it PIXIE}-like experiment. 
More generally, Fig.~\ref{fig:fX_limit} shows the $1\sigma$-detection limits for $\mu$, $\mu_1$, $\mu_2$, and $\mu_3$, as a function of the particle lifetime. CMB SDs are sensitive to decaying particles with $\epsilon_{\rm X}=f_{\rm X}/z_{\rm X}$ as low as $\simeq 10^{-2}\,{\rm eV}$ for particle lifetimes $10^7\,{\rm sec}\lesssim t_{\rm X}\lesssim 10^{10}\,{\rm sec}$. 
For {\it PRISM} the detection limit will be as low as $\epsilon_{\rm X}\simeq 10^{-3}\,{\rm eV}$ in this range.
To directly constrain $t_{\rm X}$, at least a measurement of $\mu_1$ is needed. At {\it PIXIE} sensitivity this means that the lifetime of particles with $\pot{2}{9}\,{\rm sec}\lesssim t_{\rm X}\lesssim \pot{6}{10}\,{\rm sec}$ for $\epsilon_{\rm X}\gtrsim 0.1\,{\rm eV}$ and $\pot{3}{8}\,{\rm sec}\lesssim t_{\rm X}\lesssim {10}^{12}\,{\rm sec}$ for $\epsilon_{\rm X}\gtrsim 1\,{\rm eV}$ will be directly measurable.
Most of this parameter space is completely unconstrained [see upper limit from measurements of the primordial $^3{\rm He}/{\rm D}$ abundance ratio\footnote{In the particle physics community the abundance yield, $Y_{\rm X}=N_{\rm X}/S$, and deposited particle energy, $E_{\rm vis}$ [GeV], are commonly used. Here, $N_{\rm X}$ is the particle number density at $t\ll t_{\rm X}$ and $S=\frac{4}{3}\,\frac{\rho}{kT}\simeq 7\,N_\gamma\simeq \pot{2.9}{3}\,(1+z)^3\,{\rm cm^{-3}}$ denotes the total entropy density. We thus find $\epsilon_{\rm X}\equiv (E_{\rm vis}\,Y_{\rm X})\,10^9 S/[N_{\rm H}\,(1+z_{\rm X})]\simeq \pot{1.5}{19}(E_{\rm vis}\,Y_{\rm X})/(1+z_{\rm X})$.} \citep[from Fig.~42 of][]{Kawasaki2005} in Fig.~\ref{fig:fX_limit}].
Higher sensitivity will allow cutting deeper into the parameter space and widen the range over which the particle lifetime can be directly constrained.

\begin{figure}
\centering
\includegraphics[width=\columnwidth]{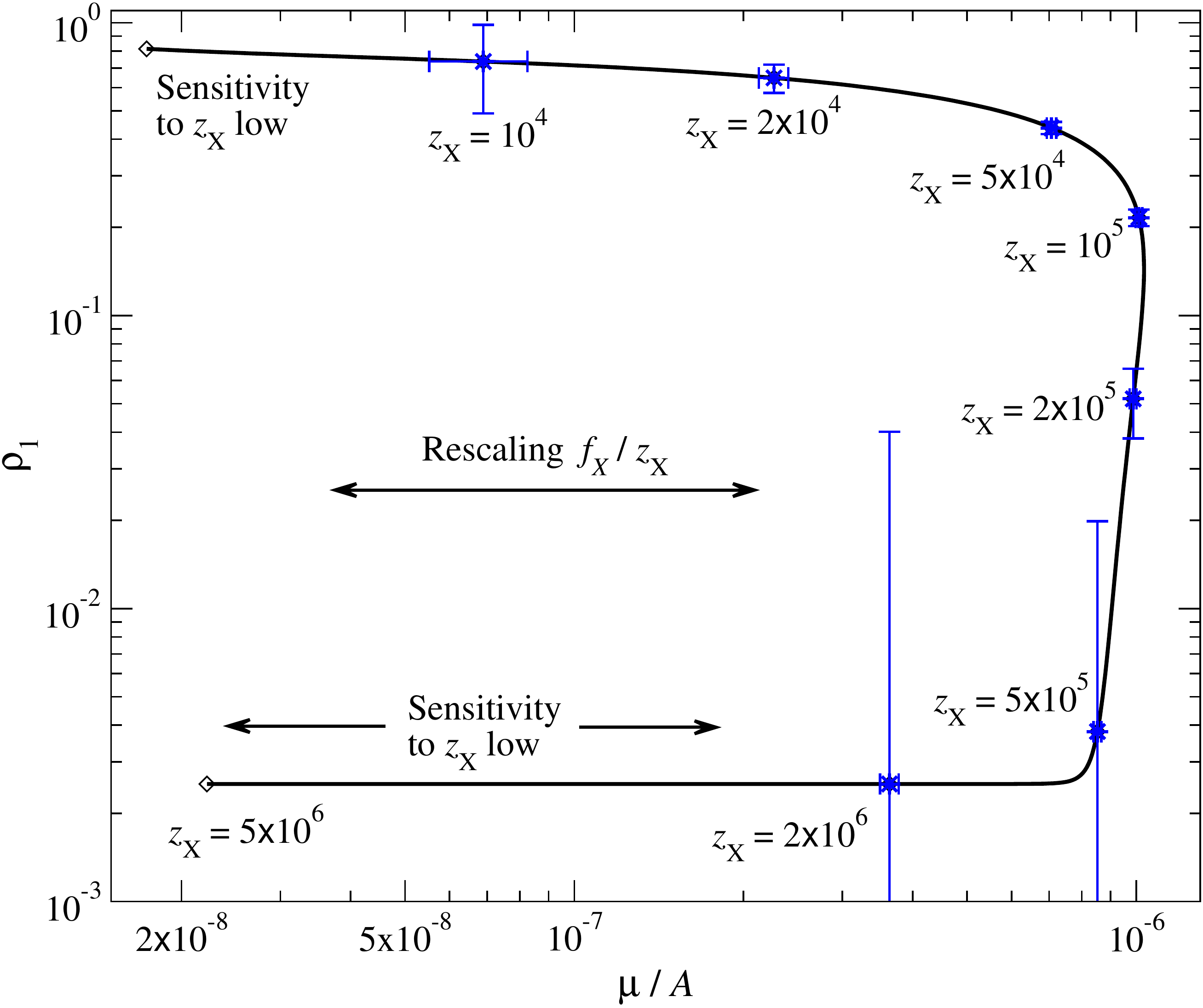}
\\[2mm]
\includegraphics[width=\columnwidth]{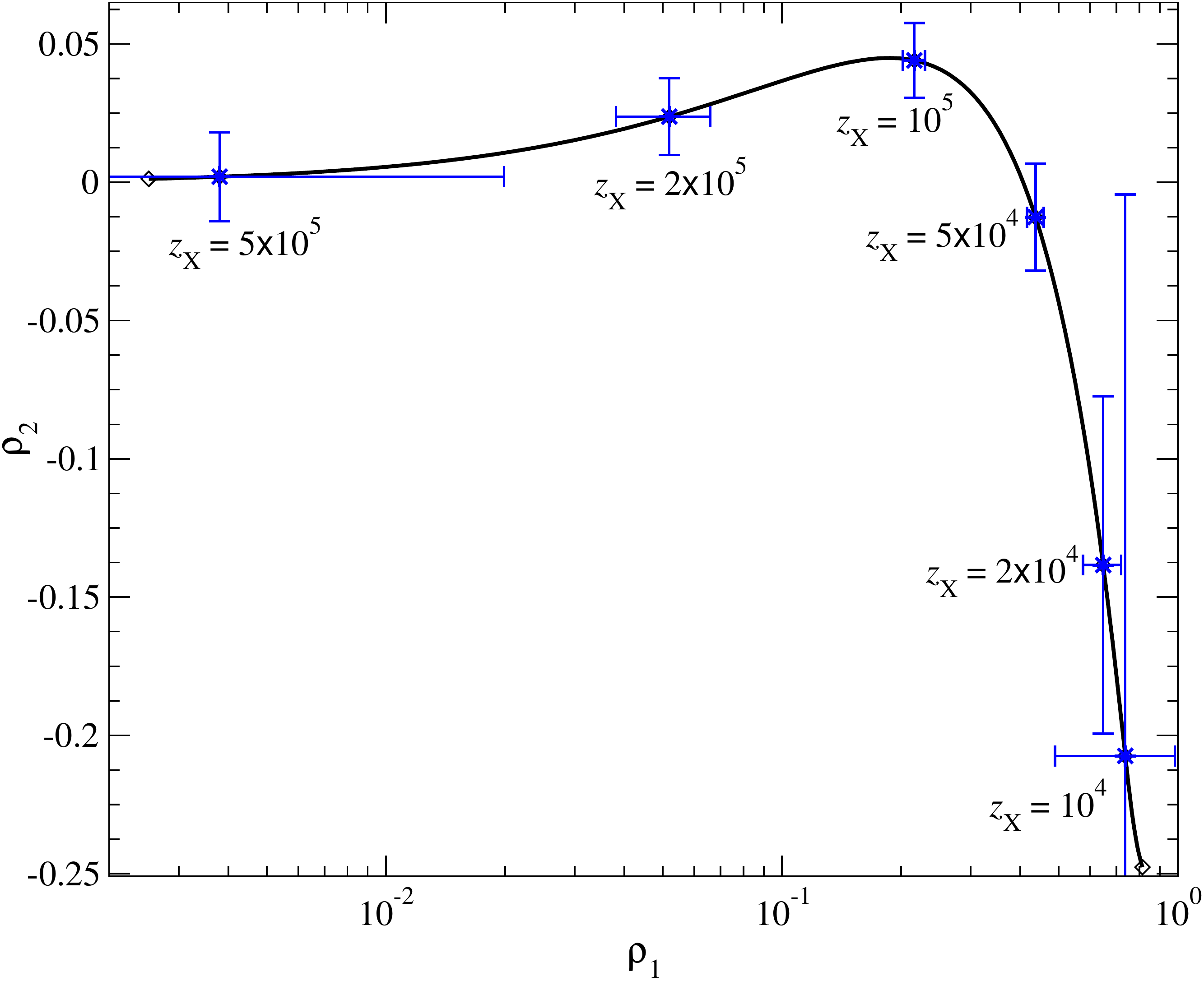}
\caption{Parameter range of $\mu$, $\mu_1$ and $\mu_2$ for decaying particle scenarios. We assumed {\it PIXIE} settings and sensitivity, and $\epsilon_{\rm X}=f_{\rm X}/z_{\rm X}=1\,{\rm eV}$ (i.e. $A\equiv \epsilon_{\rm X}/1\,{\rm eV}$). 
The blue symbols with error bars are for $z_{\rm X}$ as labeled.
Measurements in the $\mu-\rho_1$ plane can be used to constrain $z_{\rm X}$ with the most sensitive range around $z_{\rm X}\simeq \pot{5}{4}-10^5$. The $\rho_1-\rho_2$ plane can be used to further improve this measurement, but also for model comparison.}
\label{fig:Dec.mu_rho1}
\end{figure}

To illustrate this aspect even further, we can again study the $\mu-\rho_k$-parameter space covered by decaying particles. The projections into the $\mu-\rho_1$ and $\rho_1-\rho_2$-plane are shown in Fig.~\ref{fig:Dec.mu_rho1} for decay efficiency $\epsilon_{\rm X}=1\,{\rm eV}$ and {\it PIXIE} settings. Varying $\epsilon_{\rm X}$ would move the $\mu-\rho_1$ trajectory left or right, as indicated in the upper panel of Fig.~\ref{fig:Dec.mu_rho1}. Furthermore, all error bars of $\rho_k$ would have to be rescaled by $f=[\epsilon_{\rm X}/1\,{\rm eV}]^{-1}$ under this transformation. 
Measuring $\mu$ and $\rho_1$ is in principle sufficient for independent determination of $\epsilon_{\rm X}$ and the particle lifetime, $t_{\rm X}\approx[\pot{4.9}{9}/(1+z_{\rm X})]^2\,{\rm sec}$, with most sensitivity around $z_{\rm X}\simeq \pot{5}{4}-10^5$ or $t_{\rm X}\simeq \pot{2}{9}-10^{10}\,{\rm sec}$ for the shown scenario. For shorter lifetime, the SD signal is very close to a pure $\mu$-distortion, with little information in the residual ($\rho_1$ and $\rho_2$ are both very small and also show very little variation with redshift). Similarly, for longer lifetimes the particle signature is close to a $y$-distortion. In both cases the sensitivity to the lifetime is very weak and only an overall integral constraint can be derived, with large degeneracy between $\epsilon_{\rm X}$ and $z_{\rm X}$ \citep[see discussion in][]{Chluba2013fore}.

\begin{figure}
\centering
\includegraphics[width=\columnwidth]{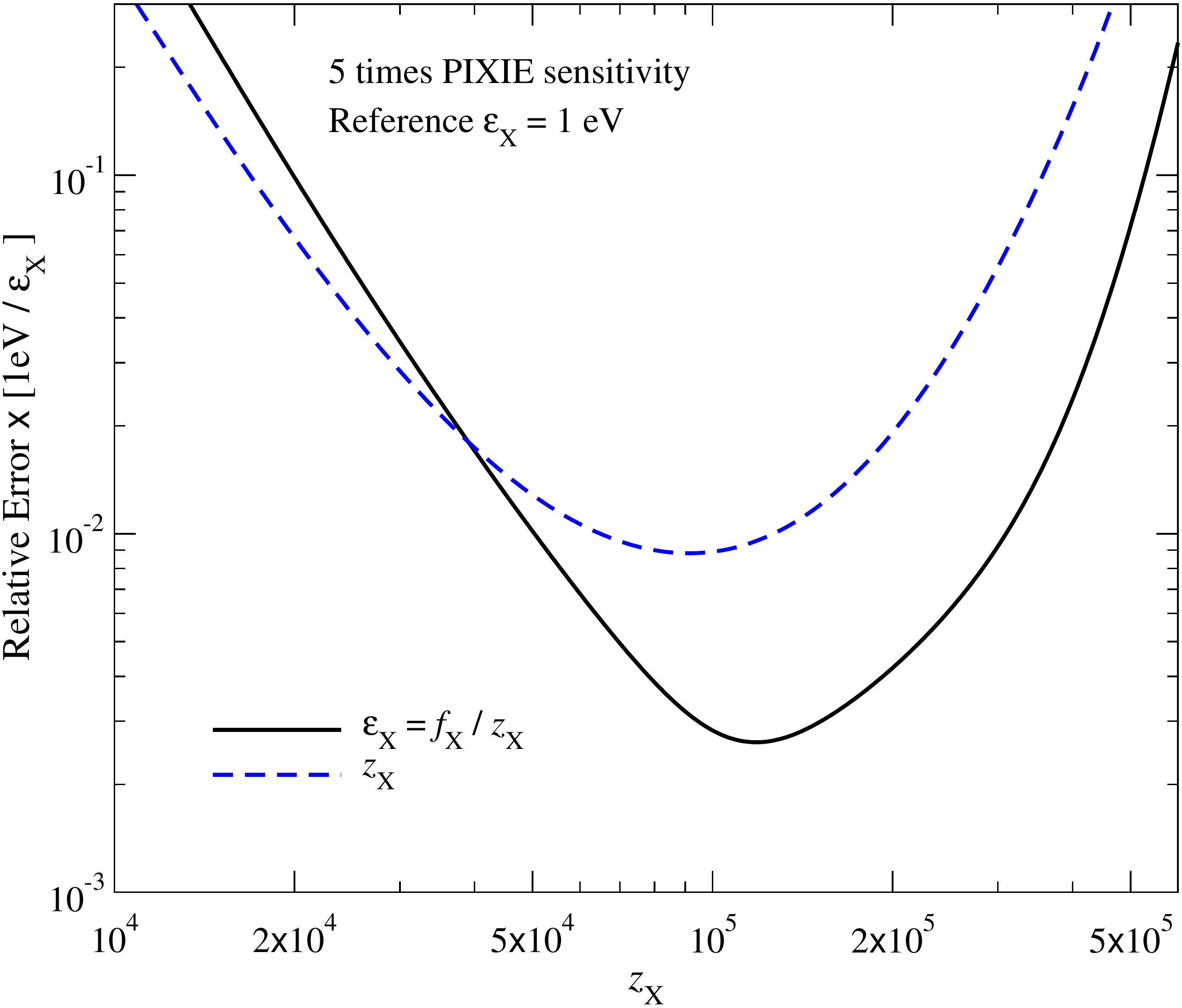}
\caption{Relative error for determination of $\epsilon_{\rm X}=f_{\rm X}/z_{\rm X}$ and $z_{\rm X}$ using measurements of $\mu$ to $\mu_2$. We assumed 5 times the sensitivity of {\it PIXIE} and $\epsilon_{\rm X}=1\,{\rm eV}$ as reference value (other cases can be estimated by simple rescaling). 
The corresponding error in the particle lifetime is $\Delta t_{\rm X}/t_{\rm X}\simeq 2 \Delta z_{\rm X}/z_{\rm X}$.}
\label{fig:Dec.error}
\end{figure}

We can again estimate the expected $1\sigma$-errors on $\epsilon_{\rm X}$ and $z_{\rm X}$ around the fiducial value using the Fisher information matrix, $\mathcal{F}_{ij}=\Delta\mu^{-2}\,\partial_{p_i} \mu\,\partial_{p_j} \mu+\sum_k \Delta\mu_k^{-2}\,\partial_{p_i} \mu_k\partial_{p_j} \mu_k$, with parameters $p\equiv \{\epsilon_{\rm X}, z_{\rm X}\}$.
In Fig.~\ref{fig:Dec.error} we show the corresponding Fisher-forecasts assuming {\it PIXIE}-setting but with 5 times its sensitivity. We included information from $\mu$, $\mu_1$ and $\mu_2$, because adding $\mu_3$ did not change the forecast significantly.
For $\pot{1.7}{4}\lesssim z_{\rm X}\lesssim \pot{3.5}{5} \, (\pot{2}{9}\,{\rm sec}\lesssim t_{\rm X}\lesssim \pot{8.3}{10}\,{\rm sec})$, the particle lifetime can be constrained to better than $\simeq 20\%$ and $\epsilon_{\rm X}$ can be measured with uncertainty $\lesssim 10\%$ .
These findings are in good agreement with those of \citet{Chluba2013fore}, where direct MCMC simulations were performed.
CMB SD are thus a powerful probe of early-universe particle physics, providing tight constraints that are independent and complementary to those derived from light element abundances \citep[e.g.,][]{Kawasaki2005, Kohri2010, Pospelov2010}.

We emphasize that the CMB spectrum can be utilized to directly probe the particle lifetime, a measurement that cannot be obtained by other means. CMB SDs furthermore provide a calorimetric constraint, which is sensitive to the total heat that is generated in the decay process. For very light relic particles (mass smaller than a few MeV), measurements of light element abundances will not allow placing constraints, while the CMB spectrum should still be sensitive, assuming that the particle is abundant enough.

We also mention, that the Fisher estimates become crude, once the error reaches much more than $\simeq 15\%-20\%$. In this case, the likelihood becomes non-Gaussian and non-linear dependences are important. We also find that the solutions can be multi modal, with regions of low probability far away from the fiducial value. This means that MCMC sampling has to be performed in several steps, using wide priors to find regions of interest, followed by re-simulations around different maximum likelihood points. In this case, we refer to the methods developed in \citet{Chluba2013fore}.

\begin{figure}
\centering
\includegraphics[width=0.98\columnwidth]{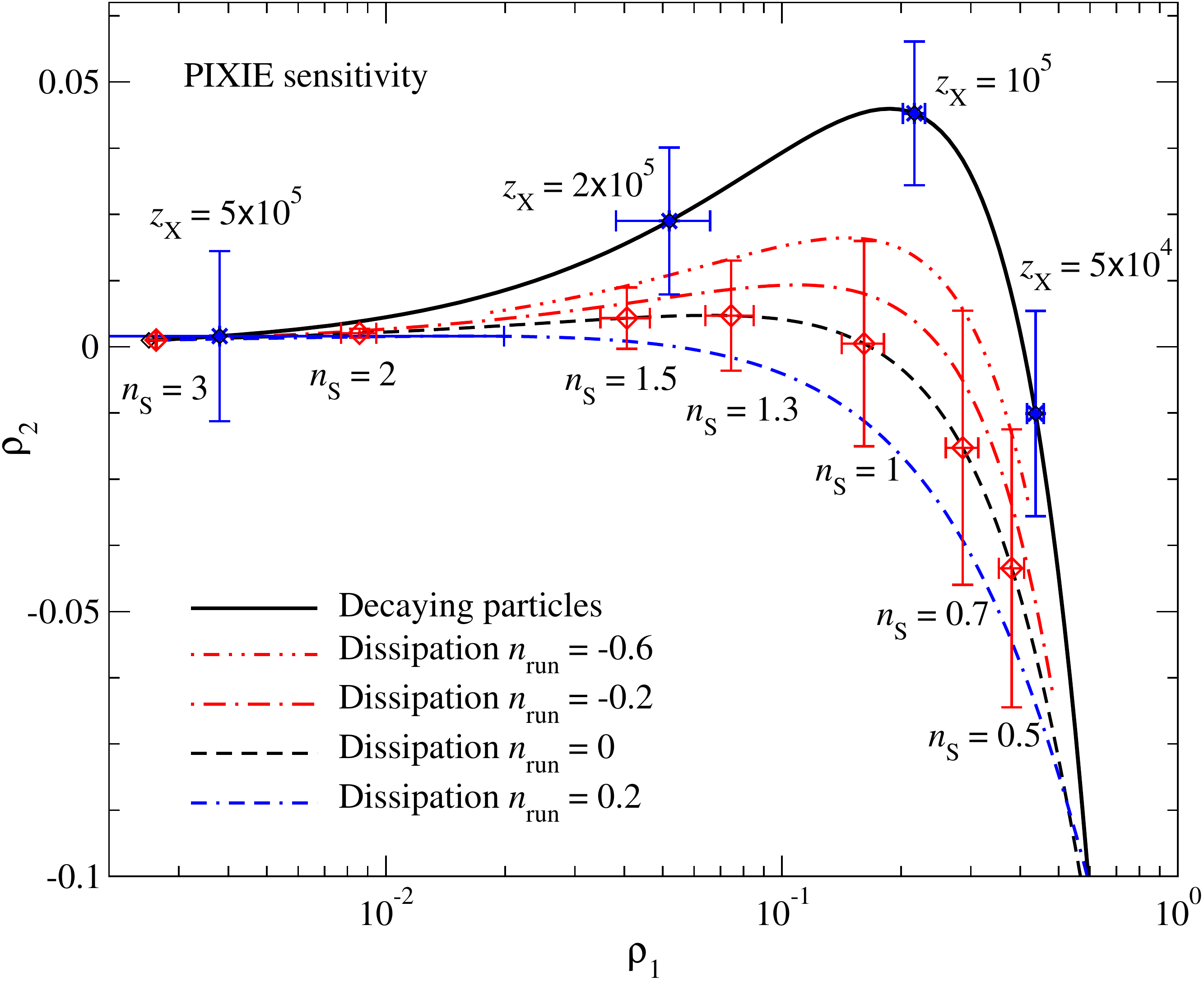}
\\[2mm]
\includegraphics[width=0.98\columnwidth]{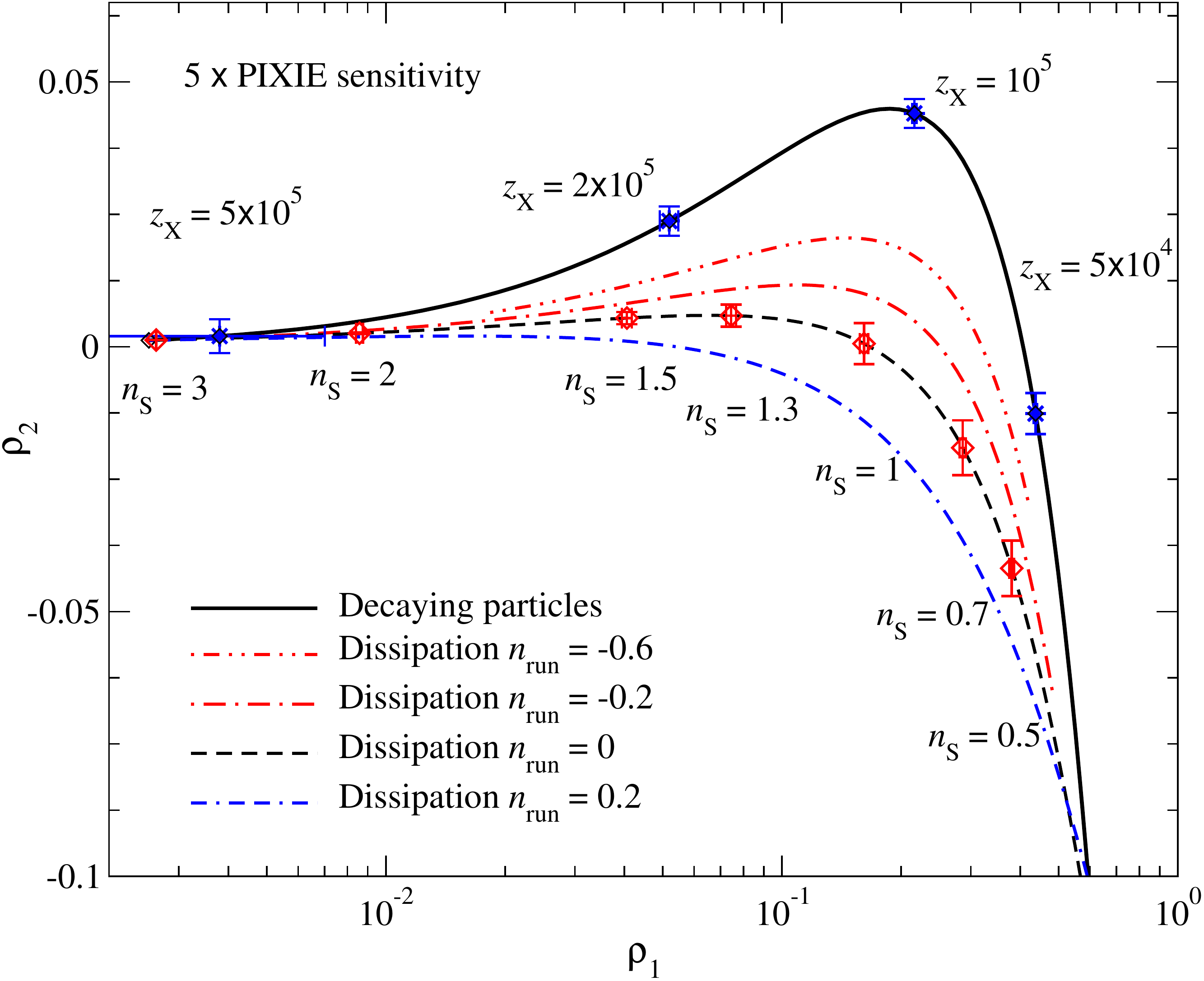}
\caption{Model comparison for dissipation and decaying particle scenarios in the $\rho_1-\rho_2$ plane. We assumed $A_\zeta(k_0=45\,\Mpc^{-1})=\pot{5}{-8}$ and $\epsilon_{\rm X}=1\,{\rm eV}$. The upper panel is for {\it PIXIE} sensitivity, while the lower is for 5 times higher sensitivity.}
\label{fig:Dec_Diss.rho1_rho_2}
\end{figure}

\subsection{Comparing models using distortion eigenmodes}
\label{sec:model_comparison}
In the previous section, we presented parameter estimation cases using eigenmodes in a model-by-model basis. Furthermore, since each model has rather unique predictions for the observable $\mu_k$s, the eigenmode
analysis opens a new possibility of distinguishing different ERSs. In this section, we shall illustrate this point by some solid examples. First, let us assume that the time dependence of the energy release is fixed. In that case, the shape of eigenspectrum does not change and only the overall amplitude is free, and we can directly use the examples given in Table~\ref{tab:two}. 
If only $\mu$ can be constrained then different models cannot be distinguished unless some other constraint can be invoked. For example, finding $\mu\simeq 10^{-7}$ is unlikely to be caused by s-wave annihilation, which is bound to much smaller annihilation efficiencies by CMB anisotropy measurements. It could, however, be caused by a decaying relic particle or the dissipation of small-scale perturbations.

Once some of the $\mu_k$, which directly probe the time dependence of the energy-release history, can be determined with signal-to-noise $S/N>1$, one can in principle distinguish between different scenarios.
For instance, from Table~\ref{tab:two} the dissipation scenario with $(\nS, \nrun)=(1,0)$ has a $\rho$-vector $\rho_{\rm diss}\approx (0.161, \pot{5.86}{-4}, \pot{4.31}{-3})$, while for the s-wave annihilation scenarios we find $\rho_{\rm ann, s}\approx (0.159, \pot{1.18}{-3}, \pot{4.17}{-3})$. Comparing the entries of these vectors indicates that the two cases are quasi-degenerate. The small differences stem from the late-time behavior of $\mathcal{Q}(z)$ at $z\lesssim 10^4$, but very high precision is indeed needed to discern them. In addition, by slightly adjusting $\nS$ to $\simeq 1.01$ one can align these two $\rho$-vectors nearly perfectly.
For the p-wave scenario, we find $\rho_{\rm ann, p}\approx (\pot{2.33}{-2}, \pot{4.37}{-3}, \pot{1.38}{-3})$, which clearly is different from the s-wave annihilation and scale-invariant dissipation scenarios. However, adjusting $\nS\simeq 1.67$ practically aligns $\rho_{\rm diss}$ with $\rho_{\rm ann, p}$. This is expected since for $\nrun=0$ one has $\mathcal{Q}_{\rm ac}\propto (1+z)^{3(\nS-1)/2}$ \citep[e.g.,][]{Chluba2012}, which becomes $\mathcal{Q}_{\rm ac}\propto (1+z)$ for $\nS\simeq 5/3$.
Similarly, we find $\rho_{\rm diss}\approx (0.235, -\pot{8.07}{-3}, \pot{6.25}{-3})$ for $(\nS, \nrun)=(0.96, -0.02)$, which is distinguishable from the s- and p-wave annihilation scenario, if $\rho_1$ could be measured with $\simeq 10\%$ precision. However, as soon as the dissipation model parameters are varied, degeneracies reappear, unless even higher experimental precision is achieved. Therefore, s-wave annihilation scenarios and quasi-scale-invariant dissipation scenarios with small running are observationally hard to distinguish using CMB SD data. Similarly, p-wave and dissipation scenarios with $\nS\simeq 1.67$ and small running are degenerate. For values of $\nS\neq \{1, 1.67\}$ the degeneracies with annihilation scenarios are less severe and at high spectral sensitivity they in principle can be discerned.

For comparison of dissipation and decaying particle scenarios, let us consider the general case with all model parameters varying. As mentioned above, if only $\mu$ is measurable, no distinction can be made, unless priors are used (e.g., we assume that the primordial small-scale power spectrum is determined by extrapolation from large scales but find $\mu\simeq 10^{-7}$, which cannot be explained in this case). Assuming that $\mu$ and $\mu_1$ are measurable, by comparing the upper panels of Figs.~\ref{fig:Diss_mu_rho_1_rho_2} and \ref{fig:Dec.mu_rho1}, it is evident that due to freedom in the overall amplitude ($A_\zeta$ and $\varepsilon_{\rm X}$ can be re-scaled), dissipation and decaying particle scenarios again cannot be distinguish in a model-independent way (moving the curves left and right one can make then coincide). 

The situation changes when also $\mu_2$ can be measured. Figure~\ref{fig:Dec_Diss.rho1_rho_2} shows the trajectories of dissipation and decaying particle scenarios in the $\rho_1-\rho_2$-plane for two spectral sensitivities. Assuming that the small-scale power spectrum is quasi-scale-invariant with small running a {\it PIXIE}-type experiment will already be able to directly distinguish this from a decaying particle scenario. Allowing large negative running does increase the degeneracy and higher spectral sensitivity is needed to discern these cases. The lower panel illustrates the improvement for 5 times the sensitivity of {\it PIXIE}. Clearly, measurements of $\mu_1$ and $\mu_2$ allow discerning dissipation and decaying particle scenarios over a wide range of the parameter space, with degeneracies appearing for large negative running ($\nrun\ll -0.6$) and in the limit of large and small spectral index.

We mention, however, that when allowing more complex shapes of the small-scale power spectrum, e.g., with bumps caused by particle production during inflation \citep{2000PhRvD..62d3508C, Neil2009, Barnaby2010}, closer resemblance of the energy-release history with the one of a decaying particle can be achieved. In this case, a distinction of the two scenarios will be more challenging. Also, a combination of decaying particle and dissipation scenarios could be possible but would be hard to distinguish from the single scenarios. Nevertheless, CMB SD measurements provide a unique way to study different ERSs allowing direct model comparisons and distinction in certain situations.

\section{Conclusions}
\label{sec:conclusions}
In this work, we derived a decomposition of the CMB SD signal into temperature shift, $y$, $\mu$ and residual distortion. The residual distortion was defined to be orthogonal to the temperature shift, $y$- and $\mu$-distortion, taking experimental settings into account.
Using this decomposition, we can explicitly show how much energy, at a given instance, is transferred to the various components of the CMB spectrum (Fig.~\ref{fig:J_z}). 
The $y$-distortion part of the CMB spectrum cannot be used in a model-independent way to learn about the primordial energy-release (occurring at $z\gtrsim 10^3$), since it is degenerate with $y$-distortions introduced at later times, by reionization and the formation of structures.
The $\mu$-distortion component only provides a measure for the overall (integrated) energy release at $z\simeq \pot{\rm few}{4}-\pot{\rm few}{6}$, which can again only be interpreted in a model-dependent way.
Adding the information in the residual distortion allows us to directly constraint the time dependence of the energy-release history, and thus provides a way to discern different scenarios \citep[see also,][]{Chluba2011therm,Chluba2013fore}.

We took a step forward towards the analysis of future CMB SD data. The information contained in the residual distortion can be compressed into a few numbers. This compression is achieved by performing a principal component analysis to determine residual-distortion and energy-release eigenmodes (see Sect.~\ref{sec:R_eigenmodes}).
It introduces a new set of distortion parameters, $\mu_k$, which parametrize the shape of the residual distortion.
We demonstrated that the eigenmodes can be used to simplify the analysis of future distortion data, providing a model-independent way to extract all useful information from the average CMB spectrum. 
Using this method we discussed annihilating and decaying particle scenarios, as well as energy release caused by the dissipation of small-scale acoustic modes (corresponding to wave numbers $1\,\Mpc^{-1}\lesssim k\lesssim \pot{\rm few}{4}\,\Mpc^{-1}$) for different experimental sensitivities. 
We showed that future CMB SD measurements will allow direct detection of s-wave annihilation signals if the annihilation efficiency is $p_{\rm ann, s}\gtrsim \pot{4.6}{-7} \,{\rm m^3 \, kg^{-1} \,s^{-2}}$ using {\it PRISM}. Detection limits for dissipation and decaying particle scenarios are shown in Figs.~\ref{fig:Limit_nS_nrun} and \ref{fig:fX_limit}, respectively.

CMB SD measurement provide a unique probe of the primordial small-scale power spectrum, which can be utilized to directly constraint inflationary models. Especially, if the small-scale power spectrum is close to scale-invariant with small running, very robust constraints can be expected from {\it PIXIE} and {\it PRISM}, if the amplitude of curvature perturbations is $A_\zeta(k_0=45\,\Mpc^{-1})\simeq 10^{-8}-10^{-7}$. These conclusions are in good agreement with those of \citet{Chluba2013fore} where an MCMC analysis was used.

For decaying particle models with $\pot{2}{9}\,{\rm sec}\lesssim t_{\rm X}\lesssim \pot{8.3}{10}\,{\rm sec}$ and total energy release $\Delta\rho_\gamma/\rho_\gamma\simeq \pot{6.4}{-7}$, the particle lifetime can be constrained to better than $\simeq 20\%$ and $\epsilon_{\rm X}$ could be measured with uncertainty $\lesssim 10\%$ using a {\it PIXIE}-type experiment with 5 times its sensitivity  (see Fig.~\ref{fig:Dec.mu_rho1} for details).
These findings are in good agreement with those of \citet{Chluba2013fore}, where direct MCMC simulations were performed.
CMB SD are thus a powerful probe of early-universe particle physics, providing tight limits that are independent and complementary to those derived from light element abundances \citep[e.g.,][]{Kawasaki2005, Kohri2010, Pospelov2010} and the CMB anisotropies \citep{Chen2004, Zhang2007, Giesen2012}.

Finally, we demonstrated how the eigenmode decomposition of the residual distortion can be used for direct model comparison. The dissipation caused by a quasi-scale-invariant power spectrum gives rise to a distortion signature that is degenerate with a s-wave annihilation scenario \citep[see also,][]{Chluba2011therm, Chluba2013fore}. However, a combination of future CMB anisotropy constraints with CMB SD measurements might provide the means to disentangle these cases. In particular, a detection of an annihilating particle signature with CMB anisotropy measurements could be independently confirmed using CMB SDs.
Furthermore, decaying particle scenarios have distortion eigenspectra that are distinct from the one caused by the dissipation of small-scale acoustic modes, if the power spectrum is neither too blue nor too red and does not show too much negative running (see Fig.~\ref{fig:Dec_Diss.rho1_rho_2}).
This again demonstrates the potential of future CMB distortion measurements and we look forward to extending our method to include more realistic instrumental effects and foregrounds. 
The principal component decomposition can furthermore be used to determine the optimal experimental settings for the detection of different SD signatures, another application we plan for the future.

\small

\section*{Acknowledgments}
The authors specially thank Yacine Ali-Ha{\"i}moud and the referee for insightful comments and suggestions.  
JC furthermore thanks Kazunori Kohri and Josef Pradler for useful discussions about particle physics scenarios. 
The authors also thank Rishi Khatri and Rashid Sunyaev for comments on the manuscript,
and Silvia Galli for providing simulated covariance matrices for {\it PRISM}.
Use of the GPC supercomputer at the SciNet HPC Consortium is acknowledged. SciNet is funded by: the Canada Foundation for Innovation under the auspices of Compute Canada; the Government of Ontario; Ontario Research Fund - Research Excellence; and the University of Toronto. This work was supported by DoE SC-0008108 and NASA NNX12AE86G. 

\small 
\begin{appendix}
\section{Orthogonal basis}
\label{app:basis}
To define the residual distortion, $\vek{R}(z)$, that is perpendicular to the space spanned by $\vek{G}_{\rm T}$, $\vek{Y}_{\rm SZ}$, and $\vek{M}$, we simply follow the Gram-Schmidt orthogonalization procedure.
Aligning one axis with $\vek{Y}_{\rm SZ}$, this space is given by the orthonormal basis $\vek{e}_{y}= \vek{Y}_{\rm SZ}/ |\vek{Y}_{\rm SZ}|$, $\vek{e}_{\mu}=\vek{M}_\perp/|\vek{M}_\perp|$, and $\vek{e}_{\rm T}=\vek{G}_{\rm T, \perp}/|\vek{G}_{\rm T, \perp}|$, where $\vek{M}_\perp= \vek{M}-M_y\,\vek{e}_{y}$ and $\vek{G}_{\rm T, \perp} =\vek{G}_{\rm T}-G_y \,\vek{e}_y-G_\mu \,\vek{e}_{\mu}$. With $\vek{a}\cdot\vek{b}=\sum_i a_i\,b_i$, the required projections are $M_y=\vek{e}_{y}\cdot \vek{M}$, $G_y=\vek{e}_{y}\cdot\vek{G}_{\rm T}$, and $G_\mu=\vek{e}_{\mu}\cdot\vek{G}_{\rm T}$.
Assuming {\it PIXIE}-type settings ($\{\nu_{\rm min}, \nu_{\rm max},\Delta\nu_{\rm s}\}=\{30, 1000, 15\}\,{\rm GHz}$), we find $\{|\vek{Y}_{\rm SZ}|, |\vek{M}_\perp|, |\vek{G}_{\rm T, \perp}|\}\simeq\{73.3, 7.99, 21.4\}\,10^{-18}\,\UnitInu$ and $\{M_y, G_y, G_\mu\}\simeq\{7.66, 16.8, 41.5\}\,10^{-18}\,\UnitInu$. From Eq.~\eqref{eq:G_approx} we furthermore obtain
\beal
\label{eq:R_Jfuncs}
\vek{R}(z)&=\vek{G}_{\rm th}(z)
- \vek{G}_{\rm T}\,\mathcal{J}_{T}(z)/4
-  \vek{Y}_{\rm SZ} \,\mathcal{J}_{y}(z)/4 
- \alpha\,\vek{M} \,\mathcal{J}_{\mu}(z)
\nonumber\\
\mathcal{J}_{T}(z) &= 4 \,\vek{e}_{\rm T}\cdot\vek{G}_{\rm th}(z)/|\vek{G}_{\rm T, \perp}|
\nonumber\\
\mathcal{J}_{\mu}(z) &= 
\alpha^{-1}\,[\vek{e}_{\mu}\cdot\vek{G}_{\rm th}(z)-G_\mu\,\mathcal{J}_{T}(z)/4]/|\vek{M}_{\perp}|
\nonumber\\
\mathcal{J}_{y}(z) &= 
4\,[\vek{e}_y\cdot\vek{G}_{\rm th}(z)-\alpha\,M_y\,\mathcal{J}_{\mu}(z) -G_y\,\mathcal{J}_{T}(z)/4]/|\vek{Y}_{\rm SZ}|
\nonumber\\
\mathcal{J}_{R}(z) &=1-\mathcal{J}_{T}(z)-\mathcal{J}_{y}(z) - \mathcal{J}_{\mu}(z),
\end{align}
where we also introduced $\mathcal{J}_{R}(z)$, which determines the amount of energy found in the residual distortion. All $\mathcal{J}_{k}(z)$ are illustrated in Fig.~\ref{fig:J_z}.

\end{appendix}

\bibliographystyle{mn2e}
\bibliography{Lit}

\end{document}